\documentclass[longauth, english, twocolumn]{aa}
\usepackage[varg]{txfonts}
\usepackage{graphicx}
\usepackage{mathrsfs}
\usepackage{amsmath}
\usepackage{txfonts}
\usepackage{enumerate}
\usepackage{caption}
\usepackage{subcaption}
\usepackage{hyperref}
\usepackage{bm}
\usepackage{babel}
\usepackage[switch]{lineno}

 \usepackage{natbib}
\bibpunct{(}{)}{;}{a}{}{,} 
\graphicspath{{./}}
\usepackage{etoolbox}
\makeatletter
\patchcmd\@combinedblfloats{\box\@outputbox}{\unvbox\@outputbox}{}{%
   }
 \makeatother

\begin{document} 

\author{\small 
H.~Abdalla \inst{\ref{UNAM}}
\and F.~Aharonian \inst{\ref{DIAS},\ref{MPIK},\ref{RAU}}
\and F.~Ait~Benkhali \inst{\ref{MPIK}}
\and E.O.~Ang\"uner \inst{\ref{CPPM}}
\and C.~Arcaro \inst{\ref{NWU}}
\and C.~Armand \inst{\ref{LAPP}}
\and T.~Armstrong \inst{\ref{Oxford}}
\and H.~Ashkar \inst{\ref{IRFU}}
\and M.~Backes \inst{\ref{UNAM},\ref{NWU}}
\and V.~Baghmanyan \inst{\ref{IFJPAN}}
\and V.~Barbosa~Martins \inst{\ref{DESY}}
\and A.~Barnacka \inst{\ref{UJK}}
\and M.~Barnard \inst{\ref{NWU}}
\and Y.~Becherini \inst{\ref{Linnaeus}}
\and D.~Berge \inst{\ref{DESY}}
\and K.~Bernl\"ohr \inst{\ref{MPIK}}
\and B.~Bi \inst{\ref{IAAT}}
\and M.~B\"ottcher \inst{\ref{NWU}}
\and C.~Boisson \inst{\ref{LUTH}}
\and J.~Bolmont \inst{\ref{LPNHE}}
\and M.~de~Bony~de~Lavergne \inst{\ref{LAPP}}
\and M.~Breuhaus \inst{\ref{MPIK}}
\and F.~Brun \inst{\ref{IRFU}}
\and P.~Brun \inst{\ref{IRFU}}
\and M.~Bryan \inst{\ref{GRAPPA}}
\and M.~B\"{u}chele \inst{\ref{ECAP}}
\and T.~Bulik \inst{\ref{UWarsaw}}
\and T.~Bylund \inst{\ref{Linnaeus}}
\and S.~Caroff \inst{\ref{LAPP}}
\and A.~Carosi \inst{\ref{LAPP}}
\and S.~Casanova \inst{\ref{IFJPAN},\ref{MPIK}}
\and T.~Chand \inst{\ref{NWU}}
\and S.~Chandra \inst{\ref{NWU}}
\and A.~Chen \inst{\ref{WITS}}
\and G.~Cotter \inst{\ref{Oxford}}
\and M.~Cury{\l}o \inst{\ref{UWarsaw}}
\and J.~Damascene~Mbarubucyeye \inst{\ref{DESY}}
\and I.D.~Davids \inst{\ref{UNAM}}
\and J.~Davies \inst{\ref{Oxford}}
\and C.~Deil \inst{\ref{MPIK}}
\and J.~Devin \inst{\ref{APC}}
\and L.~Dirson \inst{\ref{HH}}
\and A.~Djannati-Ata\"i \inst{\ref{APC}}
\and A.~Dmytriiev \inst{\ref{LUTH}}
\and A.~Donath \inst{\ref{MPIK}}
\and V.~Doroshenko \inst{\ref{IAAT}}
\and L.~Dreyer  \inst{\ref{NWU}}
\and C.~Duffy \inst{\ref{Leicester}}
\and J.~Dyks \inst{\ref{NCAC}}
\and K.~Egberts \inst{\ref{UP}}
\and F.~Eichhorn \inst{\ref{ECAP}}
\and S.~Einecke \inst{\ref{Adelaide}}
\and G.~Emery \inst{\ref{LPNHE}}
\and J.-P.~Ernenwein \inst{\ref{CPPM}}
\and K.~Feijen \inst{\ref{Adelaide}}
\and S.~Fegan \inst{\ref{LLR}}
\and A.~Fiasson \inst{\ref{LAPP}}
\and G.~Fichet~de~Clairfontaine \inst{\ref{LUTH}}
\and G.~Fontaine \inst{\ref{LLR}}
\and S.~Funk \inst{\ref{ECAP}}
\and M.~F\"u{\ss}ling \inst{\ref{DESY}}
\and S.~Gabici \inst{\ref{APC}}
\and Y.A.~Gallant \inst{\ref{LUPM}}
\and G.~Giavitto \inst{\ref{DESY}}
\and L.~Giunti \inst{\ref{APC}, \ref{IRFU}} {\large *}
\and D.~Glawion \inst{\ref{ECAP}}
\and J.F.~Glicenstein \inst{\ref{IRFU}}
\and M.-H.~Grondin \inst{\ref{CENBG}}
\and J.~Hahn \inst{\ref{MPIK}}
\and M.~Haupt \inst{\ref{DESY}}
\and G.~Hermann \inst{\ref{MPIK}}
\and J.A.~Hinton \inst{\ref{MPIK}}
\and W.~Hofmann \inst{\ref{MPIK}}
\and C.~Hoischen \inst{\ref{UP}}
\and T.~L.~Holch \inst{\ref{DESY}}
\and M.~Holler \inst{\ref{LFUI}}
\and M.~H\"{o}rbe \inst{\ref{Oxford}}
\and D.~Horns \inst{\ref{HH}}
\and D.~Huber \inst{\ref{LFUI}}
\and M.~Jamrozy \inst{\ref{UJK}}
\and D.~Jankowsky \inst{\ref{ECAP}}
\and F.~Jankowsky \inst{\ref{LSW}}
\and A.~Jardin-Blicq \inst{\ref{MPIK}}
\and V.~Joshi \inst{\ref{ECAP}}
\and I.~Jung-Richardt \inst{\ref{ECAP}}
\and E.~Kasai \inst{\ref{UNAM}}
\and M.A.~Kastendieck \inst{\ref{HH}}
\and K.~Katarzy{\'n}ski \inst{\ref{NCUT}}
\and U.~Katz \inst{\ref{ECAP}}
\and D.~Khangulyan \inst{\ref{Rikkyo}}
\and B.~Kh\'elifi \inst{\ref{APC}} {\large *}
\and S.~Klepser \inst{\ref{DESY}}
\and W.~Klu\'{z}niak \inst{\ref{NCAC}}
\and Nu.~Komin \inst{\ref{WITS}}
\and R.~Konno \inst{\ref{DESY}}
\and K.~Kosack \inst{\ref{IRFU}} {\large *}
\and D.~Kostunin \inst{\ref{DESY}} 
\and M.~Kreter \inst{\ref{NWU}}
\and G.~Lamanna \inst{\ref{LAPP}}
\and A.~Lemi\`ere \inst{\ref{APC}}
\and M.~Lemoine-Goumard \inst{\ref{CENBG}}
\and J.-P.~Lenain \inst{\ref{LPNHE}}
\and F.~Leuschner \inst{\ref{IAAT}}
\and C.~Levy \inst{\ref{LPNHE}}
\and T.~Lohse \inst{\ref{HUB}}
\and I.~Lypova \inst{\ref{DESY}}
\and J.~Mackey \inst{\ref{DIAS}}
\and J.~Majumdar \inst{\ref{DESY}}
\and D.~Malyshev \inst{\ref{IAAT}}
\and D.~Malyshev \inst{\ref{ECAP}}
\and V.~Marandon \inst{\ref{MPIK}}
\and P.~Marchegiani \inst{\ref{WITS}}
\and A.~Marcowith \inst{\ref{LUPM}}
\and A.~Mares \inst{\ref{CENBG}}
\and G.~Mart\'i-Devesa \inst{\ref{LFUI}}
\and R.~Marx \inst{\ref{LSW}, \ref{MPIK}}
\and G.~Maurin \inst{\ref{LAPP}}
\and P.J.~Meintjes \inst{\ref{UFS}}
\and M.~Meyer \inst{\ref{ECAP}}
\and A.~Mitchell \inst{\ref{MPIK}}
\and R.~Moderski \inst{\ref{NCAC}}
\and L.~Mohrmann \inst{\ref{ECAP}}
\and A.~Montanari \inst{\ref{IRFU}}
\and C.~Moore \inst{\ref{Leicester}}
\and P.~Morris \inst{\ref{Oxford}}
\and E.~Moulin \inst{\ref{IRFU}}
\and J.~Muller \inst{\ref{LLR}}
\and T.~Murach \inst{\ref{DESY}}
\and K.~Nakashima \inst{\ref{ECAP}}
\and A.~Nayerhoda \inst{\ref{IFJPAN}}
\and M.~de~Naurois \inst{\ref{LLR}}
\and H.~Ndiyavala  \inst{\ref{NWU}}
\and J.~Niemiec \inst{\ref{IFJPAN}}
\and L.~Oakes \inst{\ref{HUB}}
\and P.~O'Brien \inst{\ref{Leicester}}
\and H.~Odaka \inst{\ref{Tokyo}}
\and S.~Ohm \inst{\ref{DESY}}
\and L.~Olivera-Nieto \inst{\ref{MPIK}}
\and E.~de~Ona~Wilhelmi \inst{\ref{DESY}}
\and M.~Ostrowski \inst{\ref{UJK}}
\and S.~Panny \inst{\ref{LFUI}}
\and M.~Panter \inst{\ref{MPIK}}
\and R.D.~Parsons \inst{\ref{HUB}}
\and G.~Peron \inst{\ref{MPIK}}
\and B.~Peyaud \inst{\ref{IRFU}}
\and Q.~Piel \inst{\ref{LAPP}}
\and S.~Pita \inst{\ref{APC}}
\and V.~Poireau \inst{\ref{LAPP}}
\and A.~Priyana~Noel \inst{\ref{UJK}}
\and D.A.~Prokhorov \inst{\ref{GRAPPA}}
\and H.~Prokoph \inst{\ref{DESY}}
\and G.~P\"uhlhofer \inst{\ref{IAAT}}
\and M.~Punch \inst{\ref{APC},\ref{Linnaeus}}
\and A.~Quirrenbach \inst{\ref{LSW}}
\and S.~Raab \inst{\ref{ECAP}}
\and R.~Rauth \inst{\ref{LFUI}}
\and P.~Reichherzer \inst{\ref{IRFU}}
\and A.~Reimer \inst{\ref{LFUI}}
\and O.~Reimer \inst{\ref{LFUI}}
\and Q.~Remy \inst{\ref{MPIK}}
\and M.~Renaud \inst{\ref{LUPM}}
\and F.~Rieger \inst{\ref{MPIK}}
\and L.~Rinchiuso \inst{\ref{IRFU}}
\and C.~Romoli \inst{\ref{MPIK}}
\and G.~Rowell \inst{\ref{Adelaide}}
\and B.~Rudak \inst{\ref{NCAC}}
\and E.~Ruiz-Velasco \inst{\ref{MPIK}}
\and V.~Sahakian \inst{\ref{YPI}}
\and S.~Sailer \inst{\ref{MPIK}}
\and H.~Salzmann \inst{\ref{IAAT}}
\and D.A.~Sanchez \inst{\ref{LAPP}}
\and A.~Santangelo \inst{\ref{IAAT}}
\and M.~Sasaki \inst{\ref{ECAP}}
\and M.~Scalici \inst{\ref{IAAT}}
\and J.~Sch\"afer \inst{\ref{ECAP}}
\and F.~Sch\"ussler \inst{\ref{IRFU}}
\and H.M.~Schutte \inst{\ref{NWU}}
\and U.~Schwanke \inst{\ref{HUB}}
\and M.~Seglar-Arroyo \inst{\ref{IRFU}}
\and M.~Senniappan \inst{\ref{Linnaeus}}
\and A.S.~Seyffert \inst{\ref{NWU}}
\and N.~Shafi \inst{\ref{WITS}}
\and J.N.S.~Shapopi  \inst{\ref{UNAM}}
\and K.~Shiningayamwe \inst{\ref{UNAM}}
\and R.~Simoni \inst{\ref{GRAPPA}}
\and A.~Sinha \inst{\ref{APC}}
\and H.~Sol \inst{\ref{LUTH}}
\and A.~Specovius \inst{\ref{ECAP}}
\and S.~Spencer \inst{\ref{Oxford}}
\and M.~Spir-Jacob \inst{\ref{APC}}
\and {\L.}~Stawarz \inst{\ref{UJK}}
\and L.~Sun \inst{\ref{GRAPPA}}
\and R.~Steenkamp \inst{\ref{UNAM}}
\and C.~Stegmann \inst{\ref{UP},\ref{DESY}}
\and S.~Steinmassl \inst{\ref{MPIK}}
\and C.~Steppa \inst{\ref{UP}}
\and T.~Takahashi  \inst{\ref{KAVLI}}
\and T.~Tavernier \inst{\ref{IRFU}}
\and A.M.~Taylor \inst{\ref{DESY}}
\and R.~Terrier \inst{\ref{APC}} {\large *}
\and J.~H.E.~Thiersen \inst{\ref{NWU}}
\and D.~Tiziani \inst{\ref{ECAP}}
\and M.~Tluczykont \inst{\ref{HH}}
\and L.~Tomankova \inst{\ref{ECAP}}
\and C.~Trichard \inst{\ref{LLR}}
\and M.~Tsirou \inst{\ref{MPIK}}
\and R.~Tuffs \inst{\ref{MPIK}}
\and Y.~Uchiyama \inst{\ref{Rikkyo}}
\and D.J.~van~der~Walt \inst{\ref{NWU}}
\and C.~van~Eldik \inst{\ref{ECAP}}
\and C.~van~Rensburg \inst{\ref{UNAM}}
\and B.~van~Soelen \inst{\ref{UFS}}
\and G.~Vasileiadis \inst{\ref{LUPM}}
\and J.~Veh \inst{\ref{ECAP}}
\and C.~Venter \inst{\ref{NWU}}
\and P.~Vincent \inst{\ref{LPNHE}}
\and J.~Vink \inst{\ref{GRAPPA}}
\and H.J.~V\"olk \inst{\ref{MPIK}}
\and Z.~Wadiasingh \inst{\ref{NWU}}
\and S.J.~Wagner \inst{\ref{LSW}}
\and J.~Watson \inst{\ref{Oxford}}
\and F.~Werner \inst{\ref{MPIK}}
\and R.~White \inst{\ref{MPIK}}
\and A.~Wierzcholska \inst{\ref{IFJPAN},\ref{LSW}}
\and Yu Wun Wong \inst{\ref{ECAP}}
\and A.~Yusafzai \inst{\ref{ECAP}}
\and M.~Zacharias \inst{\ref{NWU},\ref{LUTH}}
\and R.~Zanin \inst{\ref{MPIK}}
\and D.~Zargaryan \inst{\ref{DIAS},\ref{RAU}}
\and A.A.~Zdziarski \inst{\ref{NCAC}}
\and A.~Zech \inst{\ref{LUTH}}
\and S.J.~Zhu \inst{\ref{DESY}}
\and J.~Zorn \inst{\ref{MPIK}}
\and S.~Zouari \inst{\ref{APC}}
\and N.~\.Zywucka \inst{\ref{NWU}}
(H.E.S.S. Collaboration)
\and F. Acero\inst{\ref{IRFU}}
}

\institute{
Centre for Space Research, North-West University, Potchefstroom 2520, South Africa \label{NWU} \and 
Laboratoire Leprince-Ringuet, École Polytechnique, CNRS, Institut Polytechnique de Paris, F-91128 Palaiseau, France \label{LLR} \and
Dublin Institute for Advanced Studies, 31 Fitzwilliam Place, Dublin 2, Ireland \label{DIAS} \and 
Max-Planck-Institut f\"ur Kernphysik, P.O. Box 103980, D 69029 Heidelberg, Germany \label{MPIK} \and 
High Energy Astrophysics Laboratory, RAU,  123 Hovsep Emin St  Yerevan 0051, Armenia \label{RAU} \and
Aix Marseille Universit\'e, CNRS/IN2P3, CPPM, Marseille, France \label{CPPM} \and
Laboratoire d'Annecy de Physique des Particules, Univ. Grenoble Alpes, Univ. Savoie Mont Blanc, CNRS, LAPP, 74000 Annecy, France \label{LAPP} \and
University of Oxford, Department of Physics, Denys Wilkinson Building, Keble Road, Oxford OX1 3RH, UK \label{Oxford} \and
IRFU, CEA, Universit\'e Paris-Saclay, F-91191 Gif-sur-Yvette, France \label{IRFU} \and
University of Namibia, Department of Physics, Private Bag 13301, Windhoek 10005, Namibia \label{UNAM} \and
Instytut Fizyki J\c{a}drowej PAN, ul. Radzikowskiego 152, 31-342 Krak{\'o}w, Poland \label{IFJPAN} \and
DESY, D-15738 Zeuthen, Germany \label{DESY} \and
Obserwatorium Astronomiczne, Uniwersytet Jagiello{\'n}ski, ul. Orla 171, 30-244 Krak{\'o}w, Poland \label{UJK} \and
Department of Physics and Electrical Engineering, Linnaeus University,  351 95 V\"axj\"o, Sweden \label{Linnaeus} \and
Institut f\"ur Astronomie und Astrophysik, Universit\"at T\"ubingen, Sand 1, D 72076 T\"ubingen, Germany \label{IAAT} \and
Laboratoire Univers et Théories, Observatoire de Paris, Université PSL, CNRS, Université de Paris, 92190 Meudon, France \label{LUTH} \and
Sorbonne Universit\'e, Universit\'e Paris Diderot, Sorbonne Paris Cit\'e, CNRS/IN2P3, Laboratoire de Physique Nucl\'eaire et de Hautes Energies, LPNHE, 4 Place Jussieu, F-75252 Paris, France \label{LPNHE} \and
GRAPPA, Anton Pannekoek Institute for Astronomy, University of Amsterdam,  Science Park 904, 1098 XH Amsterdam, The Netherlands \label{GRAPPA} \and
Friedrich-Alexander-Universit\"at Erlangen-N\"urnberg, Erlangen Centre for Astroparticle Physics, Erwin-Rommel-Str. 1, D 91058 Erlangen, Germany \label{ECAP} \and
Astronomical Observatory, The University of Warsaw, Al. Ujazdowskie 4, 00-478 Warsaw, Poland \label{UWarsaw} \and
School of Physics, University of the Witwatersrand, 1 Jan Smuts Avenue, Braamfontein, Johannesburg, 2050 South Africa \label{WITS} \and
Universit\'e Bordeaux, CNRS/IN2P3, Centre d'\'Etudes Nucl\'eaires de Bordeaux Gradignan, 33175 Gradignan, France \label{CENBG} \and
School of Physical Sciences, University of Adelaide, Adelaide 5005, Australia \label{Adelaide} \and
Universit\"at Hamburg, Institut f\"ur Experimentalphysik, Luruper Chaussee 149, D 22761 Hamburg, Germany \label{HH} \and 
Université de Paris, CNRS, Astroparticule et Cosmologie, F-75013 Paris, France \label{APC} \and
Department of Physics and Astronomy, The University of Leicester, University Road, Leicester, LE1 7RH, United Kingdom \label{Leicester} \and
Nicolaus Copernicus Astronomical Center, Polish Academy of Sciences, ul. Bartycka 18, 00-716 Warsaw, Poland \label{NCAC} \and
Institut f\"ur Physik und Astronomie, Universit\"at Potsdam,  Karl-Liebknecht-Strasse 24/25, D 14476 Potsdam, Germany \label{UP} \and
Laboratoire Univers et Particules de Montpellier, Universit\'e Montpellier, CNRS/IN2P3,  CC 72, Place Eug\`ene Bataillon, F-34095 Montpellier Cedex 5, France \label{LUPM} \and
Landessternwarte, Universit\"at Heidelberg, K\"onigstuhl, D 69117 Heidelberg, Germany \label{LSW} \and
Institut f\"ur Physik, Humboldt-Universit\"at zu Berlin, Newtonstr. 15, D 12489 Berlin, Germany \label{HUB} \and
Institut f\"ur Astro- und Teilchenphysik, Leopold-Franzens-Universit\"at Innsbruck, A-6020 Innsbruck, Austria \label{LFUI} \and
Department of Physics, Rikkyo University, 3-34-1 Nishi-Ikebukuro, Toshima-ku, Tokyo 171-8501, Japan \label{Rikkyo} \newpage\and
Institute of Astronomy, Faculty of Physics, Astronomy and Informatics, Nicolaus Copernicus University,  Grudziadzka 5, 87-100 Torun, Poland \label{NCUT} \and
Department of Physics, University of the Free State,  PO Box 339, Bloemfontein 9300, South Africa \label{UFS} \and
Department of Physics, The University of Tokyo, 7-3-1 Hongo, Bunkyo-ku, Tokyo 113-0033, Japan \label{Tokyo} \and
Yerevan Physics Institute, 2 Alikhanian Brothers St., 375036 Yerevan, Armenia \label{YPI} \and
Kavli Institute for the Physics and Mathematics of the Universe (WPI), The University of Tokyo Institutes for Advanced Study (UTIAS), The University of Tokyo, 5-1-5 Kashiwa-no-Ha, Kashiwa, Chiba, 277-8583, Japan \label{KAVLI} 
}

\offprints{H.E.S.S.~collaboration,
\protect\\\email{\href{mailto:contact.hess@hess-experiment.eu}{contact.hess@hess-experiment.eu}};
\protect\\\protect * Corresponding authors
}

\makeatletter
\renewcommand*{\@fnsymbol}[1]{\ifcase#1\or*\or$\dagger$\or$\ddagger$\or**\or$\dagger\dagger$\or$\ddagger\ddagger$\fi}
\makeatother

\date{Received: 01/04/2021; Accepted: 08/06/2021}

  \title{Evidence of 100 TeV \mbox{$\gamma$-ray} emission from \mbox{HESS~J1702-420}: A new PeVatron candidate}
  \abstract{}{\small  The identification of  PeVatrons,  hadronic particle accelerators reaching the knee of the cosmic ray spectrum ($\text{few}\,{\small \times}\,10^{15}\,\mathrm{eV}$), is crucial to understand the origin of cosmic rays in the Galaxy. We provide an update on the unidentified source \mbox{HESS~J1702-420}, a promising PeVatron candidate.}{\small We present new  observations of  \mbox{HESS~J1702-420} made with the High Energy Stereoscopic System (H.E.S.S.), and processed using improved analysis techniques. The analysis configuration was optimized to enhance the collection area at the highest energies. We applied a three-dimensional (3D) likelihood analysis to model the source region and adjust non thermal radiative spectral models to the \mbox{$\gamma$-ray} data. We also analyzed archival \emph{Fermi} Large Area Telescope (LAT) data to constrain the source spectrum at \mbox{$\gamma$-ray} energies $>10\,\text{GeV}$.}{\small We report  the detection of  \mbox{$\gamma$-rays}  up to $100\,\text{TeV}$ from a specific region of \mbox{HESS~J1702-420}, which is well described by a new source component called \mbox{HESS J1702-420A} that was separated from the bulk of TeV emission at a $5.4\sigma$ confidence level.  The power law \mbox{$\gamma$-ray} spectrum of \mbox{HESS J1702-420A} extends  with an index of \mbox{$\Gamma=1.53 \pm 0.19_\text{stat} \pm 0.20_\text{sys}$} and  without curvature up to the energy band $64-113\,\text{TeV}$, in which it was detected by H.E.S.S.\ at a $4.0\sigma$ confidence level. This makes \mbox{HESS J1702-420A} a compelling candidate site for the presence of extremely high energy cosmic rays.  With a flux above $2\,\text{TeV}$ of \mbox{$ (2.08\pm 0.49_\text{stat} \pm 0.62_\text{sys})\times 10^{-13}\,\text{cm}^{-2}\,\text{s}^{-1}$} and a  radius of $(0.06\pm 0.02_\text{stat} \pm 0.03_\text{sys})^{\,\text{o}}$, \mbox{HESS J1702-420A} is  outshone --- below a few tens of TeV --- by the companion \mbox{\mbox{HESS J1702-420B}}. The latter has a steep spectral index of $\Gamma=2.62 \pm 0.10_\text{stat} \pm 0.20_\text{sys}$ and an elongated shape, and it accounts for most of the low-energy \mbox{HESS~J1702-420} flux. Simple hadronic and leptonic emission models can be well adjusted to the spectra of both components. Remarkably, in a hadronic  scenario, the cut-off energy of the particle distribution powering \mbox{HESS J1702-420A} is found to be higher than $0.5\,\text{PeV}$ at a $95\%$ confidence level.}{\small For the first time, H.E.S.S.\ resolved two components with significantly different morphologies and  spectral indices, both detected at $>5\sigma$ confidence level, whose combined emissions result in the source \mbox{HESS~J1702-420}. We detected \mbox{HESS~J1702-420A}  at a $4.0\sigma$  confidence level in the energy band $64-113\,\text{TeV}$, which brings evidence for the source emission up to $100\,\text{TeV}$. 
 In a hadronic emission scenario, the hard $\gamma$-ray spectrum of \mbox{HESS J1702-420A} implies that the source likely harbors  PeV protons, thus becoming one of the most solid PeVatron candidates detected so far in H.E.S.S. data. However, a leptonic origin of the observed TeV emission cannot be ruled out either.
 }

\maketitle

%
\section{Introduction}\label{introduction}
The acceleration sites of cosmic rays\footnote{Throughout this paper, unless otherwise specified, this term always refers to hadronic cosmic rays.} are a century-old unknown in modern astrophysics \citep{hess1912}. The current understanding is that the bulk of cosmic rays reaching Earth --- mostly  energetic protons --- originate within our Galaxy, outside of the solar system, at unknown sites where they are accelerated   up to the energy of the   knee feature in the cosmic ray spectrum \citep{berezinskii, gaisser}. Since the measured knee energy is around $3-4$ PeV \citep{kascade}, the Galactic  accelerators responsible for cosmic rays up to the knee are  called \mbox{PeVatrons}. Several source populations have been proposed as potential \mbox{PeVatron} candidates: among them,  supernova remnants (SNRs)  and young massive stellar clusters   stand out as well motivated   cases~\citep{bellsnr, stellarclusters}. However, to date no observation has definitively linked any particular source class to the acceleration of PeV protons. 
The H.E.S.S.\ Collaboration  has already reported evidence for the acceleration of PeV protons in the central molecular zone around \mbox{Sgr A$^{*}$} \citep{hessgc, hessgc2}, at a level that is presently insufficient to sustain the  flux of PeV cosmic rays observed at Earth. Recent searches for a high-energy cut-off in the spectrum of the diffuse emission around \mbox{Sgr A$^{*}$} have led to unclear conclusions, with MAGIC reporting a $2\sigma$ hint for a spectral turnover around \mbox{$\approx20 \,\text{TeV}$} and VERITAS measuring a straight power law up to $40 \,\text{TeV}$~\citep{magicgc, veritasgc}.

Observations of the very high energy (VHE; \mbox{$ 0.1\lesssim E_\gamma \lesssim 100$ TeV}) \mbox{$\gamma$-ray} sky with ground-based  telescope arrays such as the High Energy Stereoscopic System (H.E.S.S.),  providing relatively good angular and energy resolution as well as high sensitivity~\citep{hgps},   represent a unique tool to improve our understanding of cosmic ray physics. Charged cosmic particles radiate in this energy band due to interactions with the interstellar medium (ISM) or to the up-scattering of diffuse low-energy radiation fields.
The    H.E.S.S.\  Galactic  Plane  Survey  (HGPS) catalog~\citep{hgps} lists 78 VHE $\gamma$-ray sources, most of which have been identified as of today --- or at least likely associated --- with multi wavelength counterparts. However, a handful of sources remain  completely unidentified. Having no clear counterpart at other wavelength, they are categorized as dark TeV sources~\citep{dark}. Experience has shown that several such objects, despite being unidentified at first, were  later classified as evolved pulsar wind nebulae (PWNe), based on the discovery of energy-dependent morphologies and compact X-ray counterparts~\citep{j1303, j1825}. However, a leptonic scenario, in which the VHE \mbox{$\gamma$-ray} emission is powered by relativistic electrons up-scattering   ambient radiation fields, might not necessarily suit all of the remaining dark sources.

\mbox{HESS J1702-420} is a long-known but poorly understood  VHE \mbox{$\gamma$-ray} source. It was discovered during the first Galactic plane survey campaign  with a significance of $4\sigma$, based on a $5.7\,\mathrm{hr}$ observation livetime~\citep{first_hgps}. In~\citet{dark}, a dedicated analysis revealed a hard power law\footnote{Hearafter, the term power law refers to the functional form $dN/dE(E)=\Phi_\text{ref}(E/E_\text{ref})^{-\Gamma}$, where $\Phi_\text{ref}$ is the spectral normalization at the reference energy $E_\text{ref}$ --- usually chosen to correspond with the pivot energy which minimizes correlations between the spectral parameters ---, and $\Gamma$ is the spectral index.} spectral index of $\Gamma=2.07 \pm 0.08_\text{stat} \pm 0.20_\text{sys}$, with no sign of cut-off, 
 and a significantly extended morphology which is well described by a $0.30^{\,\text{o}}\times\,0.15^{\,\text{o}}$ elliptical Gaussian template. With better reconstruction and data selection algorithms, the HGPS catalog confirmed the spectral hardness of the source, \mbox{$\Gamma=2.09 \pm 0.07_\text{stat}\pm 0.20_\text{sys}$}, and estimated a source significance of $15\sigma$ based on $9.5\,\text{hr}$ of observations. It simplified however the source morphology to a $0.2^{\,\text{o}}$ symmetric Gaussian, due to the noninclusion of elongated shapes in the semi-automated survey analysis chain.

The physical origin of the \mbox{$\gamma$-ray} emission from \mbox{HESS J1702-420} is unknown. The remnant \mbox{SNR G344.7-0.1} and the pulsar \mbox{PSR J1702-4128} are within a $\approx 0.5^{\,\text{o}}$ aperture from the centroid of the TeV emission. The former is a $3\,\text{kyr}$ old~\citep{giancani} and small-sized --- $8\,\text{arcmin}$ in diameter ---  SNR, whose centrally-peaked radio shell~\citep{dubner, MOST, giancani} is also emitting thermal X-rays~\citep{yamaguchi, combi}, with the brightest X-ray and radio features close to each other~\citep{giancani}. Recently, the \emph{Fermi}-LAT association  \mbox{2FHL J1703.4-4145}, with the hard spectral index $\Gamma\approx1.2$, was discovered on the western edge of the SNR~\citep{eagle}. The core-collapse origin of the supernova is debated, due to the absence of a compact remnant. Also controversial is the SNR distance~\citep{eagle}, for which a limit based on the high absorbing hydrogen column density is $d\gtrsim 8\,\text{kpc}$~\citep{yamaguchi}. The cosmic ray diffusion time from the SNR to the VHE peak is compatible with the remnant age~\citep{eagle}, which suggests that both \mbox{2FHL J1703.4-4145} and \mbox{HESS J1702-420} may be associated with \mbox{SNR G344.7-0.1}. However, the detection of an extended and bright TeV source at $d\gtrsim 8\,\text{kpc}$ in the Galactic plane is unlikely given H.E.S.S.\   sensitivity~\citep{hgps}.  Moreover, the surrounding ISM does not exhibit any clear morphological association with the VHE \mbox{$\gamma$-ray} source~\citep{lau}, a fact that challenges a hadronic interpretation of the TeV emission.  With a spin-down luminosity $\dot{E}=3.4\times10^{35}\,\text{erg}\,\text{s}^{-1}$,  the large-offset pulsar \mbox{PSR J1702-4128} would need  a conversion efficiency  $> 10\%$ in order to power the whole TeV source, higher than all other PWNe identified by H.E.S.S.~\citep{gallant, hesspwn}. This fact, together with the unconclusive searches for an asymmetric X-ray PWN around the pulsar~\citep{chang}, tend to disfavor an association of \mbox{PSR J1702-4128} with \mbox{HESS J1702-420}. Finally, deep X-ray observations of the VHE source with \emph{Suzaku} revealed the presence of two   faint point-like sources close to the line of sight of \mbox{HESS J1702-420}, and the absence of extended emission, with an X-ray flux at least 12 times lower than the TeV flux in the \emph{Suzaku} field of view (FoV)~\citep{fujinaga}.

This paper reports on  new H.E.S.S.\ observations of \mbox{HESS J1702-420} that have been processed with improved techniques.  Additionally, archival \emph{Fermi}-LAT   data were analyzed, to perform a broadband modeling of the TeV source.  The paper is structured as follows: first, the H.E.S.S.\ data analysis and results are presented (Section~\ref{hess}), with focus on the three-dimensional (3D) likelihood anaysis (Section~\ref{3d}) and a morphological study made with a more classical background estimation technique (Section~\ref{traditional}). Then, Section~\ref{mwl} reports on the analysis of archival multi wavelength observations of the region, while Section~\ref{dddd} describes the adjustment of physically-motivated non thermal radiative models to the data and discusses possible interpretations of the new H.E.S.S.\ results in a broadband context. Finally, Section~\ref{conclusions} summarizes the main conclusions of the paper.

\section{H.E.S.S. data analysis and results}\label{hess}
H.E.S.S.\ is an array of five imaging atmospheric Cherenkov telescopes (IACTs) located in the Khomas Highland of Namibia, 1800 m above sea level. The design of the original array (H.E.S.S.\ I), operating since 2003, involved four $12\,\text{m}$ diameter telescopes --- CT1-4, whose cameras were upgraded in 2017~\citep{HESSIU} --- at the corners of a $120\, \text{m}\times 120\, \text{m}$ square. In 2012, a second phase began (H.E.S.S.\ II) with   the addition of a $28\,\text{m}$ diameter telescope (CT5) at the center of the array. 

All results presented in this paper make use of data collected from 2004 to 2019, using CT1-4 observations that have been carried out in multiple contexts: dedicated pointings on \mbox{HESS J1702-420}, observations of other nearby objects --- mainly \mbox{RX J1713-3946} and \mbox{PSR B1706-44} --- and Galactic plane scan observations for  the HGPS campaign. This led to the accumulation of an acceptance-corrected livetime of $ 44.9\,\text{hr}$, obtained after selecting all runs with pointing direction within $3^{\,\text{o}}$ from the HGPS position of \mbox{HESS J1702-420} ($l\,$=$\,344.30^{\,\text{o}}$, $b\,$=$\,-0.18^{\,\text{o}}$) and averaging over a $0.5^{\,\text{o}}$ circle centered at the same location. 

Observations were processed with the H.E.S.S. analysis package (HAP), applying a Hillas-type shower reconstruction \citep{hillas} and the multi-variate analysis (MVA) technique~\citep{becherini} for efficient $\gamma/$hadron discrimination. Preselection and MVA discrimination cuts were optimized to improve the collection area at high energies ($E>1\,\text{TeV}$), assuming a \mbox{$\gamma$-ray} spectral index of $2.3\,$. The reduced data and instrument response functions\footnote{They are the effective area, exposure livetime, point-spread function, energy dispersion and background model.} (IRFs) were exported to \href{https://fits.gsfc.nasa.gov/}{FITS} files  complying with the standard format developed by~\citet{deil}. Finally, all high-level analysis results --- that is sky maps and spectra ---  were obtained using \texttt{gammapy} (version 0.17), an open source python library for \mbox{$\gamma$-ray} astronomy~\citep{gammapy:2017, gammapy:2019, christoph_deil_2020_4701492}. The analysis cross-check was also performed with \texttt{gammapy}, by   applying the same high-level analysis pipeline to data that were reduced according to an alternative low-level chain of calibration, reconstruction and $\gamma$-hadron separation methods~\citep{ohm}.  Independent crosschecks for the classical morphological and spectral analyses presented in Section~\ref{traditional} and Appendix~\ref{spectral} were also carried out using the standard HAP software. 

\subsection{Three-dimensional likelihood analysis}\label{3d}
Three-dimensional  likelihood analysis, routinely used for high energy (HE; $ 0.2\lesssim E_\gamma \lesssim 100$ GeV) \mbox{$\gamma$-ray} data processing~\citep{mattox, abdo}, has been recently introduced in the VHE \mbox{$\gamma$-ray} astronomy domain~\citep{lars, donath, skyprism}. In its binned version, this technique allows the adjustment of a spectro-morphological model to a data cube, which carries information on the number of reconstructed events within 3D  bins. The term 3D refers to the fact that the data are distributed along 2 spatial dimensions (e.g., Galactic longitude and latitude) plus 1 energy dimension. The model can be seen as a collection of  spectral and spatial parametric shapes that are assumed to describe all \mbox{$\gamma$-ray} sources in the  (source model), plus the residual hadronic background of $\gamma$-like events (background model). The model is  convolved with the IRFs to predict  the number of photons that would be detected by the telescope array within each spatial and spectral bin, based on the assumed model and its given parameter values.  The 3D analysis allows the fine-tuning of all free parameters of the model, in such a way that the cube of model-predicted counts mimics as closely as possible the measured data cube. This approach is known as ``forward-folding''~\citep{piron}.

We performed a 3D binned likelihood analysis of the region surrounding \mbox{HESS J1702-420}, in order to determine the best spectro-morphological model to describe   the observed TeV emission. The next sections discuss the analysis setup (Section~\ref{setup}) and results (Section~\ref{results}), while Section~\ref{alphabeta} focuses on the most relevant components of the model, that is those describing \mbox{HESS J1702-420}.

\subsubsection{Background model and analysis setup}\label{setup}

\begin{table*}
\label{bins}
\begin{center}
\resizebox{\textwidth}{!}{\begin{tabular}{||c|c|c|c|c||}
    \hline
    \text{\textbf{Number of runs}}&\text{\textbf{Zenith [deg]}}&\text{\textbf{Pointing offset from the source [deg]}}& \text{\textbf{On-source livetime [hr]}} {\boldmath$^{*}$}&\text{\textbf{Observation period}}\\
    \hline
    26 & 17.8 - 35.1& 0.2 - 1.4& 9.7 & 2004 - 2011\\
    166 & 16.3 - 35.9 & 1.6 - 2.8 & 1.8 & 2004 - 2014\\
    88 & 36.1 - 62.4 & 2.2 - 2.8 & 0.0 & 2004 - 2014\\
    80 & 37.5 - 59.1 & 0.3 - 0.8 & 33.4 & 2017 - 2019\\
        \hline
    \multicolumn{5}{l}{\footnotesize {\boldmath$^{*}$} Obtained by averaging over a $0.5^{\,\text{o}}$-radius circle centered at  the center of the HGPS position of \mbox{HESS J1702-420}. We observe that runs belonging to}\\
\multicolumn{5}{l}{\footnotesize    the second and third groups  were   included to better contrain the background level in the source region. The fourth group contains all (and only) runs}\\
\multicolumn{5}{l}{\footnotesize   taken after the 2017 camera upgrade~\citep{HESSIU}.}\\
\end{tabular}}
\end{center}
\vspace*{-0.3cm}
\caption{Details on the four groups of H.E.S.S.\ observations, with similar pointing zenith angle and source offsets from the pointing direction, that were used for the 3D likelihood analysis. }
\label{binning}
\end{table*}

H.E.S.S. data-taking consists of consecutive observations (also called runs), usually of 28 minutes duration. The background model was produced from a large set of empty-field --- that is devoid of known \mbox{$\gamma$-ray} sources --- observations, following an approach similar to the one described in~\citet{lars}. We first produced a general model in the form of a lookup table, describing the residual hadronic background as a function of few observational parameters --- namely, the zenith angle and optical efficiency. We then assigned each observation a specific background model, called its FoV background model, based on a multi-variable interpolation of the general model. In addition, the FoV background model was renormalized run-by-run, to account for possible differences in the level of night sky background (NSB) and atmospheric absorption with respect to the observations that were used to build the general background model.

As a result of the region's observation history (see the introduction of Section~\ref{hess}), standard run selection criteria led to a   heterogeneous set of observations  in terms of array response, zenith angles and source offsets from the pointing direction. We therefore se\-pa\-rated observations obtained before and after the 2017 camera upgrade, for which different IRFs have to be used, and grouped together observations with similar zenith and offset values. More details on   the four groups of observations that were defined are reported in Table~\ref{binning}. We stacked observations within each group, thus obtaining four independent datasets. This choice represents a good compromise between the time-saving, due to a decrease in the number of degrees of freedom,  and information loss, due to the IRFs averaging, that are connected with the data stacking procedure. 

For each of the four observation groups, the list of reconstructed events was reduced to a binned data cube, with spatial dimensions corresponding to an analysis  region of interest (RoI) of $4^{\,\text{o}}\times\,4^{\,\text{o}}$ centered at the HGPS position of \mbox{HESS J1702-420}. This choice represents a compromise between   a sufficiently large RoI, to get enough off-source regions for the background estimation,  and a sufficiently small RoI, to minimize   the number of unrelated sources needing to be modeled. A spatial pixel size of $0.02^{\,\text{o}}\times\,0.02^{\,\text{o}}$ was adopted, to ensure sufficient per-pixel statistics while still providing good spatial resolution. The third axis of the cube, encoding the reconstructed energy of incident photons, was divided into 20 equally-spaced --- in logarithmic scale --- bins between 0.5 and $150\,\mathrm{TeV}$. In order to reject poorly reconstructed data, for each observation all events with offset $\geq 2^{\,\text{o}}$ from the pointing direction were excluded, together with those whose reconstructed energy was below the safe threshold 
\begin{linenomath}
\begin{equation}
E_\text{threshold}=\text{max}\{E_\text{bkg}, E_{A_\text{eff}}\}\,\,,
\end{equation} 
\end{linenomath}
where $E_\text{bkg}$ represents the energy at which the maximum rate of hadronic background events occurs and $ E_{A_\text{eff}}$ is the energy at which the effective area at the center of the FoV drops to 10\% of its maximal value (e.g.,~\citet{lars}).

With this setup, we performed a joint-likelihood analysis of the four independent datasets, each one having its own FoV background model but all sharing the same source model. This means that we summed the four dataset-specific log-likelihood values, and maximized the total resulting likelihood with respect to the model parameters. For the fit, the Cash statistic~\citep{cash} for Poisson-distributed data with perfectly known background model was used. The 3D analysis was performed in the energy range $2-150 \,\text{TeV}$. 
All events with reconstructed energy below $2\,\text{TeV}$ were excluded from the likelihood computation, to avoid threshold effects arising from the high energy optimization (above $1\,\text{TeV}$) of the analysis configuration and to ensure that the power law assumption made for the background model spectrum was valid --- which is true only well above the background peak.
A $0.25^{\,\text{o}}$ band around the borders of the RoI was excluded from the analysis, in order to limit  possible contamination   due to non-modeled sources outside the RoI. Additionally, a $0.3^{\,\text{o}}$-radius circular region centered at $l\,$=$\,343.35^{\,\text{o}}$ and $b\,$=$\,-0.93^{\,\text{o}}$, containing a $\approx3\sigma$ significance hotspot, was excluded from the likelihood calculation instead of being modeled. This choice was motivated by the high offset between the hotspot and \mbox{HESS J1702-420}, and the necessity of limiting the number of nuisance parameters of the source model.

\subsubsection{Source model derivation and results}\label{results}

\begin{figure*}
\centering
\includegraphics[width=18cm]{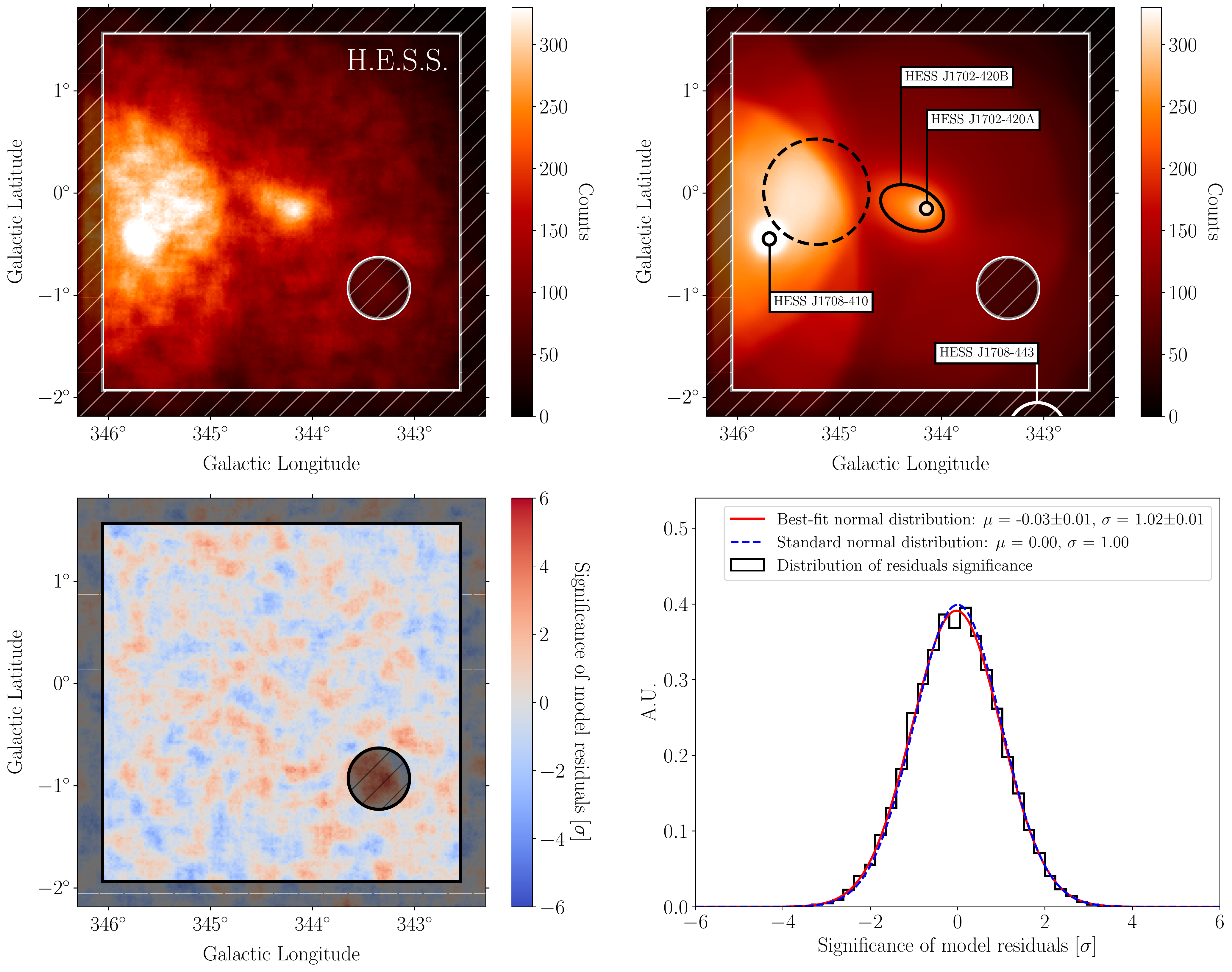}
\caption{\emph{Upper left panel}: Image of the RoI  obtained by integrating the binned cube of measured counts over the energy axis ($E>2\,\text{TeV}$), and correlating it with a $0.1^{\,\text{o}}$-radius top-hat kernel. The hatched regions were excluded from the likelihood computation. The   bright area around $l\gtrsim 345^{\,\text{o}}$ results from deep observations of RX~J1713-3946 and HESS~J1708-410. \emph{Upper right panel}: Energy-integrated ($E>2\,\text{TeV}$) map of model-predicted counts, with names and $1\sigma$ shapes of all model components overlaid. The large-scale discarded component is indicated by the dashed circle --- see the main text for more details. \emph{Lower left panel}: Spatial distribution of model residuals, showing the statistical significance --- in units of Gaussian standard deviations --- of \emph{counts - model} fluctuations. The image was obtained assuming Cash statistic   for Poisson-distributed signals with perfectly known background model~\citep{cash}.  \emph{Lower right panel}: Histogram containing the number of occurrences of each significance value (assuming Cash statistic), from the lower left panel. The adjustment of a Gaussian function to the histogram is shown, together with a reference standard normal distribution.}
\label{3dmodel}
\end{figure*}

The optimal source model for the RoI was determined using a statistical approach based on the improvement of a first-guess model with the iterative addition of new components. As a starting point, we defined a source model including all known VHE sources within the $4^{\,\text{o}}\times\,4^{\,\text{o}}$ RoI, with the exception of \mbox{HESS J1702-420}. Each iteration   then consisted  either in the addition of a new source component --- described by a symmetric Gaussian morphology and power law spectrum ---, or the test of a  different assumption on the spatial or spectral shape of an already existing component. Specifically, we looked for the presence of  high energy spectral cut-offs or elongated shapes for all  components.

Step-by-step, the improvement of the source model was assessed looking at two indicators. Firstly, we used the likelihood-ratio test, which allows  to estimate the relative significance of nested hypotheses taking into account the number of additional degrees of freedom added at each step. For example, the presence of an additional model component $\mu$ with 1 (5) free parameter(s) was considered significant at $5\sigma$ confidence level  only if $\mathrm{TS}\geq 25$ ($\mathrm{TS}\geq 37.1$), where TS is the test statistic  
\begin{linenomath}
\begin{equation}\label{ts}
\text{TS}=-2\ln\,\left(\frac{\mathscr{L}^\text{\,max}_{0}}{\mathscr{L}^\text{\,max}_{\mu}}\right)\,\,.
\end{equation}
\end{linenomath}
Here, $\mathscr{L}^\text{\,max}_{0}$ ($\mathscr{L}^\text{\,max}_{\mu}$) represents the maximum likelihood of the model under the null (alternative) hypothesis --- that is the absence (presence) of $\mu$. More details can be found in Appendix~\ref{wilks}. Secondly, we assessed --- by visual inspection --- the flattening of spatial and spectral residuals toward zero, as a result of the addition of new model components.

The final results of the   procedure  are shown in Figure~\ref{3dmodel}. The upper left (right) panel of the figure shows the measured (model-predicted)  counts map, obtained after stacking the four individual datasets and integrating over the energy axis above $2\,\text{TeV}$. Diagonal line hatches represent portions of the RoI that were excluded from the likelihood computation (Section~\ref{setup}). The  measured counts map is well matched by the prediction, since the spatial distribution of the significance of model residuals (lower left panel) does not contain significant structures. The histogram of significance values (lower right panel) closely matches a standard normal distribution, as expected if residuals are only due to statistical Poisson fluctuations.  Additionally, the spatial distribution of model residuals in three independent energy bands is shown in Figure~\ref{res-bands}.

The upper right panel of Figure~\ref{3dmodel} also shows the $1\sigma$ contours of all components found in the final source model. There are  two overlapping objects, called \mbox{HESS J1702-420A} and \mbox{HESS J1702-420B}, that together describe the emission from \mbox{HESS J1702-420}. Being the most relevant model components  for the scope of this paper, their details are discussed in the dedicated Section~\ref{alphabeta}. Two other model components represent the nearby sources \mbox{HESS~J1708-410} and \mbox{HESS~J1708-443}. Due to their large angular distance from the center of the RoI, the details of their modeling  do not have a strong impact on \mbox{HESS J1702-420}. The fitted model for \mbox{HESS~J1708-410} was found to be consistent with that reported in the HGPS catalog, while the model for \mbox{HESS~J1708-443}, being only partially contained in the RoI, was directly fixed to the catalog one. Finally, we found a large-scale component, indicated by the dashed circle in Figure~\ref{3dmodel} (upper right panel), whose presence was not confirmed by the crosscheck analysis. More details can be found in Appendix~\ref{largescale}. 
During the analysis, all parameters describing \mbox{HESS J1702-420A} and \mbox{HESS J1702-420B}, together with the nuisance parameters of all other components and the background model, were left free to vary. Details on the final parameters for the most relevant model components are provided in Tables~\ref{3dtable-spatial},~\ref{3dtable-spectral} and~\ref{3dtable-detection}.

\subsubsection{\mbox{HESS J1702-420A} and \mbox{HESS J1702-420B}}\label{alphabeta}

\begin{figure}
\centering
\includegraphics[width=\linewidth]{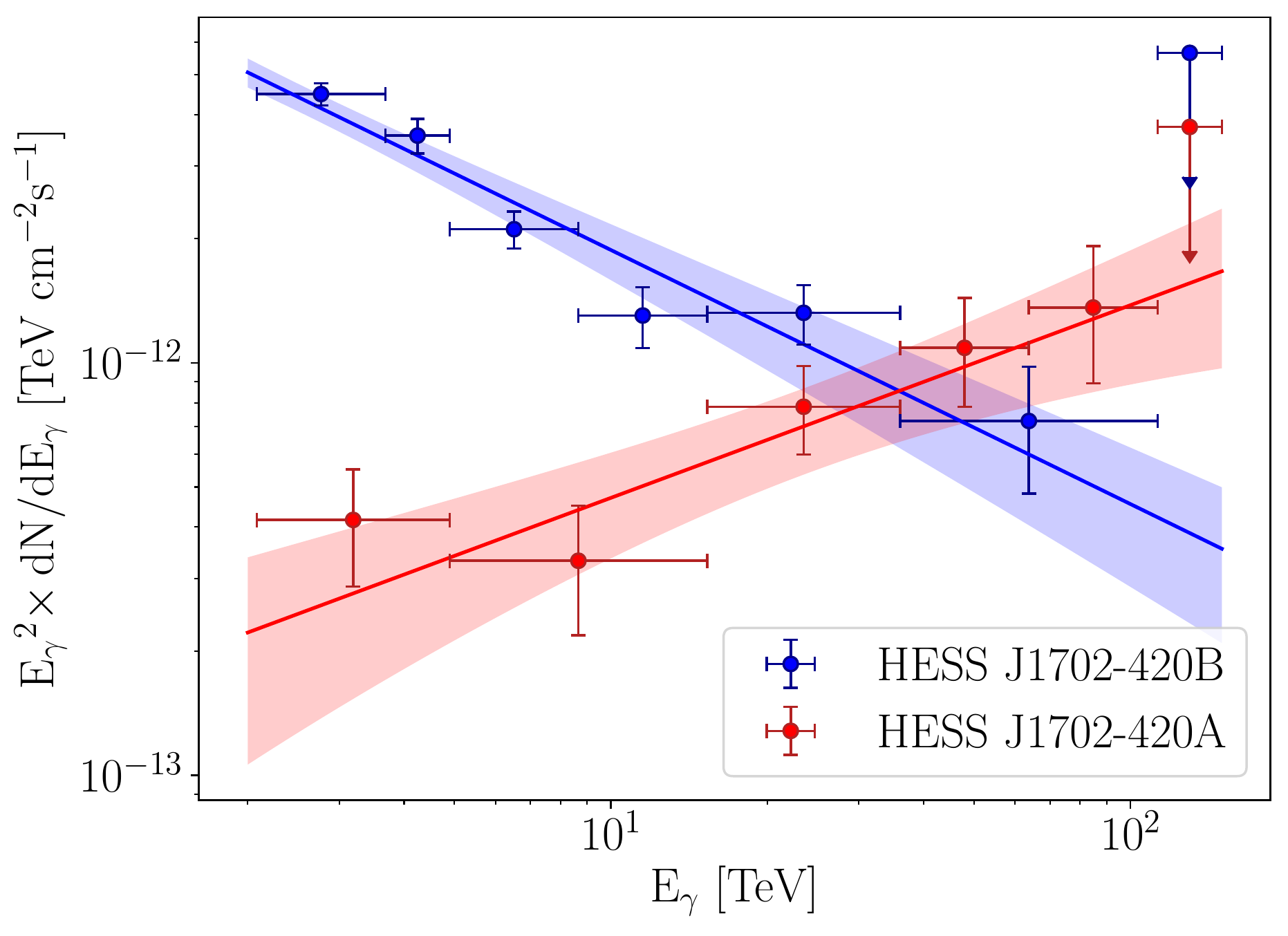}
\caption{Power law spectra of \mbox{HESS J1702-420A} (red solid line) and \mbox{HESS J1702-420B} (blue solid line), as a function of the incident photon energy E$_\gamma$. The butterfly envelopes indicate the $1\sigma$ statistical uncertainty on the spectral shape. They have been obtained from a 3D fit of the H.E.S.S.\ data with \texttt{gammapy} (more details in the main text). The spectral points, shown for reference purpose only, have been obtained by rescaling the amplitude of the reference spectral model within each energy bin, re-optimizing at the same time all free nuisance parameters of the model. In the energy bins with less than $3\sigma$ excess significance, the $3\sigma$ confidence level upper limits are shown.}
\label{fp}
\end{figure}

The  most relevant result for the identification of Galactic Pevatrons is the discovery --- with a \mbox{TS-based} confidence level corresponding to $5.4\,\sigma$ --- of a new source component, \mbox{HESS J1702-420A}, hidden under the bulk emission formerly  associated  with  \mbox{HESS J1702-420}. This object has a spectral index of \mbox{$\Gamma=1.53 \pm 0.19_\text{stat} \pm 0.20_\text{sys}$} and a \mbox{$\gamma$-ray} spectrum that, extending  with no sign of curvature up to at least $64\,\text{TeV}$ (possibly $100\,\text{TeV}$), makes it a compelling candidate site for the presence of extremely high energy cosmic rays.  With a flux above $2\,\text{TeV}$ of \mbox{$ (2.08\pm 0.49_\text{stat} \pm 0.62_\text{sys})\times 10^{-13}\,\text{cm}^{-2}\,\text{s}^{-1}$ } and a $1\sigma$ radius of $(0.06\pm 0.02_\text{stat} \pm 0.03_\text{sys})^{\,\text{o}}$, \mbox{HESS J1702-420A} is outshone below $\approx40\,\text{TeV}$   by the companion \mbox{HESS J1702-420B}. The test of a point-source hypothesis for \mbox{HESS J1702-420A} resulted in a non-convergence of the fit. \mbox{HESS J1702-420B} has a steep spectral index of $\Gamma=2.62 \pm 0.10_\text{stat} \pm 0.20_\text{sys}$, elongated shape and a flux above $2\,\text{TeV}$ of $ (1.57\pm0.12_\text{stat}\pm0.47_\text{sys})\,10^{-12}\,\text{cm}^{-2}\,\text{s}^{-1}$ that accounts  for most of the low-energy \mbox{HESS J1702-420} emission. 
 By comparing results obtained with the main and crosscheck analysis configurations, we verified that all discrepancies were consistent with the expected level of H.E.S.S.\  systematic  uncertainties~\citep{hgps}. 

For neither of the two sources did an exponential cut-off function statistically improve  the fit with respect to a simple power law (cut-off significance $\ll 1\sigma $). The \mbox{$\gamma$-ray} spectra of both components are shown in Figure~\ref{fp}, together with spectral points computed under a power law assumption and re-optimizing all the nuisance parameters of the model --- see Table~\ref{tab:fp} for details. We adapted the binning of the spectral energy distributions  to obtain approximately equal counts in each bin. \mbox{HESS J1702-420B} is the brightest component up until roughly $40\,\text{TeV}$, where \mbox{HESS J1702-420A} eventually starts dominating with its   $\Gamma\approx1.5$ power law spectrum  up to $100\,\text{TeV}$.  The second to last spectral point of \mbox{HESS J1702-420A} (\mbox{HESS J1702-420B}), covering the reconstructed energy range $64-113\,\text{TeV}$ ($36-113\,\text{TeV}$), is significant at $4.0\sigma$ ($3.2\sigma$) confidence level. 

We explicitly point out that, based on this dataset, it is impossible to tell whether \mbox{HESS J1702-420A} and \mbox{HESS J1702-420B} actually represent two separate sources --- superimposed on the same line of sight ---  or rather different emission zones of a  single complex object. Moreover,   any morphology assumption based on exact geometric shapes ---   in this case, two overlapping Gaussian components --- represents an idealization, that might differ from the real underlying astrophysical model. In particular, a model assumption based on the energy-dependent morphology of a single source might also be well suited to describe the emission of \mbox{HESS J1702-420}. 
To address this point, we performed dedicated studies that   ultimately provided a confirmation of the 3D analysis results, in that they brought no evidence of energy-dependent variations of the 3D model or  spectral softening as a function of the distance from \mbox{HESS J1702-420A} (see Appendix~\ref{3d-energy-bands} and~\ref{spectral}). 
Therefore, a model describing \mbox{HESS J1702-420} with  a single energy-dependent component is disfavored, even if it cannot be definitively ruled out.

\begin{table*}[ht]
\small
\begin{center}
\resizebox{1.04\textwidth}{!}{\hspace*{-0.7cm}\begin{tabular}{||c|c|c|c|c|c||}
    \hline
    \text{\textbf{Component name}}&\text{\textbf{Galactic longitude}}&\text{\textbf{Galactic latitude}}&\text{\textbf{Major  semi-axis}}&\text{\textbf{Minor  semi-axis}}
    &\text{\textbf{Rotation angle}} {\boldmath$^{*}$}\\
     &\text{\textbf{[deg]}}&\text{\textbf{[deg]}}&\text{\textbf{[deg]}}
    &\text{\textbf{[deg]}}&\text{\textbf{[deg]}}\\
\hline
\mbox{HESS J1702-420A}&$344.15\pm0.02_\text{stat}\pm0.01_\text{sys}$&$-0.15\pm0.02_\text{stat}\pm0.01_\text{sys}$&\multicolumn{2}{c|}{$0.06\pm0.02_\text{stat}\pm0.03_\text{sys}$}& ---\\
\mbox{HESS J1702-420B}&$344.29\pm0.03_\text{stat}\pm0.01_\text{sys}$&$-0.15\pm0.02_\text{stat}\pm0.01_\text{sys}$&$0.32\pm0.02_\text{stat}\pm0.03_\text{sys}$&$0.20\pm0.02_\text{stat}\pm0.03_\text{sys}$&$67.0\pm5.4_\text{stat}\pm9.7_\text{sys}$\\
\hline
\multicolumn{5}{l}{\footnotesize {\boldmath$^{*}$} Measured counterclockwise starting from the $l=0$, $b>0$ axis.}\\
\end{tabular}}
\end{center}

\vspace*{-0.3cm}
\caption{Best-fit morphology parameters of \mbox{HESS J1702-420A} and \mbox{HESS J1702-420B}. }
\label{3dtable-spatial}
\end{table*}

\begin{table*}[ht]
\small
\begin{center}
\begin{tabular}{||c|c|c|c|c||}
    \hline
\text{\textbf{Component name}}&\text{\textbf{Spectral index}}&\text{\textbf{Decorrelation
     energy}}&$\mathbf{dN/dE(E=E_\text{\textbf{decorr}})}$&$\mathbf{F(E>2\,\text{\textbf{TeV}})}$\\
    & & \text{\textbf{[TeV]}}&$\mathbf{[\text{\textbf{TeV}}^{-1\,}\text{\textbf{cm}}^{-2\,}\text{\textbf{s}}^{-1}]}$&$\boldsymbol{[\text{\textbf{cm}}^{-2\,}\text{\textbf{s}}^{-1}]}$\\
    \hline
\mbox{HESS J1702-420A}&$1.53 \pm 0.19_\text{stat} \pm 0.20_\text{sys}$&$24.53$&$(1.19\pm0.28_\text{stat}\pm0.34_\text{sys})\,10^{-15}$&$ (2.08\pm 0.49_\text{stat}\pm 0.62_\text{sys})\times 10^{-13}$\\
\mbox{HESS J1702-420B}&$2.62 \pm 0.10_\text{stat} \pm 0.20_\text{sys}$&$2.67$&$(5.93\pm0.46_\text{stat}\pm1.78_\text{sys})\,10^{-13}$&$ (1.57\pm 0.12_\text{stat}\pm 0.47_\text{sys})\times 10^{-12}$\\
    \hline
\end{tabular}
\end{center}\vspace*{-0.3cm}
\caption{Best-fit spectral parameters of \mbox{HESS J1702-420A} and \mbox{HESS J1702-420B}.}
\label{3dtable-spectral}
\end{table*}

\begin{table*}[ht]
\small
\begin{center}
\begin{tabular}{||c|c|c|c|c||}
    \hline
\text{\textbf{Component name}}&\text{\textbf{Surface brightness above $\mathbf{2}\,\text{TeV}$}}&\text{\textbf{Test statistic (TS)}}&\text{\textbf{Number of d.o.f.}}&\text{\textbf{Significance}}\\
    & $\boldsymbol{[\text{\textbf{cm}}^{-2\,}\text{\textbf{s}}^{-1\,}\text{\textbf{sr}}^{-1}]}$ & & &$\mathbf{[\mbox{\boldmath$\sigma$}]}$ \\
    \hline
\mbox{HESS J1702-420A}& $(6.2\pm 2.6_\text{stat})\times 10^{-8}$& 42 & 5 & 5.4\\
\mbox{HESS J1702-420B}& $(2.5\pm 0.4_\text{stat})\times 10^{-8}$ & 606 & 7 & 23.9\\
    \hline
\end{tabular}
\end{center}\vspace*{-0.3cm}
\caption{Surface brightness and detection significance of \mbox{HESS J1702-420A} and \mbox{HESS J1702-420B}.}
\label{3dtable-detection}
\end{table*}

\subsection{Flux maps and source morphology}\label{traditional}

\begin{figure*}

\centering

\begin{subfigure}[b]{9.1cm}
\centering
\includegraphics[width=\linewidth]{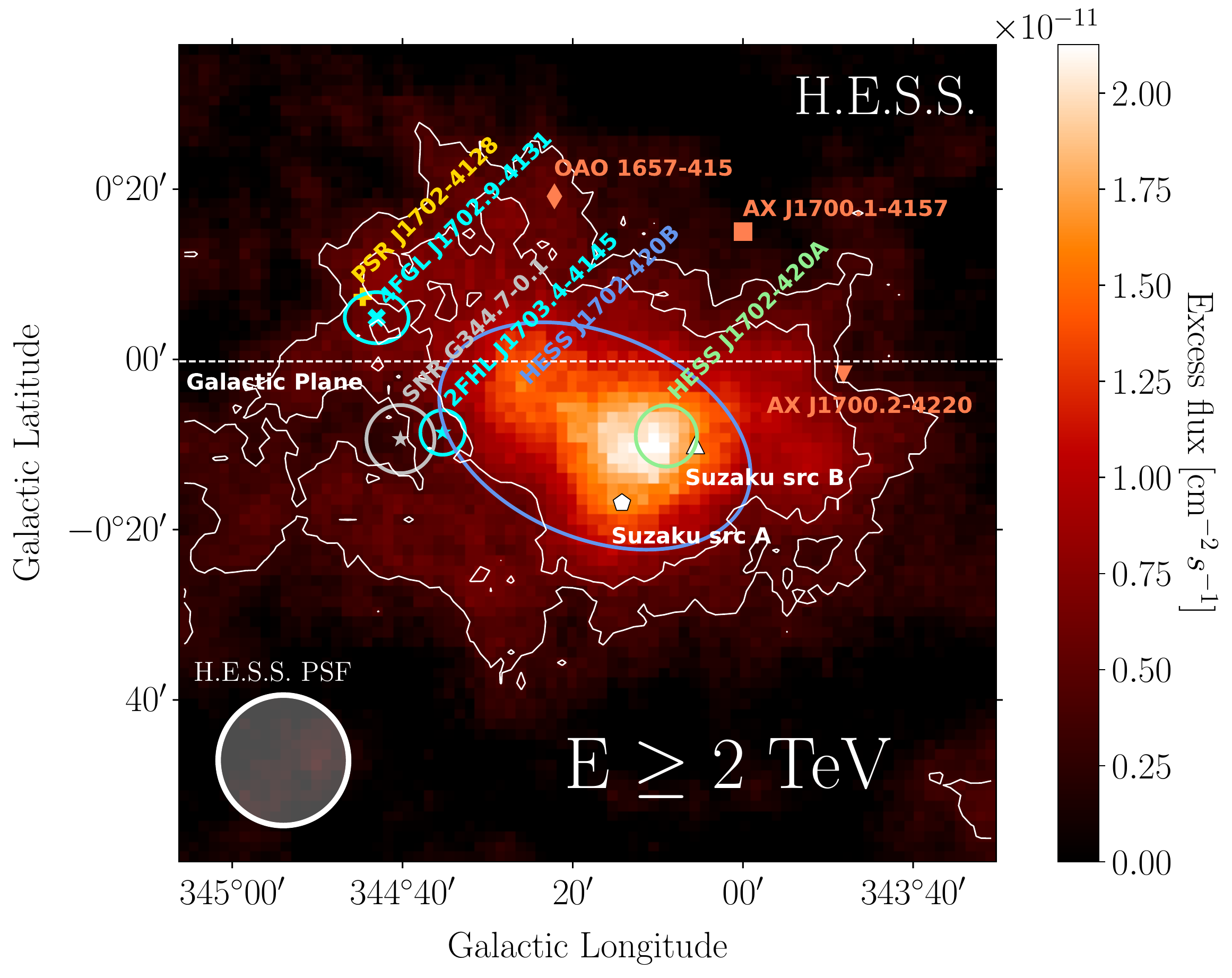}
\label{excess_2}
\end{subfigure}
\begin{subfigure}[b]{9.1cm}
\centering
\includegraphics[width=\linewidth]{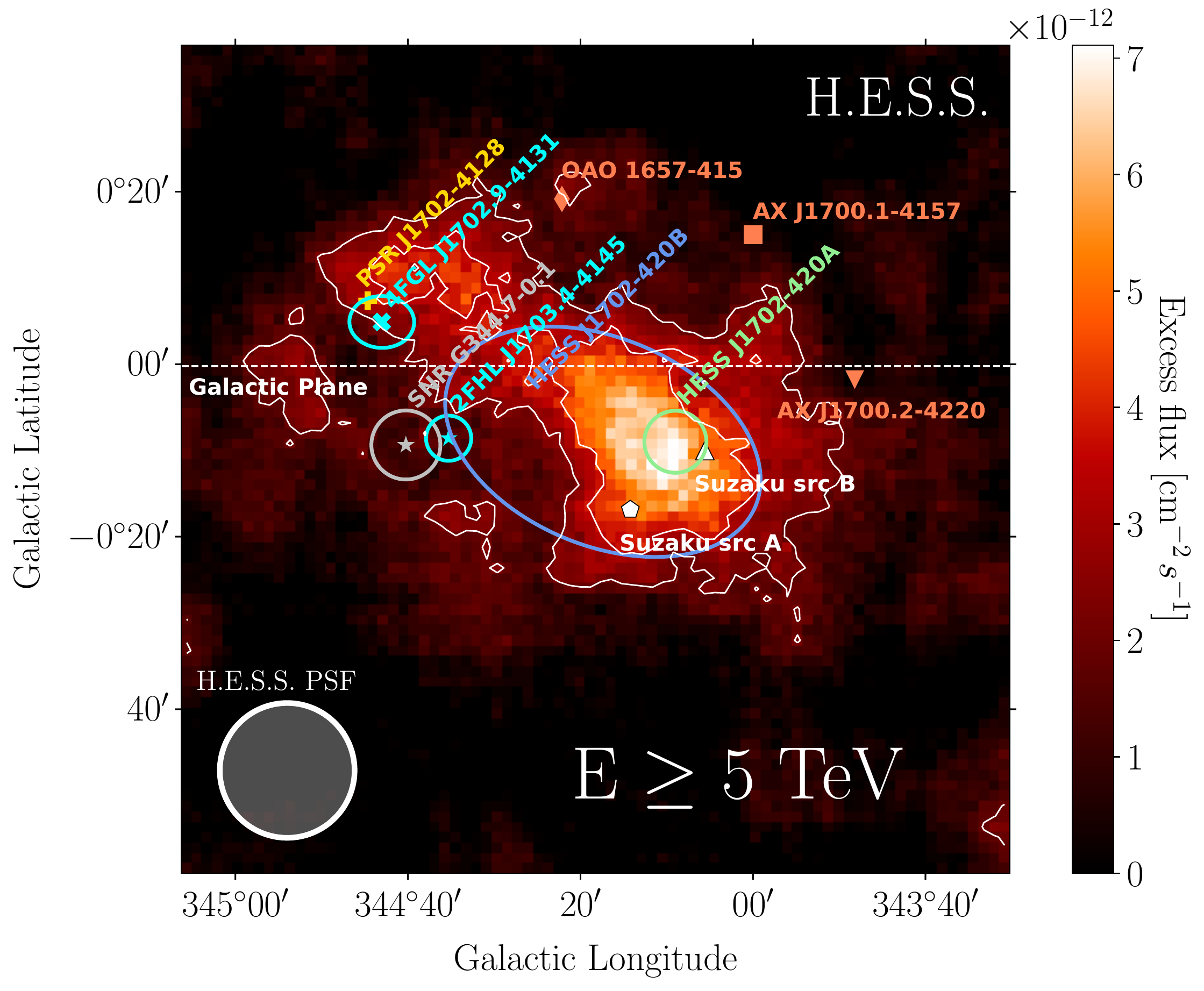}
\label{excess_5}
\end{subfigure}
\begin{subfigure}[b]{9.1cm}
\centering
\includegraphics[width=\linewidth]{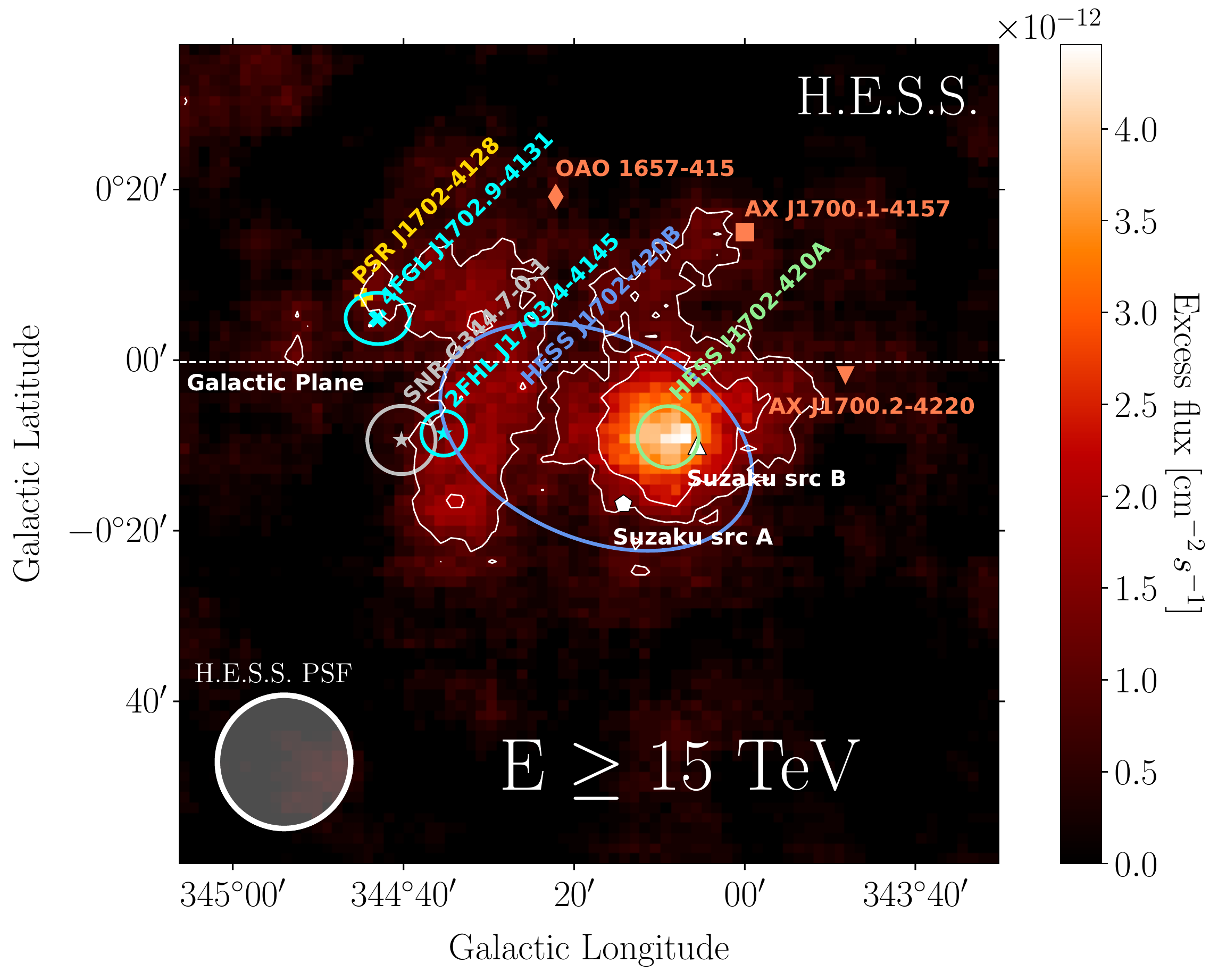}
\label{excess_15}
\end{subfigure}
\begin{subfigure}[b]{9.1cm}
\centering
\includegraphics[width=\linewidth]{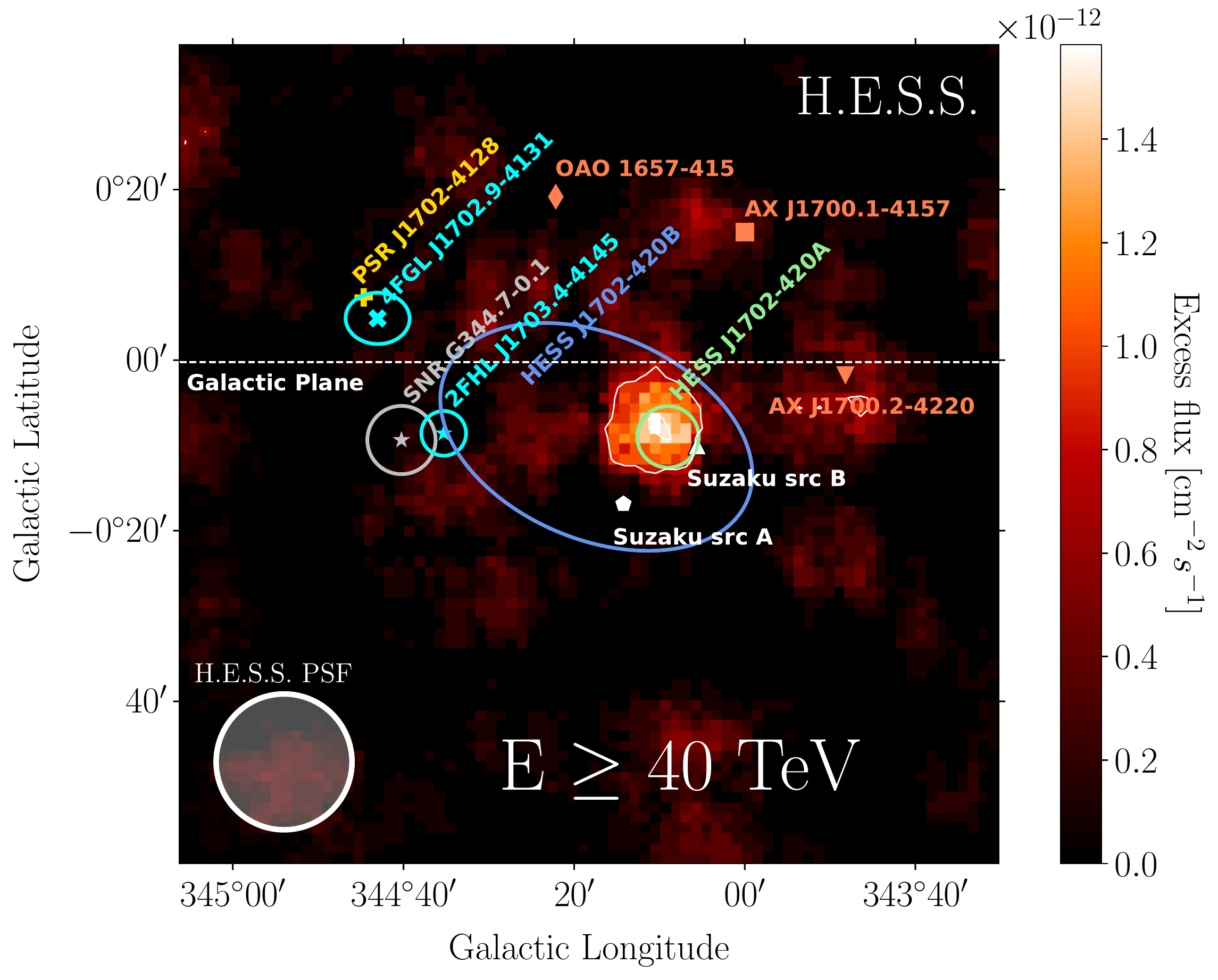}
\label{excess_40}
\end{subfigure}

\caption{$\gamma$-ray flux maps of the \mbox{HESS J1702-420} region, computed with the Ring Background Method, above 2 (\emph{top left}), 5 (\emph{top right}), 15 (\emph{bottom left}) and 40 (\emph{bottom right}) TeV. All maps are correlated with a $0.1^{\,\text{o}}$-radius top-hat kernel, and the color code is in unit of $\gamma$-ray flux (cm$^{-2\,}$s$^{-1}$) per smoothing area. The white contours indicate the $3\sigma$ and $5\sigma$  H.E.S.S.\ significance levels~\citep{liema}. The cyan markers indicate the position (surrounded by uncertainty ellipses) of the \emph{Fermi}-LAT sources 4FGL~J1702.9-4131 and 2FHL~J1703.4-4145. The former is associated with the PSR~J1702-4128 (in yellow),  the latter with the SNR~G344.7-0.1 (in gray). The circle around the SNR represents its angular extension~\citep{giancani}. The white pentagon and upward-pointing triangle represent unidentified X-ray \emph{Suzaku} sources. The orange markers show the positions of nearby X-ray binaries. Finally, the center and $1\sigma$ extension of \mbox{HESS J1702-420A} (\mbox{HESS J1702-420B}) are indicated in green (blue). In the bottom-left corner of each panel the 68\% containment radius of the H.E.S.S.\ PSF is shown, which --- for the chosen analysis configuration --- does not have a strong dependency on the energy.}

\label{ring}
\end{figure*}

As a complementary study, we performed a 2D  analysis of the energy-integrated morphology of \mbox{HESS J1702-420}  in different energy bands. This technique is useful to assess the overall source morphology and verify the persistence of the TeV emission up to the highest energies, even if it does not allow to disentangle \mbox{HESS J1702-420A} from \mbox{HESS J1702-420B}.  The level of cosmic ray background in the region was estimated using the adaptive ring background estimation method~\citep{berge, carrigan}. We also verified that consistent flux and significance distributions can be obtained with the FoV background estimation method (Section~\ref{3d}). After subtracting the $\gamma$-like hadronic background, we  measured \mbox{$\gamma$-ray} flux integrated above 2, 5, 15 and $40\,\text{TeV}$ inside a $1.6^{\,\text{o}}\times\,1.6^{\,\text{o}}$ region encompassing \mbox{HESS J1702-420}. The result  is shown in Figure~\ref{ring}.  
 The figure suggests a shrinking of the VHE emission at high energy, with a  shift of the \mbox{$\gamma$-ray} peak toward the position of the unidentified source \emph{Suzaku} src B. Based on the 3D analysis results (Section~\ref{3d}), this effect is understood as the transition between a low energy regime --- dominated by the steep spectrum of \mbox{HESS J1702-420B} --- to a high energy one, in which \mbox{HESS J1702-420A} stands out with its exceptionally hard power law spectrum. Quantitatively, the distance between the low and high energy emission peaks --- estimated from the distance between the centroids of \mbox{HESS J1702-420A} and \mbox{HESS J1702-420B} --- amounts to $(0.14\pm 0.04_\text{stat}\pm 0.02_\text{sys})^{\,\text{o}}$.

\section{Multi wavelength observations}\label{mwl}
Even in the absence of a multi wavelength detection of \mbox{HESS J1702-420},   low-energy observations can help to constrain  the TeV emission scenarios. Section~\ref{fermi} summarizes a dedicated analysis of archival \emph{Fermi}-LAT data in the \mbox{HESS J1702-420} region, while Section~\ref{ism} reports   our considerations on the surrounding ISM and Section~\ref{xxx} discusses archival \emph{Suzaku} measurements in the context of the new H.E.S.S.\ results.

\subsection{\emph{Fermi}-LAT data analysis and results}\label{fermi}

\begin{figure}
\centering
\includegraphics[width=\linewidth]{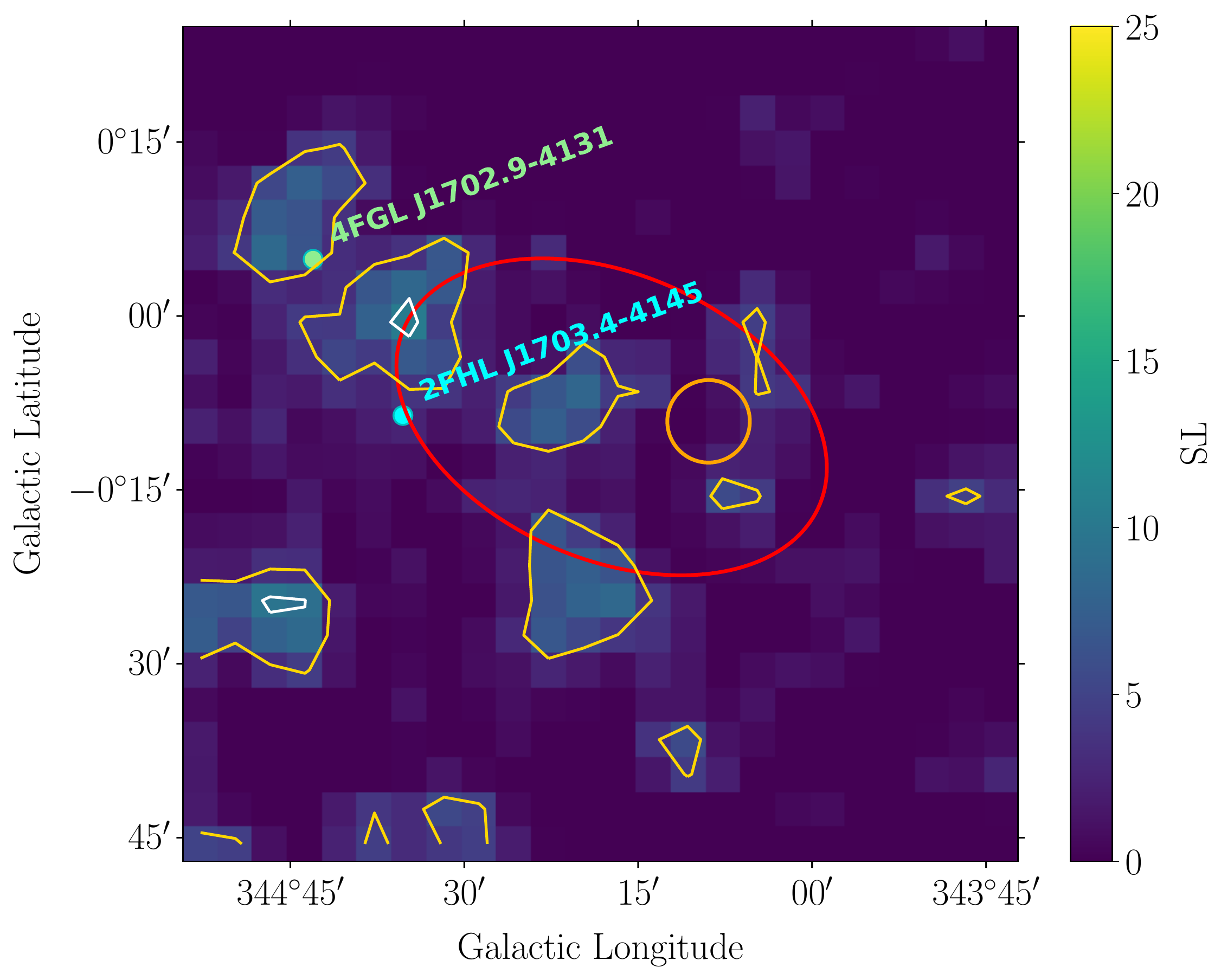}
\caption{Residual TS map after source modeling in the RoI (see the main text for details). The yellow and white contours represent the $\text{TS}=4$ ($2\sigma$) and $\text{TS}=9$ ($3\sigma$) significance levels, respectively. The red ellipse (orange circle) correspond to the $1\sigma$ shape of \mbox{HESS J1702-420B} (\mbox{HESS J1702-420A}). The positions of nearby 4FGL and 2FHL sources are shown as green and cyan circles, respectively.}
\label{fig:fermits}
\end{figure}

Launched in 2008, the \emph{Fermi}-LAT is a pair-conversion instrument  sensitive to the HE \mbox{$\gamma$-ray} domain~\citep{atwood}. 
We analyzed $\approx12 \,\text{yr}$ of events with energies in the $10-900\,\text{GeV}$ interval, within a $10^{\,\text{o}}\times\,10^{\,\text{o}}$ RoI encompassing \mbox{HESS J1702-420}. Event selection and binning criteria are detailed in Appendix \ref{fermi details}. The analysis, performed with \texttt{fermipy}~\citep{wood}, made use of Pass 8 IRFs~\citep{pass8}. 

To build the source model, we selected all sources from the Fourth \emph{Fermi} General Catalog (4FGL), \citet{4fgl}, and second \emph{Fermi}-LAT Catalog of High Energy Sources (2FHL), \citet{2fhl}, within $20^{\,\text{o}}$ from the RoI center. In the model, we also included recent diffuse \mbox{$\gamma$-ray} emission templates, both Galactic  and extra-Galactic. More details on the source modeling can be found in Appendix~\ref{fermi details}.
After the maximum likelihood fit, we produced a TS map to investigate the presence of statistically significant excesses. For each spatial bin, the algorithm compared the maximum log-likelihood obtained by fitting the model, with the addition of a point source ($\Gamma=2$ frozen, amplitude free) at that position, with that of the starting model alone (null hypothesis). We  verified that the TS map does not significantly depend on the spectral index or spatial morphology chosen for the test source.  The TS map displayed in Figure~\ref{fig:fermits}  shows that, within the source region, there is no evidence for a significant excess, but some low-significance fluctuations are present. For comparison,   Figure~\ref{fig:fermicounts} shows a TS map computed before removing the contribution from point sources.   

Finally, we included an additional model component  defined by a power law spectrum and an elliptical Gaussian morphology identical  to the spatial model of \mbox{HESS J1702-420B} (Section~\ref{alphabeta}). Its $1\sigma$ contour is indicated by the red ellipse in Figure~\ref{fig:fermits}. We left free to vary the normalization and index of source spectrum, performed a maximum likelihood fit and compared the resulting model likelihood with the null hypothesis (no source). 
We found only marginal significance ($4.3\sigma$) for a positive excess corresponding to the chosen Gaussian template. In the absence of a clear detection, we estimated the 99\% confidence-level upper limit for the HE emission, associated with the \mbox{HESS J1702-420B} template shape, at the level of 
\begin{linenomath}
\begin{equation} 
\left(E^2\,\frac{dN}{dE} \right)_{E=E_\text{ref}}\leq 7.6\times 10^{-9}\,\text{GeV}\text{cm}^{-2}\,\text{s}^{-1}\,\,,
\end{equation}
\end{linenomath}
where $E_\text{ref}\approx 95\,\text{GeV}$ is the geometric mean of the \emph{Fermi}-LAT energy range. This value was used  to constrain the low-energy extrapolation of the \mbox{HESS J1702-420B} spectrum to the \emph{Fermi}-LAT energy range (Section~\ref{betanaima}).

\subsection{The interstellar medium}~\label{ism}
Observations of the southern Galactic plane in the 109--115$\,$GHz radio band with Mopra~\citep{mopra}, together with the Southern Galactic Plane Survey of the $\lambda=21\,$cm line emission with the ATCA and Parkes telescopes~\citep{sgps}, allow the study of the molecular and atomic gas distribution in the direction of \mbox{HESS J1702-420}.

The gas densities along the line of sight were measured by~\citet{lau}, integrating the velocity peaks within a $0.30^{\,\text{o}}\times\,0.15^{\,\text{o}}$ ellipse centered at $l\,$=$\,344.30^{\,\text{o}}$ and $b\,$=$\,-0.18^{\,\text{o}}$. This choice of integration region reflected the approximate shape of the TeV source, from~\citet{dark}. Based on the new H.E.S.S.\ observations and improved analyses  presented in this paper (Section~\ref{hess}), we repeated the ISM analysis, with the same radio dataset and approach as in~\citet{lau} but adopting a smaller extraction window to focus on \mbox{HESS J1702-420A}. 
Our conclusions agree with~\citet{lau}, in that dense target material, although present at various distances along the line of sight (see Figure~\ref{fig:clouds}), does not exhibit any obvious correlation with the VHE \mbox{$\gamma$-ray} maps (see Figure~\ref{clouds-integrated}). In particular, no hydrogen cloud clearly correlates with \mbox{HESS J1702-420A} or \mbox{HESS J1702-420B}.  

\subsection{Comparison with X-ray observations of \mbox{HESS J1702-420}}\label{xxx}
In the X-ray domain, deep \emph{Suzaku} observations of the \mbox{HESS J1702-420} region revealed the presence of two   faint point-like objects (src A and src B, indicated in Figure~\ref{ring}) and the absence of diffuse X-ray emission  in the \emph{Suzaku} FoV, whose dimensions were however insufficient to fully cover the whole TeV source~\citep{fujinaga}. 
\emph{Suzaku} src B, in particular, is positionally close to the newly discovered component \mbox{HESS J1702-420A} (see Figure~\ref{ring}), which might hint at the first multi wavelength association for \mbox{HESS J1702-420}. For \emph{Suzaku} src B \citet{fujinaga}  estimated a very low flux (in the 2-$10\,\mathrm{keV}$ band) of $(1.9\pm 0.7)\times10^{-14}\,\mathrm{erg}\,\mathrm{s}^{-1}\,\mathrm{cm}^{-2}$, and did not report any evidence of source extension linked with a compact PWN. However, we point out that  the \emph{Suzaku} measurement likely suffered from strong systematics at the position of src B. Indeed, referring to Figure 2 in \citet{fujinaga}, it appears that src B  was probably not fully contained in the \emph{Suzaku} FoV, thus leading to an underestimated flux. The   higher level of X-ray fluctuations in the corner surrounding src B  suggests that the actual level of diffuse emission at that position  might be larger than elsewhere in the FoV. Therefore, an association between \mbox{src B} and  \mbox{HESS J1702-420A} cannot be ruled out at this stage. 

\section{Discussion}\label{dddd}
To model the \mbox{$\gamma$-ray} emission of \mbox{HESS J1702-420A} and \mbox{HESS J1702-420B},   we replaced the power law  spectral functions  that were used in the 3D analysis (Section~\ref{3d}) with simple physically-motivated non thermal radiative models from \href{https://naima.readthedocs.io/en/latest/}{\texttt{naima}}~\citep{naima}. We derived  the  present-age  spectral shape of the parent cosmic ray population, exploring both hadronic and leptonic one-zone emission scenarios. Owing to the \texttt{NaimaSpectralModel} class implemented in \texttt{gammapy}, we could forward-fold the \texttt{naima} radiative models  directly on the H.E.S.S.\ 3D data. This  represents a significant improvement with respect to a simple fit to precomputed flux points, which is inevitably biased by the spectral assumption made for the flux point computation. 

Because of the unclear level of association between \mbox{HESS J1702-420A} and \mbox{HESS J1702-420B}, we modeled them independently.  For the hadronic emission models, based on the analytic parametrization of $p$-$p$ interaction and $\pi^0$ decay developed in~\citet{kafexhiu}, we assumed a fixed target density $n_\text{H}=100\,\text{cm}^{-3}$. During the fit, a fiducial distance from Earth of $d=3.5\,\text{kpc}$ was assumed. We note that the gas density, as well as the source distance from Earth, do not influence the spectral shape and are both degenerate with the source intrinsic luminosity, that may be rescaled a posteriori assuming different values of $n_\text{H}$ and $d$ (see for example Eq.~\ref{protonalpha}).  In the leptonic scenario, based on the analytic approximation presented in~\citet{khangulyan_IC}, the VHE \mbox{$\gamma$-ray} emission was attributed to inverse-Compton up-scattering by electrons of the cosmic microwave background (CMB) and infrared (IR) low-energy photon fields.  The uniform CMB field was described as a black-body radiation with energy density of \mbox{$\epsilon_\text{CMB}=0.261\,\text{eV}\,\text{cm}^{-3}$ }and temperature of $T_\text{CMB}=2.73\,\text{K}$. The   starlight emission in the near IR ($\epsilon_\text{NIR}=1\,\text{eV}\,\text{cm}^{-3}$ and $T_\text{NIR}=3000\,\text{K}$) and dust re-emission in the far IR ($\epsilon_\text{FIR}=0.5\,\text{eV}\,\text{cm}^{-3}$ and $T_\text{FIR}=30\,\text{K}$) were obtained using the 3D interstellar radiation field (ISRF) model from~\citet{isrf}, at the coordinates of \mbox{HESS J1702-420} and the assumed $3.5\,\text{kpc}$ distance. We verified that the  level of fluctuations of the ISRF along the line of sight did not significantly impact the modeling conclusions.
The results are discussed in Sections~\ref{alphanaima} and~\ref{betanaima}. 

\begin{figure*}

\centering
\begin{subfigure}[b]{0.45\linewidth}
\includegraphics[width=\linewidth]{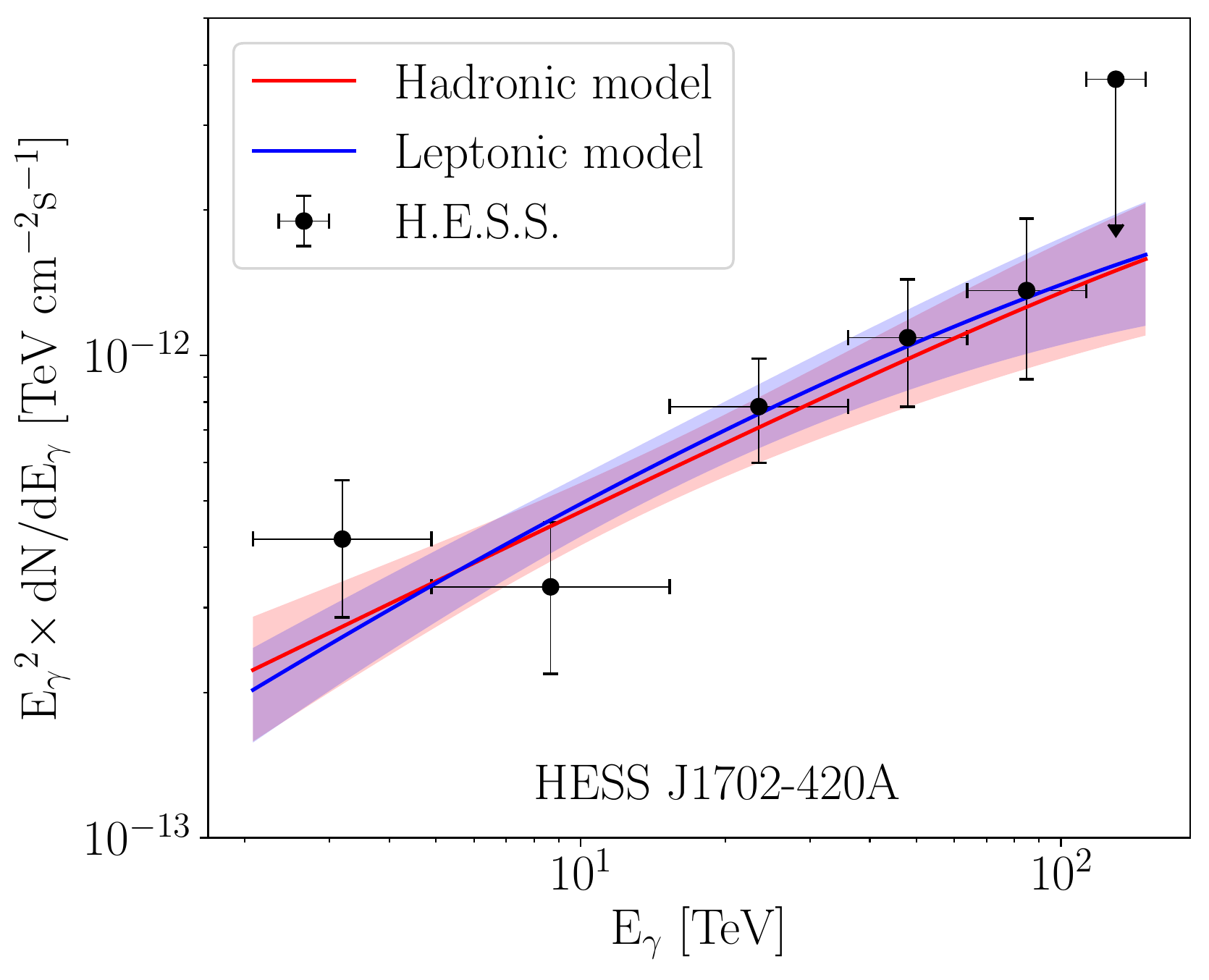}
\end{subfigure}
\hspace*{0.5cm}
\begin{subfigure}[b]{0.45\linewidth}
\includegraphics[width=\linewidth]{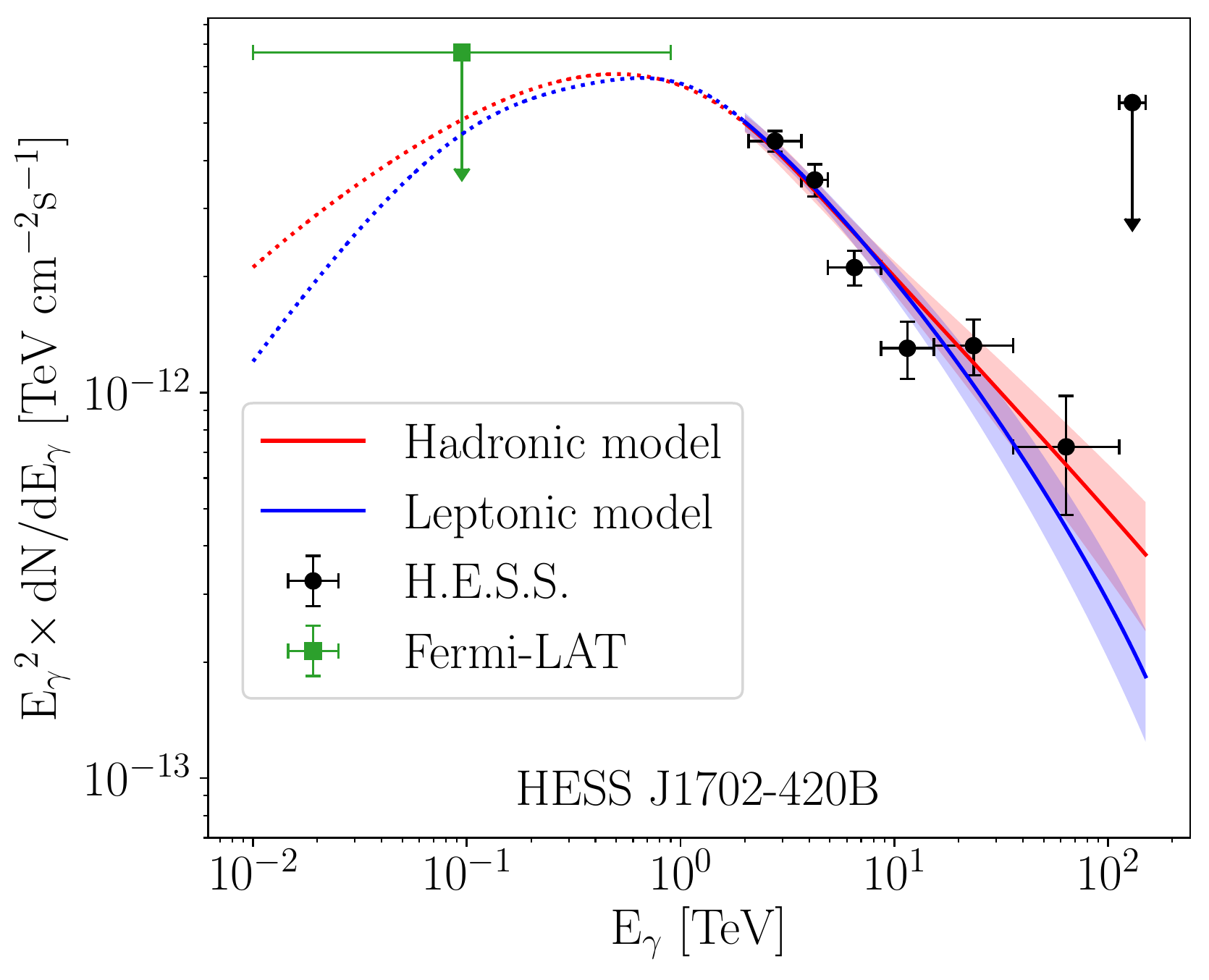}
\end{subfigure}
\caption{ Models of \mbox{$\gamma$-ray} emission based on hadronic (red) and leptonic (blue) one-zone scenarios, for \mbox{HESS J1702-420A} --- \emph{left} panel --- and \mbox{HESS J1702-420B} --- \emph{right} panel. The best-fit spectra, under the assumption of simple power law distribution of the underlying particle populations, are shown as solid lines, while the shaded areas and dotted lines represent the $1\sigma$ statistical error envelope and extrapolations outside the fit range, respectively. The H.E.S.S.\ and \text{Fermi}-LAT flux points are also shown, for reference purpose. Fit results were obtained with a 3D fit of H.E.S.S.\ data.}
\label{alphabetanaima}
\end{figure*}

\subsection{\mbox{HESS J1702-420A}}\label{alphanaima}
\mbox{HESS J1702-420A} has one of the hardest \mbox{$\gamma$-ray} spectra ever detected in a VHE \mbox{$\gamma$-ray} source. This means that the spectral indices of the underlying particle distributions, responsible for the \mbox{$\gamma$-ray} flux via hadronic or leptonic processes, have to be extremely hard themselves.
A pure power law distribution of protons (electrons) with slope $\Gamma_{p}=1.58 \pm 0.14_\text{stat}$ ($\Gamma_{e}=1.61 \pm 0.15_\text{stat}$) is well suited to produce the  \mbox{$\gamma$-ray} emission of \mbox{HESS J1702-420A}, via hadronic (leptonic) radiative processes.  
The two spectra, with their $1\sigma$ butterfly envelopes, are shown in Figure~\ref{alphabetanaima} (left panel), where the H.E.S.S.\ spectral points --- see Table~\ref{tab:fp} --- are shown for reference purpose only as they were not used for the fit.

Based on the currently available H.E.S.S.\ data, any attempt of fitting an additional parameter for the cut-off energy of the particle spectra    led either to a non-covergence of the fit or to an unphysically high  cutoff energy value.
We therefore computed   lower limits on the particle cut-off energy, following the procedure described in Appendix~\ref{ll}. To estimate the lower limits, we  modified the model likelihood adding a Gaussian prior on the particle spectral index, to prevent it from  floating toward nonphysical regions (i.e., very small or even negative values), due to the trial of  low cut-off energies and the reduced lever arm for this spectral modeling. In the case of the hadronic model, we assumed as a prior a Gaussian distribution centered at $\Gamma_{p}=2$ and with $\sigma=0.5$, based on standard diffusive shock acceleration theory (DSA; \citet{bell}). We estimated the impact of this prior choice by varying the Gaussian central values to $\Gamma_{p}=1.7$ and $\Gamma_{p}=2.3$.  We  found that for a prior centered at $\Gamma_{p}=2$ (1.7, 2.3) the 95\% confidence-level lower limit on the proton cut-off energy is 0.82 (0.55, 1.16)$\,$PeV. The fact that --- independenly of the chosen prior --- the cut-off energy lower limit is found at $E_{p}>0.5\,\text{PeV}$  means that in a hadronic scenario the source likely harbors PeV cosmic rays.  In the leptonic case, we tested three different Gaussian priors, all having width $\sigma=0.5$. Based on \citet{sironi} and \citet{werner}, we chose a prior centered at $\Gamma_{e}=1.5$ to probe shock-driven magnetic reconnection in conditions of moderate wind magnetization,  $\Gamma_{e}=2.5$ to account for  Fermi-like acceleration at the termination shock in conditions of low upstream magnetization, and finally $\Gamma_{e}=2.0$ as an intermediate scenario. Our results showed that assuming $\Gamma_{e}=2.0$ (1.5, 2.5) the 95\% confidence-level lower limit on the electron cut-off energy is 106 (64, 152)$\,$TeV.

The energy contents in protons and electrons, necessary to sustain the \mbox{$\gamma$-ray} emission of \mbox{HESS J1702-420A}, were computed integrating the particle spectra above $1\,\text{TeV}$:
\begin{equation}\label{energ}
W_{p/e}(E_{p/e}>1\,\text{TeV}) = \int_{1\,\text{TeV}}^\infty E_{p/e} \frac{dN}{dE_{p/e}}\,dE_{p/e}\,\,.
\end{equation}
Given the best-fit proton and electron distributions found for \mbox{HESS J1702-420A}, with power law indices $\Gamma_{p/e}\approx1.6$, Eq.~\ref{energ} would diverge unless the presence of a high energy cut-off is assumed. We therefore adopted the 95\% confidence level lower limits on the cut-off energies, thus obtaining lower limits on the integrated particle energetics. We verified that the results are not strongly influenced by the choice of spectral index prior. They are:
\begin{linenomath}
\begin{equation}
W_p(E_p>1\,\text{TeV})\gtrsim  1.8\times 10^{47}\left(\frac{d}{3.5\,\text{kpc}}\right)^2\left(\frac{n_\text{H}}{100\,\text{cm}^{-3}}\right)^{-1}\,\text{erg} 
\label{protonalpha} 
\end{equation}
\end{linenomath}

\begin{linenomath}
\begin{equation}
W_e(E_e>1\,\text{TeV})\gtrsim  8.1\times 10^{45}\left(\frac{d}{3.5\,\text{kpc}}\right)^2\,\text{erg}\,\,.
\label{electronalpha}
\end{equation}
\end{linenomath}

In a leptonic scenario,  \mbox{HESS J1702-420A} would be powered by an electron popultion with unusually hard spectral index, $\Gamma_e\approx1.6\,$, and the electron energy required to power the $\gamma$-ray emission (see Eq.~\ref{electronalpha}) would be high compared to the typical   values for TeV detected PWNe~\citep{hesspwn}. A simple one-zone leptonic model is therefore challenged, also because it would imply the unlikely presence of inverse-Compton emitting electrons with $E_e\approx100\,\mathrm{TeV}$. Indeed, given the $\propto 1/E_e$ dependence of the synchrotron loss  timescale   in the Thomson regime (see Eq.~\ref{tausyn}),  such energetic electrons  would cool down extremely fast creating a high energy spectral curvature or break, which is not observed for \mbox{HESS J1702-420A}.   To further understand whether a pulsar-PWN association between \emph{Suzaku} src B (see Section~\ref{xxx}) and \mbox{HESS J1702-420A} is plausible, we made use of the simple one-zone leptonic model derived in this Section to match the synchrotron emission of \mbox{HESS J1702-420A} with the measured X-ray flux of src B, thus estimating the magnetic field value in the vicinity of the source. This turns out to be unrealistically low: $B\approx 0.3\,\mu G$ (see Figure~\ref{fig:suzaku}). In other words, if the \emph{Suzaku} measurement is reliable (see Section~\ref{xxx}) an association between \mbox{HESS J1702-420A} and \emph{Suzaku} src B is very unlikely in a simple one-zone leptonic scenario. For all these reasons, a standard PWN model is disfavored, but cannot be definitively ruled out   mainly due to the uncertainties on the X-ray measurement. We notice that an alternative interpretation is possible, in which the observed $\gamma$-ray emission is due to electrons that are accelerated by the reconnection electric field at X-points in the current sheets of a pulsar striped wind, where the magnetic field value is expected to be low~\citep{sironi, werner, guo}. 
In this case, a Doppler boost of the VHE emission due to relativistic plasma motions might be invoked to explain the apparent presence of $100\,\text{TeV}$ inverse-Compton emitting electrons~\citep{cerutti}.  If true, this would be the first time that a TeV measurement probes the reconnection spectrum immediately downstream of the termination shock of a pulsar wind. However, the lack of a clear multi wavelength detection of the compact object providing the necessary electron population renders this hypothesis  unlikely.

In a hadronic scenario, VHE $\gamma$-ray emission is attributed to the interaction of energetic protons with target material within a source or a nearby molecular cloud.   In this case,  the $100\,\mathrm{TeV}$ $\gamma$-ray emission from \mbox{HESS J1702-420A}, together with its proton cut-off energy lower limit at $0.55-1.16\,\mathrm{PeV}$, would make it a compelling candidate site for the presence of PeV cosmic ray protons. Therefore \mbox{HESS J1702-420A} becomes one of the most solid PeVatron candidates detected in H.E.S.S. data, also based on the modest value of the total energy in protons  that is necessary to power its $\gamma$-ray emission (see Eq.~\ref{protonalpha}) and  the excellent agreement of a simple proton  power law spectrum with the data.
However, we notice that a proton spectrum with a slope of $\Gamma_p\approx 1.6$ over two energy decades is hard to achieve in the standard DSA framework~\citep{bell}. This fact may suggest that \mbox{HESS J1702-420A}, instead of being a proton accelerator, is in fact a gas cloud that, being illuminated by cosmic rays transported from elsewhere, acts as  a passive $\gamma$-ray emitter. In that case, the hard measured proton spectrum  could result from the energy-dependent particle escape from a nearby proton PeVatron~\citep{gabici2}.  Alternatively, the $\gamma$-ray emission from \mbox{HESS J1702-420A} might be interpreted  as the hard high energy end of a concave spectrum arising from nonlinear DSA effects~\citep{kang}, or originate from the interaction of SNR shock waves with a young stellar cluster wind~\citep{bykov}.
The absence of a clear spatial correlation between the ISM and the observed TeV emission (see Section~\ref{ism}) prevents  a confirmation of the hadronic emission scenario, unless an extremely powerful hidden PeVatron  is present. In the latter case,  even a modest gas density would suffice to produce the measured \mbox{$\gamma$-ray} emission of \mbox{HESS J1702-420A}, which would explain the observed nonlinearity between the ISM and TeV maps (see Section~\ref{ism}).

\subsection{\mbox{HESS J1702-420B}}\label{betanaima}
The baseline proton  and electron  spectra, used to model the \mbox{$\gamma$-ray} emission of \mbox{HESS J1702-420B}, are broken power laws of the form
\begin{linenomath}
\begin{equation}
             \frac{dN}{dE} \propto \left \{
                     \begin{array}{ll}
                        (E / E_0) ^ {-\alpha_1} & , \text{if}\,\,E <\tilde{E} \\
                        (\tilde{E}/E_0) ^ {\alpha_2-\alpha_1}
                           (E / E_0) ^ {-\alpha_2} & , \text{if}\,\,  E > \tilde{E} \\
                     \end{array}
                   \right.
\end{equation}
\end{linenomath}
where $\tilde{E}$ and $E_0$ are the energy of the spectral break and the reference energy, respectively. The introduction of a spectral break was necessary, because a simple power law extrapolation from the VHE to the HE \mbox{$\gamma$-ray} range would have led to unrealistic energy budgets and an overshoot of the \emph{Fermi}-LAT upper limit (Section~\ref{fermi}). The first power law index, $\alpha_1$, was adjusted manually with respect to the  \emph{Fermi}-LAT upper limit --- its value is therefore not to be interpreted as a fit result, but rather as a working assumption.

In the hadronic (leptonic) scenario, the best-fit proton (electron) spectrum corresponds to a broken power-law with slopes $\alpha_1=1.6$ (1.4) and $\alpha_2=2.66\,\pm 0.11_\text{stat}$ ($3.39\,\pm 0.11_\text{stat}$), and with break energy of $\tilde{E}=(6.77\,\pm 3.64_\text{stat})\,\text{TeV}$ ($\,(4.19\,\pm 1.25_\text{stat})\,\text{TeV}\,$). The 95\% confidence-level lower limit on the proton (electron) cut-off energy --- computed as described in Appendix~\ref{ll} --- is 550 (140)$\,$TeV.

The values of proton and electron energetics, necessary to power the \mbox{$\gamma$-ray} emission of \mbox{HESS J1702-420B}, were computed integrating the broken power law particle spectra above $1\,\text{GeV}$. They are:
\begin{linenomath}
\begin{equation}
W_p(E_p>1\,\text{GeV})\approx  2.8\times 10^{48}\left(\frac{d}{3.5\,\text{kpc}}\right)^2\left(\frac{n_\text{H}}{100\,\text{cm}^{-3}}\right)^{-1}\,\text{erg}
\label{protonbeta}
\end{equation}
\end{linenomath}

\begin{linenomath}
\begin{equation}
W_e(E_e>1\,\text{GeV})\approx 4.5\times 10^{47}\left(\frac{d}{3.5\,\text{kpc}}\right)^2\,\text{erg}\,\,.
\label{electronbeta}
\end{equation}
\end{linenomath}
In a leptonic scenario, \mbox{HESS J1702-420A} and \mbox{HESS J1702-420B} could be seen as different emission zones belonging to the same PWN complex. However, we deem this interpretation unlikely for several reasons. First of all, a leptonic scenario for  \mbox{HESS J1702-420A} is disfavored by the arguments in Section~\ref{alphanaima}.  Also, the only known nearby pulsar is \mbox{PSR J1702-4128}, that to power the whole TeV source would require an extremely high conversion efficiency of its spin down luminosity into $1-10\,\text{TeV}$  \mbox{$\gamma$-rays};
\begin{equation}\label{epsilon}
\epsilon = \frac{L_{[1, 10]\,\text{TeV}}}{\dot{E}}\approx 19\%\,\,,
\end{equation}
where $L_{[1, 10]\,\text{TeV}}$ was obtained considering both \mbox{HESS J1702-420A} and \mbox{HESS J1702-420B}, and assuming the same pulsar's distance from Earth $d=5.2\,\text{kpc}$~\citep{kramer}. The result of Eq.~\ref{epsilon} is well above the efficiency of all other PWNe identified by H.E.S.S.\ in the same energy range~\citep{hesspwn}. 
Finally, several PWNe detected by H.E.S.S.\ have an energy-dependent morphology with spectral softening away from the pulsar position (e.g.,~\citet{j1303, j1825}), which seems not to be the case for \mbox{HESS J1702-420} (see Appendices~\ref{3d-energy-bands} and~\ref{spectral}). However, we point out that this might be due to insufficient statistics or spatial resolution, and that not all TeV-bright PWNe detected by H.E.S.S. have an energy-dependent morphology (e.g.,~\citet{crabext}). 
Therefore, leptonic scenarios cannot be definitively ruled out. In particular, as argued in~\citet{gallant}, the \mbox{PSR J1702-4128} might power only part of the TeV emission. Indeed, significant VHE \mbox{$\gamma$-ray} emission is detected by H.E.S.S. near the pulsar position --- see Figure~\ref{ring} (upper right panel).

In a hadronic scenario, \mbox{HESS J1702-420B} might be interpreted as a proton accelerator, whose spectral break around \mbox{$E_p\approx 7\,\text{TeV}$} is due to energy-dependent cosmic ray escape from the source. In this case, as argued in Section~\ref{alphanaima}, the hard \mbox{$\gamma$-ray} spectrum of \mbox{HESS J1702-420A} could be the signature of delayed  emission from the highest energy runaway protons, hitting target material in the ISM. This scenario is challenged however by the absence of clear TeV$-\,n_\text{H}$ correlation at the location of \mbox{HESS J1702-420A} (see Section \ref{ism}). 

\subsection{Distance from Earth and environmental parameters}\label{theory}

\begin{figure*}
\centering
\begin{subfigure}[b]{0.46\linewidth}
\includegraphics[width=\textwidth]{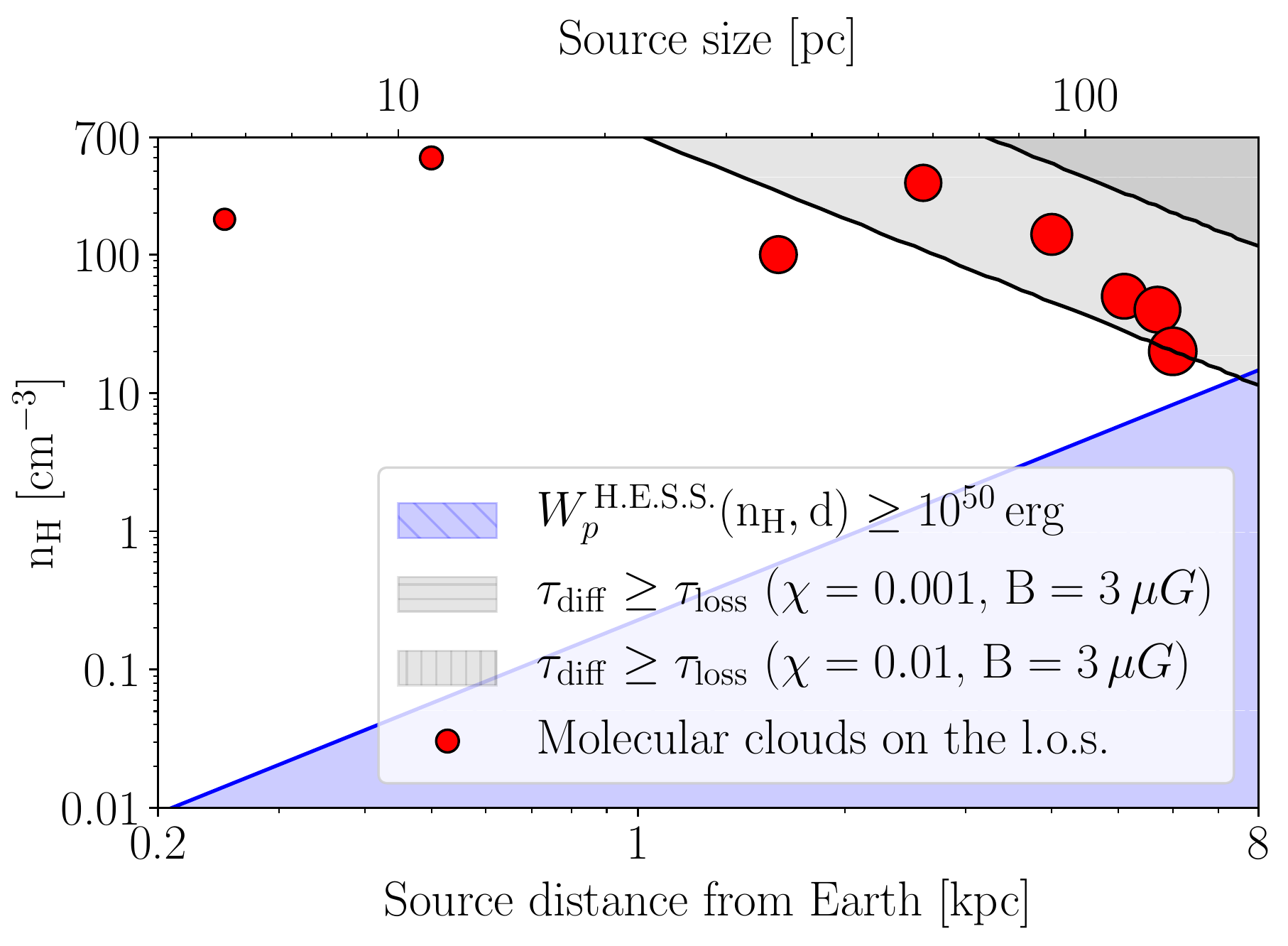}
\end{subfigure}
\begin{subfigure}[b]{0.46\linewidth}
\centering
\includegraphics[width=\textwidth]{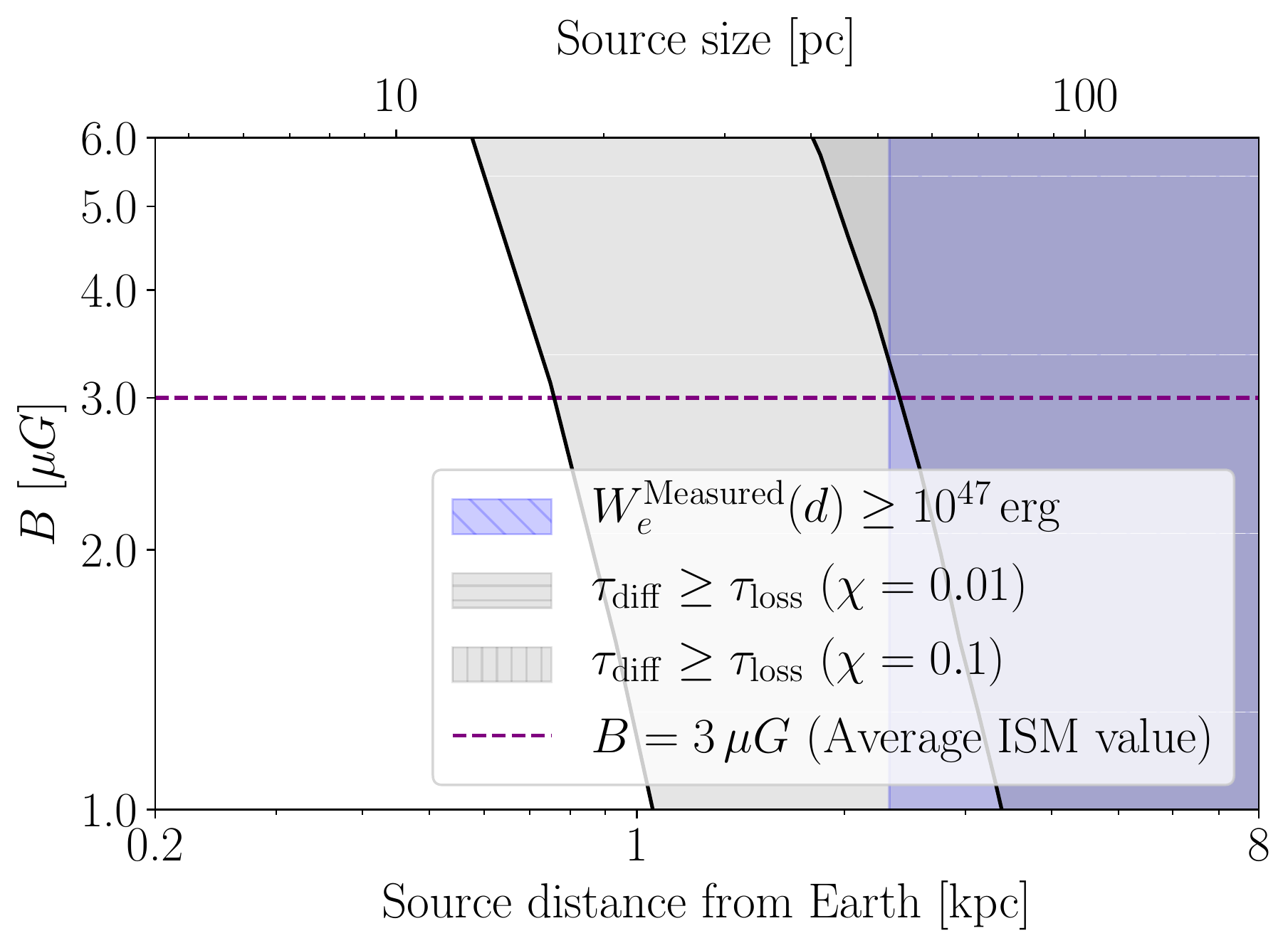}
\end{subfigure}
\caption{Possible constraints on the gas density, magnetic field and distance of \mbox{HESS J1702-420}, based on multi-wavelegth observations, under the assumption of simple one-zone hadronic (left panel) or leptonic (right panel) scenarios. More details are given in the main text.}
\label{discussion}
\end{figure*}

Even if an unequivocal identification of \mbox{HESS J1702-420} remains elusive, mostly due to the uncertain  relationship  between \mbox{HESS J1702-420A} and \mbox{HESS J1702-420B}, the new H.E.S.S.\ observations allow us to constrain the source distance from Earth $d$ and the values of the most relevant environmental parameters in a hadronic or leptonic emission scenario,  which are respectively the gas density $n_\text{H}$ and magnetic field strength $B$.  In this section, we make the assumption that the two components are associated. This means assuming that their distance from Earth is roughly the same, and their TeV emissions are connected.

The constraints we found are shown in Figure~\ref{discussion}.
The left panel focuses on hadronic scenarios: molecular clouds from~\citet{lau} are indicated by red circles, with size proportional to (the logarithm of) the proton energy necessary to power the \mbox{$\gamma$-ray} emission of \mbox{HESS J1702-420B} in each case (see Table~\ref{tab:clouds} for more details). For all clouds, the nearer kinematic distance was assumed. The  blue exclusion region in the figure was obtained requiring that the measured proton energetics above $1\,\text{GeV}$ (from Eq.~\ref{protonbeta}) do not exceed  $10^{50}\,\text{erg}$, which is the kinetic energy   transferred to cosmic rays by a typical SNR. 
Finally, the gray shaded areas exclude portions of the parameter space in which protons of energy $E_p\geq 1\,\text{TeV}$ are cooled down due to $p$-$p$ collisions before having time to diffuse across the whole size of \mbox{HESS J1702-420B}. For this calculation, we assumed a standard ISM magnetic field of $3\,\mu G$, and tested different values for the normalization factor of the diffusion coefficient $\chi$ --- defined in~\citet{gabici}.  It is clear that, if the source lies in the diluted ISM where $n_\text{H}\lesssim 1\,\text{cm}^{-3}$, it has to be relatively close --- $d\lesssim 2\,\text{kpc}$ ---, unless its proton energy budget exceeds $10^{50}\,\text{erg}$. If the normalization of the diffusion coefficient is   low ($\chi\lesssim 0.001$), only the three nearest molecular clouds would be apt to harbor the source, whose distance would again be $d\lesssim 2\,\text{kpc}$. 

In the right panel, which focuses on leptonic scenarios, the gray exclusion areas correspond to portions of the parameter space in which electrons with energy $E_e\geq 1\,\text{TeV}$ do not have time to fill the whole component \mbox{HESS J1702-420B} before being cooled down. 
From the figure, it is clear that if the normalization of the diffusion coefficient is low ($\chi\lesssim 0.01$), then the source has to be relatively close --- less than $\approx 3\,\text{kpc}$ away, for realistic values of $B$ field. 

Further details on the assumptions that were made to produce the Figure~\ref{discussion} can be found in Appendix~\ref{detailsdiscussion}.

\section{Conclusions}\label{conclusions}

We present new H.E.S.S.\ observations of the unidentified source \mbox{HESS~J1702-420}, processed using improved techniques, that bring new evidence for the presence of \mbox{$\gamma$-rays} up to $100\,\text{TeV}$. The low-level analysis configuration, used to reduce the raw telescope data to lists of $\gamma$-like events, was adapted to maximize the telescope's sensitivity at the highest energies.
 We performed a 3D likelihood analysis --- a relatively new high-level technique in the VHE \mbox{$\gamma$-ray} domain --- with \texttt{gammapy}, to determine the simplest and best suited spatial and spectral models to describe  the source and its surroundings. This allowed us to separate for the first time two components --- both detected at $>5\sigma$ confidence level --- inside \mbox{HESS~J1702-420} based on their different morphologies and \mbox{$\gamma$-ray} spectra, both of which extend  with no sign of curvature  up to several tens of TeV (possibly 100\,\text{TeV}).
 We report the $4.0\sigma$ confidence level detection of \mbox{$\gamma$-ray} emission from the hardest component, called \mbox{HESS~J1702-420A}, in the energy band $64-113\,\text{TeV}$, which is an unprecedented achievement for the H.E.S.S.\ experiment and brings evidence for the source emission up to $100\,\text{TeV}$. With a spectral index of $\Gamma=1.53 \pm 0.19_\text{stat} \pm 0.20_\text{sys}$, this object is a compelling candidate site for the presence of PeV cosmic rays.     

We adjusted physically-motivated non thermal radiative models to the H.E.S.S.\ data,  testing simple one-zone hadronic and leptonic models, and determined that the available observations do not allow us to rule out either of the two scenarios. The 95\% confidence level energy cut-off of the baseline proton (electron) distribution of  \mbox{HESS J1702-420A} was found in the range $0.55-1.16\,\mathrm{PeV}$ ($64-152\,\mathrm{TeV}$), depending on the assumption made on the particle spectral index. Remarkably, in a hadronic emission scenario the particle spectral cut-off is at $E_{p}>0.5\,\text{PeV}$, for a range of tested priors. For such a scenario, this implies that the source 
harbors  PeV protons, thus becoming one of the most solid PeVatron candidates detected in H.E.S.S.\ data. Nevertheless, a leptonic emission scenario for \mbox{HESS~J1702-420A} could not be definitively ruled out.
 We additionally measured   the particle energetics that are necessary to power the observed \mbox{$\gamma$-ray} emission.  We finally discussed possible constraints on the source distance, ambient magnetic field and surrounding gas density.

In the future, the improved angular resolution of the Cherenkov Telescope Array (CTA) and higher energy coverage  of the Southern Wide-field Gamma-ray Observatory (SWGO) will possibly close the debate on the nature of \mbox{HESS J1702-420}. In particular, deep measurements in the 100-200$\,\text{TeV}$ \mbox{$\gamma$-ray} band will constrain the spectral shape near the cut-off region, thus probing the hadronic or leptonic origin of the emission and determining  whether either of the two detected components operates as a real cosmic ray PeVatron. Observations in the X-ray band, on the other hand, will be important to search for a multi wavelength counterpart of the TeV source, and clarify the relationship between \mbox{\mbox{HESS J1702-420A}}  and the unidentified \emph{Suzaku} src B.

\section*{Acknowledgements}
{\tiny 
The 
support 
of 
the 
Namibian 
authorities
 and 
of 
the 
University 
of Namibia 
in 
facilitating
the 
construction
 and
 operation 
of 
H.E.S.S. 
is    gratefully
 acknowledged, 
as is the 
support
by 
the 
German 
Ministry 
for 
Education 
and 
Research 
(BMBF), 
the
 Max 
Planck 
Society, 
the
German 
Research 
Foundation 
(DFG), 
the 
Helmholtz 
Association,
 the
 Alexander
 von 
Humboldt 
Foundation,
the 
French 
Ministry
 of 
Higher 
Education, 
Research 
and 
Innovation, 
the
 Centre 
National 
de 
la
Recherche 
Scientifique 
(CNRS/IN2P3 
and 
CNRS/INSU), 
the
 Commissariat 
\`a    
l'\'energie
 atomique
et 
aux 
\'energies 
alternatives 
(CEA), 
the 
U.K.
 Science 
and
 Technology 
Facilities
Council 
(STFC),
the 
Knut 
and 
Alice 
Wallenberg 
Foundation, 
the 
National 
Science 
Centre,
 Poland
 grant 
no. 
2016/22/M/ST9/00382,
the 
South 
African 
Department 
of Science 
and
 Technology 
and
 National 
Research 
Foundation, 
the
University 
of Namibia, 
the 
National 
Commission 
on Research,
 Science 
\&    Technology 
of 
Namibia 
(NCRST),
the 
Austrian 
Federal 
Ministry
 of 
Education, 
Science 
and 
Research 
and
 the
 Austrian 
Science 
Fund 
(FWF),
the 
Australian
 Research 
Council 
(ARC), 
the 
Japan 
Society 
for 
the
 Promotion 
of Science 
and 
by 
the
University 
of Amsterdam. 
We
 appreciate 
the 
excellent 
work
 of 
the
 technical
 support 
staff 
in 
Berlin,
Zeuthen, 
Heidelberg, 
Palaiseau, 
Paris, 
Saclay,
 Tubingen 
and
 in 
Namibia 
in 
the 
construction 
and
operation 
of 
the 
equipment. 
This
 work 
benefited 
from
 services 
provided 
by the
H.E.S.S.
Virtual 
Organisation,
 supported 
by 
the 
national 
resource 
providers 
of 
the
 EGI
Federation.}

\bibliographystyle{aa} 
\bibliography{paper_giunti_HESSJ1702} 

\appendix

\section{Spatially-resolved spectral analysis of H.E.S.S.\ data}\label{spectral}

\begin{figure}
\centering
\begin{subfigure}[b]{\linewidth}
\includegraphics[width=\linewidth]{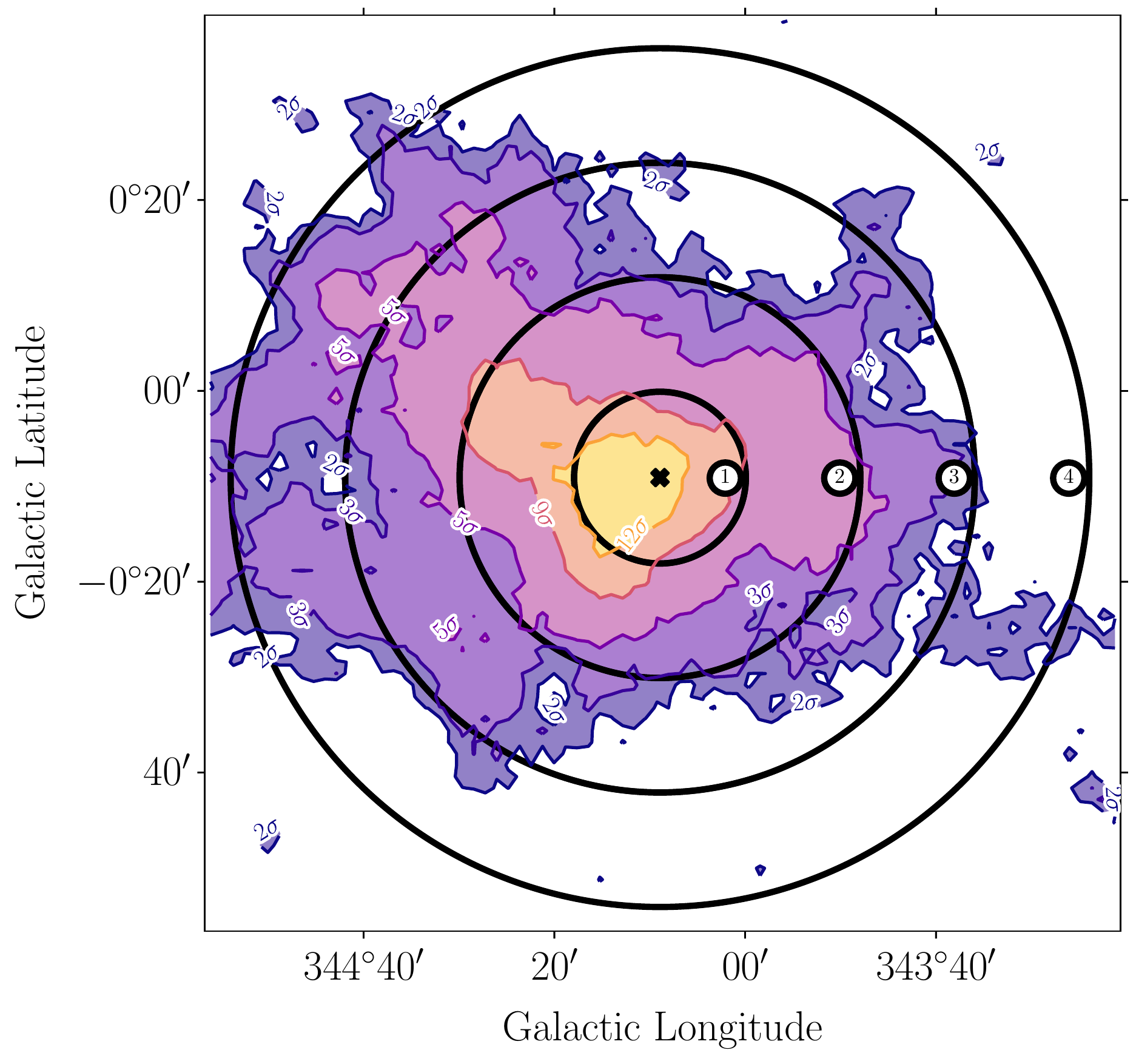}
\end{subfigure}
\begin{subfigure}[b]{\linewidth}
\includegraphics[width=\linewidth]{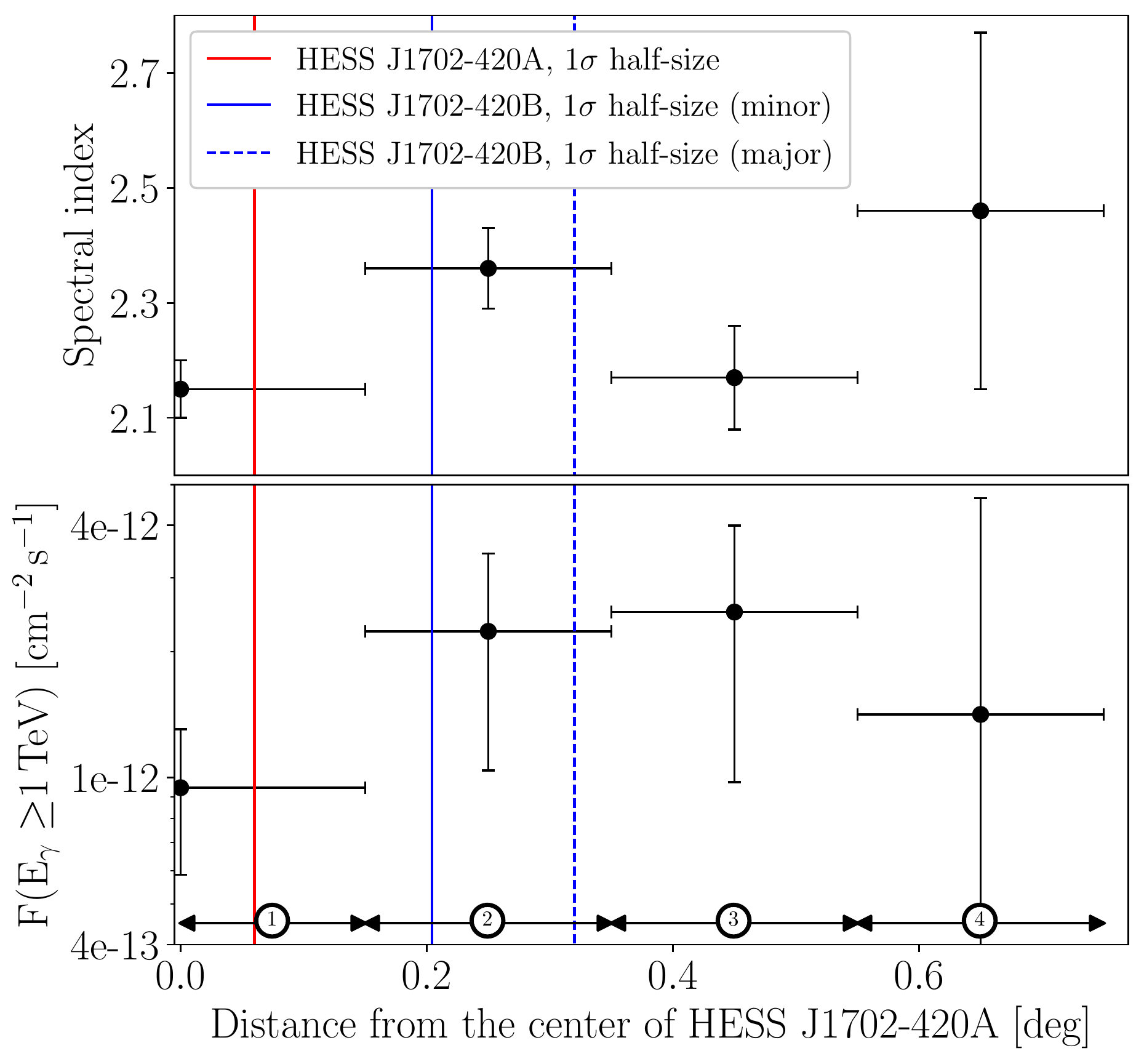}
\end{subfigure}
\caption{\emph{Upper panel:} Map of the H.E.S.S.\ $\gamma$-ray signal significance above $2\,\text{TeV}$, with contours corresponding to $2\sigma$, $3\sigma$, $5\sigma$, $9\sigma$ and $12\sigma$ significance levels~\citet{liema}. The map has been obtained with the Adaptive Ring Background estimation method, and centered at the position of \mbox{HESS J1702-420A}. Overlaid on the map are the concentric regions --- one circle and three annuli --- that were used to extract the source spectrum. \emph{Lower panel:} Results of the spatially-resolved spectral analysis, showing the spectral index and flux as a function of the distance from \mbox{HESS J1702-420A}. }
\label{annuli}
\end{figure}

\begin{table*}[ht]
\begin{center} 
\begin{tabular}{||c|c|c|c|c|c|c|c|c||}
    \hline
\text{\textbf{Region}}&\text{\textbf{Spectral index}}&$\mathbf{E_\text{\textbf{decorr}}}$&$\mathbf{F(E>1\,\text{\textbf{TeV}})}$& \text{\textbf{Livetime}}&\text{\textbf{Area}}& \text{\textbf{Excess}} & \text{\textbf{Significance}}\\
    & & \text{\textbf{[TeV]}}&$\boldsymbol{[\text{\textbf{cm}}^{-2\,}\text{\textbf{s}}^{-1}]}$ & \text{\textbf{[h]}}&\text{\textbf{[sr]}} &  \text{\textbf{[counts]}}& $\mathbf{[\mbox{\boldmath$\sigma$}]}$\\
    \hline
   1 & $2.15 \pm 0.05_\text{stat}\pm0.20_\text{sys}$ & 2.61 &$(9.47 \pm 3.59_\text{stat}\pm2.84_\text{sys})\,10^{-13}$& 32.7 &   $2.15\,10^{-5}$    & 499 & 18.8\\
   
   2 & $2.36 \pm 0.07_\text{stat}\pm0.20_\text{sys}$ & 2.21 &$(2.23 \pm 1.19_\text{stat}\pm0.67_\text{sys})\,10^{-12}$& 24.8 &  $9.57\,10^{-5}$     & 857 & 17.8\\
   
   3 & $2.17 \pm 0.09_\text{stat}\pm0.20_\text{sys}$ & 3.13 &$(2.48 \pm 1.51_\text{stat}\pm0.74_\text{sys})\,10^{-12}$& 15.4 &    $1.72\,10^{-4}$   & 552 & 10.5\\
   
   4 & $2.46 \pm 0.31_\text{stat}\pm0.20_\text{sys}$ & 1.96 &$(1.42 \pm 3.22_\text{stat}\pm0.43_\text{sys})\,10^{-12}$& 7.3 &   $2.49\,10^{-4}$    & 135 & 3.9\\
   \hline
\end{tabular}
\end{center}
\caption{Spectral  results for  the four extraction regions of Figure~\ref{annuli}, under the assumption of   power law \mbox{$\gamma$-ray} emission. }
\label{tab:annuli}
\end{table*}

With the benefit of an unprecedented level of statistics in this region, we performed a spatially-resolved spectral analysis for \mbox{HESS J1702-420}. A $0.15^{\,\text{o}}$-radius circle and three $0.2^{\,\text{o}}$-radius annuli were used to measure the VHE \mbox{$\gamma$-ray} spectrum of the source. We did not adopt narrower extraction regions, in order to limit the level of PSF-induced correlation between them. 
Figure~\ref{annuli} (upper panel) shows the four nonoverlapping regions, overlaid on a map of the $\gamma$-ray flux significance above $2\,\text{TeV}$. 
 All  regions are concentric around Galactic coordinates $l\,$=$\,344.15^{\,\text{o}}$ and $b\,$=$\,-0.15^{\,\text{o}}$, corresponding to the position of \mbox{HESS J1702-420A}. 
 The  level of cosmic ray background within each region was computed with the reflected region background estimation technique~\citep{berge}, while a forward-folding approach~\citep{piron} was adopted to determine the maximum-likelihood estimates of the spectral slope and flux, in each region, under a power-law assumption.

The detailed results of the spectral analysis are reported in Table~\ref{tab:annuli}, while the spectral variations as a function of the distance from \mbox{HESS J1702-420A} are shown in Figure~\ref{annuli} (bottom panel). The error bars in the figure represent the statistical errors on the fitted parameters. The level of systematic uncertainties, reported in Table~\ref{tab:annuli}, have been estimated following \citet{hgps}.  
The figure shows that, in this datasets, there is no  evidence for significant spectral variations around \mbox{HESS J1702-420A}. This measurement tends to support a two-component approach, with respect to a model based on a single source with energy-dependent morphology. Indeed, in the latter case significant spatially-resolved spectral variations would be expected, as seen for other well known H.E.S.S.\ sources (e.g.,~\citet{j1825}).

\section{Hypothesis testing for nested parametric models}\label{wilks}
According to Wilks' theorem~\citep{wilks}, the TS defined in Eq.~\ref{ts} is distributed as a $\chi^2_\nu$, where $\nu$ is the number of additional degrees of freedom of the alternative hypothesis with respect to the null hypothesis. The theorem is valid under the assumptions --- always satisfied in our analysis --- of high statistics and nested models. Thanks to this theorem, the statistical significance of the alternative hypothesis can be directly estimated from the TS value, by determining the corresponding right-tail $p$-value of a $\chi^2_\nu$ distribution. To convert the significance into units of Gaussian standard deviations ($\sigma$), it is then sufficient to compute z-score of a Gaussian distribution corresponding to the given $p$-value. 

\section{Non-confirmed large-scale emission component}\label{largescale}
In the main analysis, a large-scale --- $\approx0.5^{\,\text{o}}$ in radius --- model component was detected around $l\,$=$\,345.23^{\,\text{o}}$ and $b\,$=$\,-0.01^{\,\text{o}}$. At that position, the borders of several  runs partially overlap, resulting in a boosted exposure level but also strong systematics due to edge effects. In the crosscheck analysis, the exposure level at the position of the large-scale component is lower, resulting in a decreased sensitivity. Accordingly, we verified that its inclusion  or exclusion in the source model of the crosscheck analysis did not have any relevant impact on the predicted number of counts at its position. We therefore could not confirm the detection of this new large-scale  emission component nearby \mbox{HESS J1702-420}. We point out that the HGPS already reported the presence of a large-scale component with similar position and size, called \mbox{HGPSG 041},  that was similarly discarded   due to a non-detection in the crosscheck analysis. In the future, new dedicated observations of the region with more uniform exposure will ultimately probe its presence and nature. 

\section{3D analysis of H.E.S.S.\ data in independent energy bands}\label{3d-energy-bands}

\begin{figure}
\centering
\begin{subfigure}[b]{\linewidth}
\includegraphics[width=\linewidth]{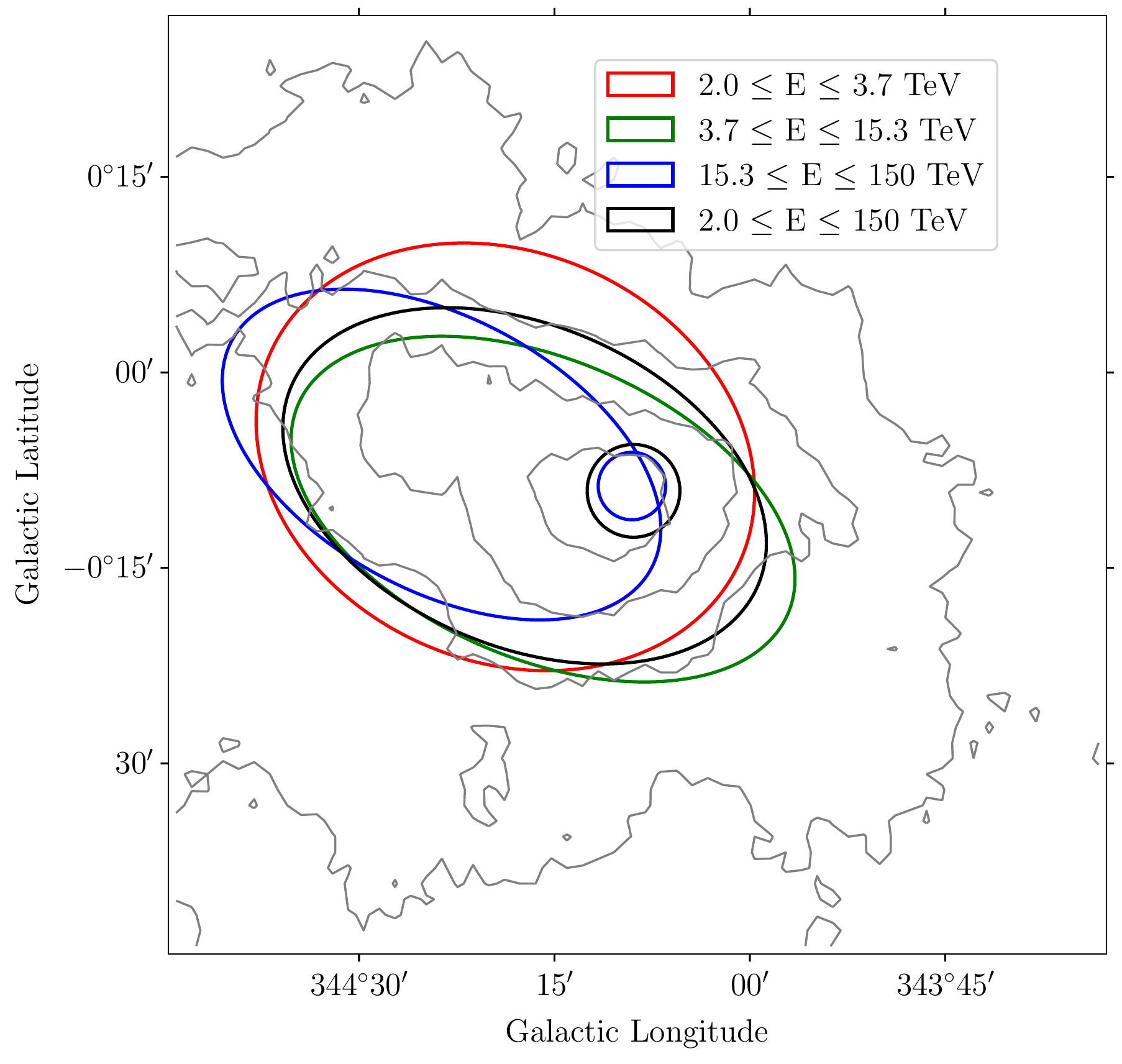}
\end{subfigure}
\begin{subfigure}[b]{\linewidth}
\includegraphics[width=\linewidth]{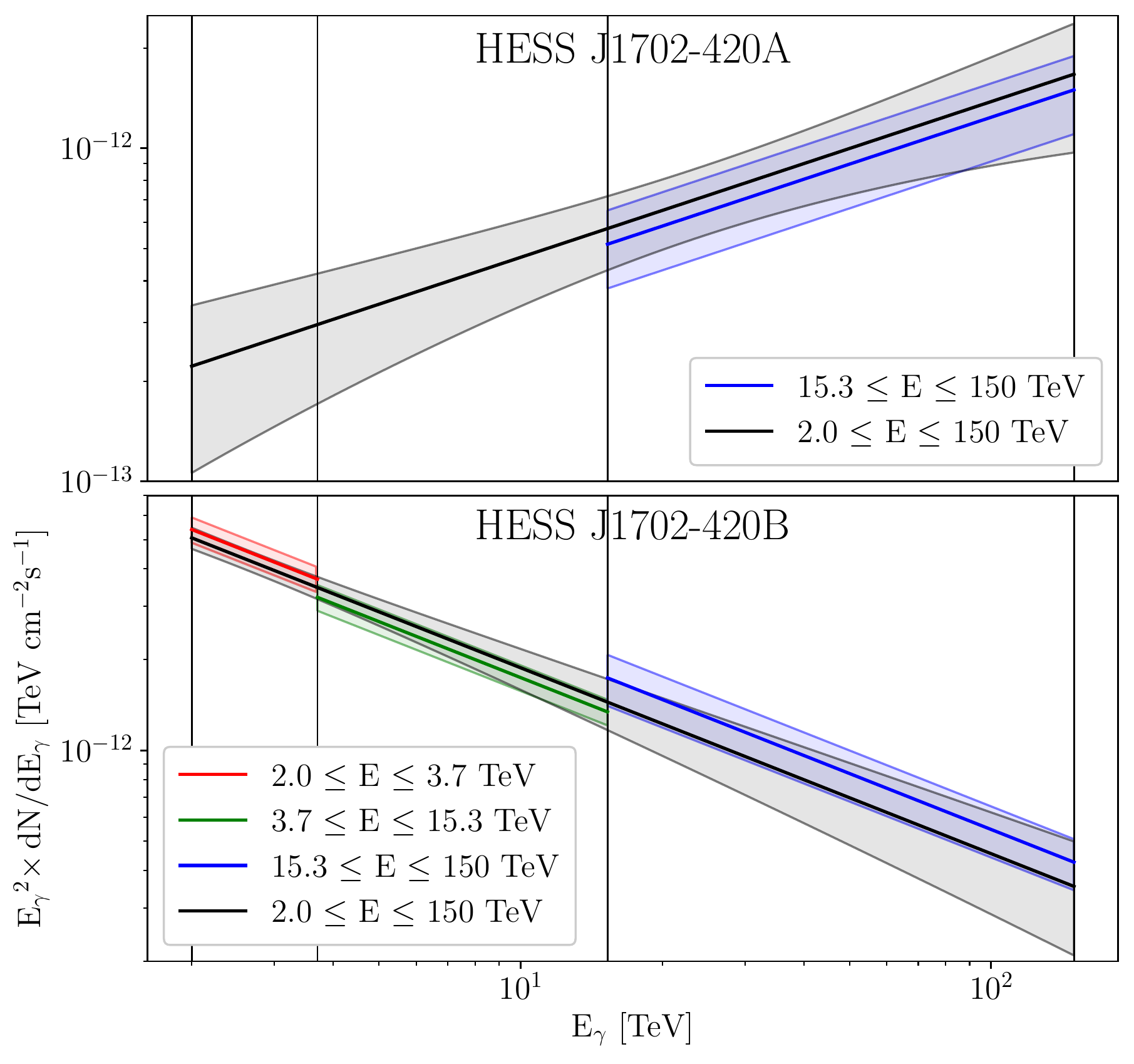}
\end{subfigure}
\caption{\emph{Upper panel:} The figure, centered at the approximate position of \mbox{HESS J1702-420}, shows contours corresponding to 150, 200, 250 and 300 counts detected by H.E.S.S. above $2\,\text{TeV}$ per smoothing area. Overlaid on the map are the $1\sigma$ extension contours of   the components \mbox{HESS J1702-420A} and \mbox{HESS J1702-420B}, as obtained from the 3D fit in separate energy bands. \emph{Lower panel:} spectral results of the energy-resolved 3D analysis, for \mbox{HESS J1702-420A} and \mbox{HESS J1702-420B}. Vertical  lines separate the energy bands that were independently used to perform the source modeling. In both panels, the reference results obtained over the full energy range (see Section~\ref{alphabeta}) are indicated in black.}
\label{fig:3d-energy-bands}
\end{figure}

\begin{table*}[ht]
\begin{center} 
\begin{tabular}{||c|c|c|c||}
    \hline
$\mathbf{E_{\text{\textbf{min}}}}$&$\mathbf{E_{\text{\textbf{max}}}}$& \text{\textbf{Significance of $\mathbf{\mbox{\boldmath\text{\textbf{H}}}_1}$ vs. $\mathbf{\mbox{\boldmath\text{\textbf{H}}}_0}$}}&\text{\textbf{Significance of $\mathbf{\mbox{\boldmath\text{\textbf{H}}}_2}$ vs. $\mathbf{\mbox{\boldmath\text{\textbf{H}}}_1}$}}\\
\text{\textbf{TeV}}&\text{\textbf{TeV}}&$\mathbf{[\mbox{\boldmath$\sigma$}]}$& $\mathbf{[\mbox{\boldmath$\sigma$}]}$\\
    \hline
   2.0 & 3.7 & 19.0 & 2.4\\
   3.7 & 15.3 & 20.2 & 2.0\\
   15.3 & 150 & 12.1 & 5.0\\
   \hline
\end{tabular}
\end{center}
\caption{Significance for the presence of zero ($\text{H}_0$), one ($\text{H}_1$) or two ($\text{H}_2$) model components in each independent energy band, as described in the text (Section~\ref{3d-energy-bands}). The significance was obtained by converting a log-likelihood ratio (see Eq.~\ref{ts}) to a confidence levels in units of Gaussian standard deviations, taking into account the number of additional degrees of freedom corresponding to each new hypothesis.}
\label{tab:3d-energy-bands}
\end{table*}

According to the 3D model developed in Section~\ref{results}, \mbox{HESS~J1702-420} is best described by the superposition of two independent  components. 
To further validate this fact, we repeated the 3D analysis within three independent --- that is nonoverlapping --- energy bands defined by the edges $2.0, 3.7, 15.3$ and $150\,\text{TeV}$. These were chosen to ensure a roughly constant level of \mbox{$\gamma$-ray} flux.
During the fit, the spectral indices of all model components  were fixed to the values obtained in the whole energy range (Section~\ref{results}). This was meant to prevent poor spectral modeling, due to the limited lever arm and insufficient number of photons reconstructed within each individual energy band. The spectral normalizations were instead left free to vary, together with all spatial and background parameters. For each energy band, we used the likelihood ratio test (see Eq.~\ref{ts}) to compare the statistical significance of three nested hypotheses: 
\begin{description}
    \item[H$_0$:]{Null hypothesis, with no model component describing \mbox{HESS~J1702-420};}
    \item[H$_1$:]{\mbox{HESS~J1702-420} is described by one Gaussian component, with the spectral index of \mbox{HESS J1702-420B}. For this component, we left the spatial eccentricity and rotation angle free to vary, for a total of six\footnote{One spectral normalization, plus five spatial parameters.} free model parameters;}
    \item[H$_2$:]{\mbox{HESS~J1702-420} is described by two Gaussian components, with the spectral indices of \mbox{HESS J1702-420B} and \mbox{HESS J1702-420A}. For the latter, we considered a strictly symmetric Gaussian morphology, for a total of four\footnote{One spectral normalization, plus three spatial parameters.} free model parameters.}
\end{description}

The relative significance of each hypothesis, for all the energy bands, is reported in Table~\ref{tab:3d-energy-bands}. 
 It turns out that \mbox{HESS J1702-420B} is significant in all energy bands, while \mbox{HESS J1702-420A} is significant only in the $15.3-150 \,\text{TeV}$ band.

The spatial and spectral shapes of the two components within each energy band are shown in Figure~\ref{fig:3d-energy-bands}, where the reference result from the whole $2-150 \,\text{TeV}$ fit range are reported in black.  
The upper panel shows the $1\sigma$ contours of \mbox{HESS J1702-420A} and \mbox{HESS J1702-420B}. \mbox{HESS J1702-420A} is not drawn in the two lowest energy bands (i.e., for $E\leq15.3\,\text{TeV}$), because it is not significant (see Table~\ref{tab:3d-energy-bands}). In each energy band, the best-fit morphologies of both components are consistent with the reference results obtained in Section~\ref{3d}. The lower panel of the figure compares   the reference spectra of \mbox{HESS J1702-420A} and \mbox{HESS J1702-420B} (in black)  with the spectra obtained in different energy bands. The energy-resolved results are well connected and in agreement with the reference power laws.

To summarize, in each independent energy band \mbox{HESS J1702-420} is well described by a simple model, based on  either one or two components with Gaussian morphologies and power law spectra. \mbox{HESS J1702-420B} is significant in all energy bands, with stable morphology and spectrum. \mbox{HESS J1702-420A} instead is significant only in the highest energy band, due to its exceptionally hard spectrum. This is precisely what would be expected in case the emission is due to two separate components.
Additionally, we summed the log-likelihood values obtained in each independent energy band, and estimated that globally a two-component model is better than a one-component model with a confidence level of $5.3\sigma$. This value is consistent with the $5.4\sigma$ significance for the presence of \mbox{HESS J1702-420A} obtained from the 3D analysis in the full $2-150 \,\text{TeV}$ energy range (see Table~\ref{3dtable-spectral}). All these facts   support the validity of the simple two-component approach that resulted naturally from the iterative procedure described in Section~\ref{results}. However, the possibility that a single source component with energy-dependent morphology could provide a better fit of the data --- at the expense of a large number of free parameters describing the variation of source size, eccentricity and center position as a function of the energy --- cannot be ruled out at this stage.

\section{\emph{Fermi}-LAT analysis details}\label{fermi details}

The data (photon event file and spacecraft file) were retrieved from the \href{https://fermi.gsfc.nasa.gov/ssc/data/}{LAT data server}, through a query defined by the parameters in Table~\ref{qq}. We adopted the event selection cuts described in Table \ref{details}. For the analysis, we defined a square $10^{\,\text{o}}\times\,10^{\,\text{o}}$ RoI, fully inscribed within the events selection circle. Events were binned spatially using $0.05^{\,\text{o}}\times\,0.05^{\,\text{o}}$ spatial pixels, and spectrally using 8 bins per energy decade. 

\begin{table*}[ht]
\begin{center}
\begin{tabular}{||c|c|c|c||}
    \hline
    \text{\textbf{Direction (Gal)}}&\text{\textbf{Radius}}&\text{\textbf{Time range (Gregorian)}}&\text{\textbf{Energy (GeV)}}\\
    \hline
    $(344.3^{\,\text{o}}, -0.2^{\,\text{o}})$&$21.21^{\,\text{o}}$&2008-08-04 --- 2020-06-26&1 --- 1000\\
    \hline
\end{tabular}
\end{center}
\caption{Query details for the \emph{Fermi}-LAT data.}
\label{qq}
\end{table*}

\begin{table*}[ht]
\begin{center}
\begin{tabular}{||c|c|c|c||}
    \hline
    \text{\textbf{zmax}}&\text{\textbf{evclass}}&\text{\textbf{evtype}}&\text{\textbf{Selection filter}}\\
    \hline
    90&120&3&\text{\texttt{(DATA\_QUAL$>$0)\&\&(LAT\_CONFIG$==$1)}}\\
    \hline
\end{tabular}
\end{center}
\caption{Events selection cuts for the \emph{Fermi}-LAT analysis.}
\label{details}
\end{table*}

During the maximum likelihood fit, the spectral index and normalization of all sources within $3^{\,\text{o}}$ from the RoI center and having a TS value higher than 25 were left free to vary. Additionally, the spectral normalization of all sources with $\mathrm{TS}>30$ within the whole $10^{\,\text{o}}\times\,10^{\,\text{o}}$ was also adjusted. The Galactic diffuse emission model (\texttt{gll\_iem\_v07.fits}) was left free to vary, while the extra-Galactic diffuse model was considered fixed to the default one (\texttt{iso\_P8R3\_SOURCE\_V2\_v1.txt}).

\section{Method for the derivation of lower limits on the particle cut-off energy}\label{ll}
The lower limits on the cut-off energies in the spectra of the proton or electron parent populations were obtained as follows:
\begin{enumerate}[i]
\item{we defined an array of trials particle cut-off energies $\{E^\text{c}_1,\,\ldots,E^\text{c}_N\}$;}
\item{for each fixed cut-off energy $E^\text{c}_i$, we adjusted a power-law with exponential cut-off to the 3D H.E.S.S.\ data. The spectral normalization and index, together with all free nuisance parameters of the model, were optimized at each step. For each trial cut-off energy, we stored the likelihood value of the fit, $\mathscr{L}^\text{\,max}_{(E^{\,\text{c}}_i)}$;}
\item{then we computed the profile of 
\begin{linenomath}
\begin{equation}
\text{TS}(E^{\,\text{c}}_i)=-2\ln\,\left(\frac{\mathscr{L}^\text{\,max}_{0}}{\mathscr{L}^\text{\,max}_{(E^{\,\text{c}}_i)}}\right)\,\,,
\end{equation}
\end{linenomath}
where $\mathscr{L}^\text{\,max}_{0}$ represents the maximum model likelihood under the null power law --- or equivalently $E^{\,\text{c}}$$\rightarrow$$\,\infty$ --- hypothesis;}
\item{we finally computed the 90\%,95\% and 99\% confidence level lower limits on the particle cut-off energy by finding the values where the TS profile increased from the minimum (which in our case was at infinity) by an amount $\text{TS}_{(90\%)}=2.706$, $\text{TS}_{(95\%)}=3.841$ and $\text{TS}_{(99\%)}=6.635$ respectively.}
\end{enumerate}
This procedure, based on~\citet{rolke}, is partly implemented in the \texttt{Fit.stat\_profile()} routine of \texttt{gammapy}.

\section{Cosmic ray diffusion model and energy loss calculation}\label{detailsdiscussion}
The source physical size $R_\text{source}$ depends directly on the distance from Earth $d$ and the measured angular size of the source $\theta_\mathrm{source}$, as \mbox{$R_\text{source}=d\,\times\,\tan(\theta_\mathrm{source})$}. Here, we assumed \mbox{$\theta_\mathrm{source}=1.28^{\,\text{o}}$}, which corresponds to the major $2\sigma$ diameter of \mbox{HESS J1702-420B}.  For both hadronic and leptonic cosmic rays, we adopted the energy-dependent  diffusion coefficient defined in~\citet{gabici}, testing different values for the normalization $\chi$.     
The relation between the diffusion coefficient and the diffusion timescale is given by:
\begin{equation}
\tau_\mathrm{diff}\approx\frac{[R_\mathrm{source}(d)]^2}{6\,D(E, B)}\,\,.
\end{equation}
Energy losses for protons due to $p$-$p$ collisions were estimated --- neglecting ionization losses that are irrelevant for relativistic protons --- as
\begin{equation}
\tau_{pp}\approx 6\times 10^{5}\left( \frac{n_\text{H}}{100\,\mathrm{cm}^{-3}}\right)^{-1}\,\,\mathrm{yr}\,\,,
\end{equation}
as in ~\citet{gabici}.
Finally, the electron energy loss timescale was computed using 
\begin{linenomath}
\begin{equation}
\tau_\mathrm{loss} = \left( \frac{1}{\tau_\mathrm{syn}} + \sum_i\frac{1}{\tau_\mathrm{IC}^{\,i}} \right)^{-1}\,\,,
\end{equation}
\end{linenomath}
where 
\begin{equation}\label{tausyn}
\tau_\mathrm{syn}\approx1.3\times 10^{5}\left( \frac{E}{1\,\mathrm{TeV}}\right)^{-1} \left( \frac{B}{10\,\mu G}\right)^{-2}\,\,\mathrm{yr}
\end{equation}
is the synchrotron loss timescale   and 
\begin{equation}
\tau_\mathrm{IC}\approx3\times 10^{7}\left( \frac{E}{10\,\mathrm{GeV}}\right)^{-1} \left( \frac{U_\mathrm{rad}}{1\,\mathrm{eV}\,\mathrm{cm}^{-3}}\right)^{-1}\,\,\mathrm{yr}
\end{equation}
is the inverse-Compton loss timescale for a given photon field~\citep{ginzburg}.

\section{Additional material}~\label{additionalmaterial}
In this section, additional figures and tables are provided.

\begin{figure*}
\centering
\begin{subfigure}[b]{18cm}
\includegraphics[width=\linewidth]{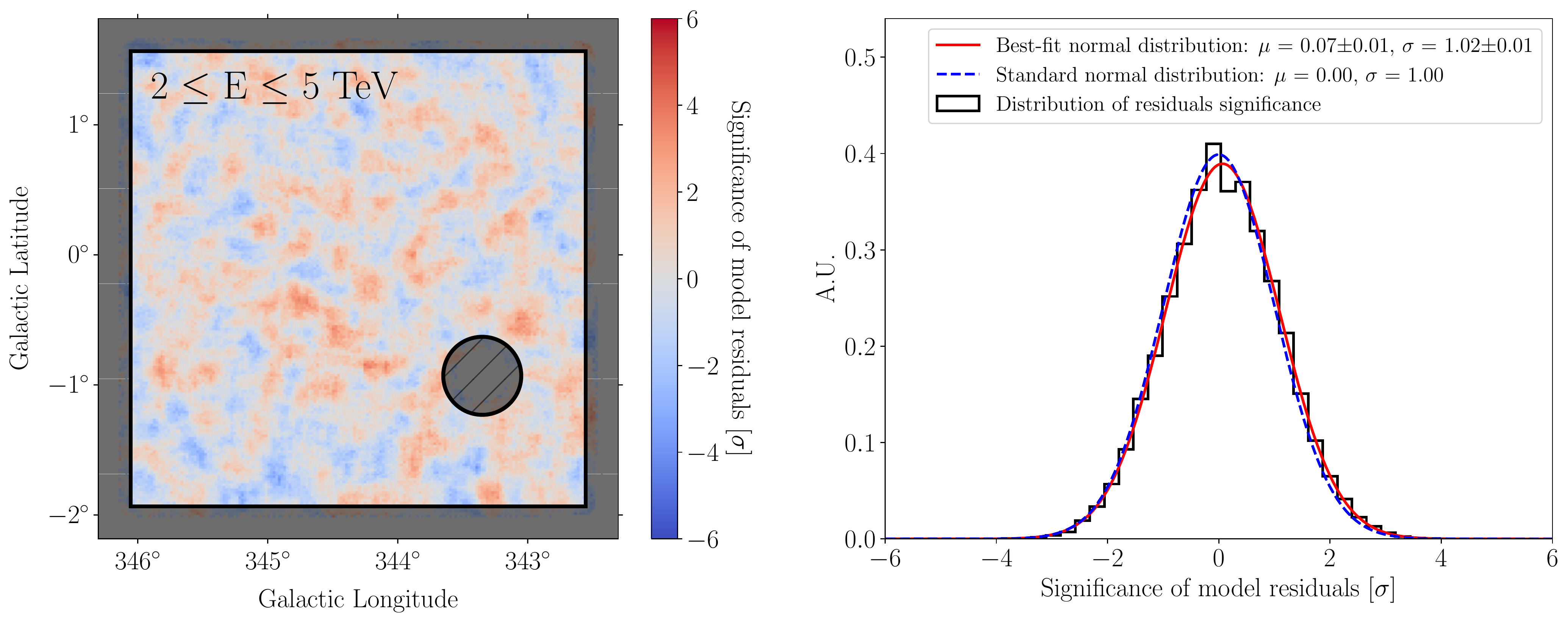}
\end{subfigure}
\begin{subfigure}[b]{18cm}
\includegraphics[width=\linewidth]{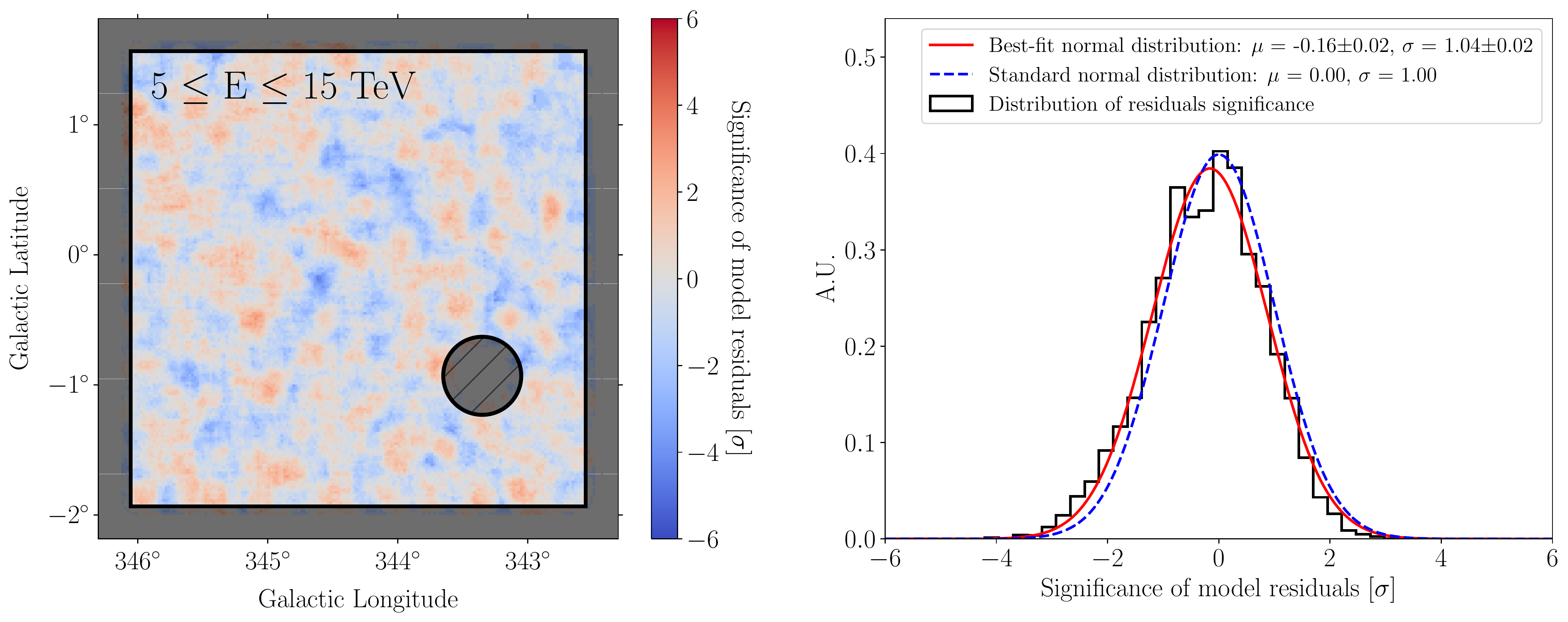}
\end{subfigure}
\begin{subfigure}[b]{18cm}
\includegraphics[width=\linewidth]{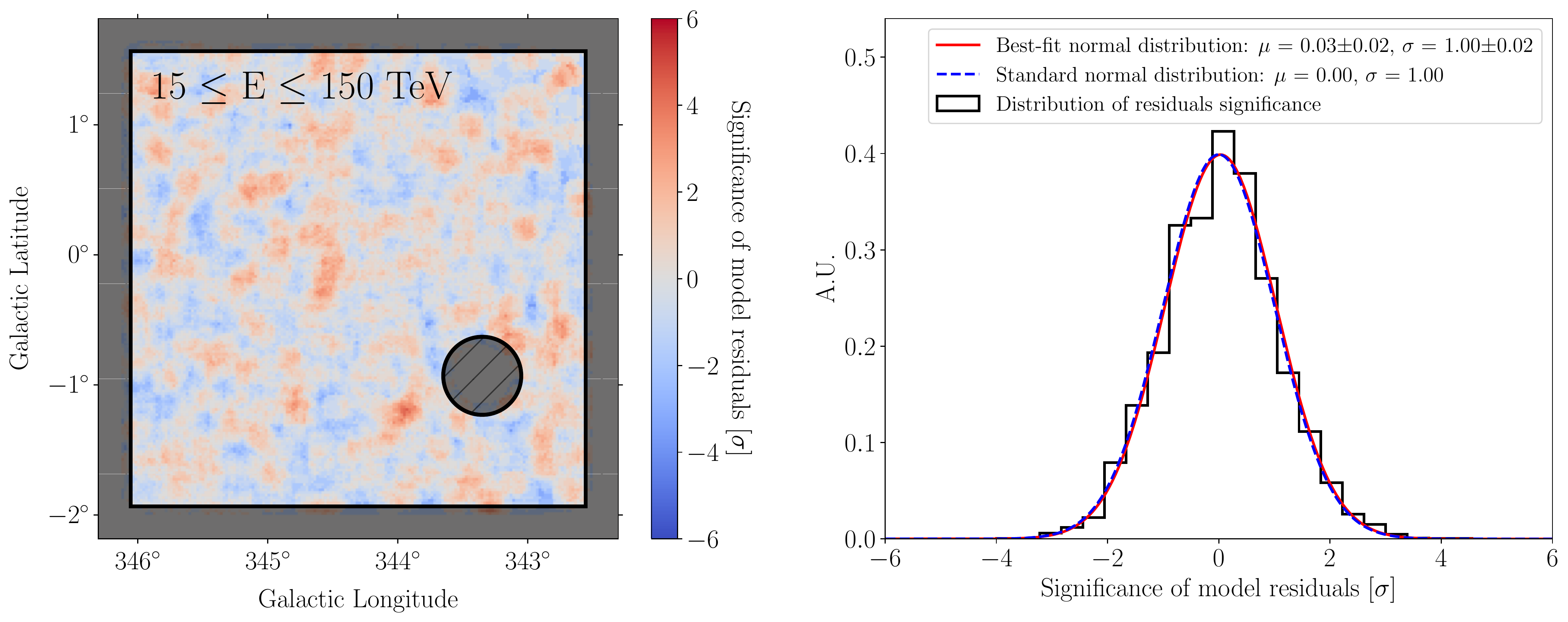}
\end{subfigure}
\caption{Spatial distributions of the significance of model residuals (\emph{left} column) and histograms of significance values (\emph{right} column), computed in the energy bands $2.0 - 5.0 \,\text{TeV}$ (\emph{first} row), $5.0 - 15.0 \,\text{TeV}$ (\emph{second} row) and $15.0 - 150 \,\text{TeV}$ (\emph{third} row).}
\label{res-bands}
\end{figure*}

\begin{table*}[ht]
\begin{center}
\begin{subtable}{1\textwidth}
\begin{tabular}{||c|c|c|c|c|c|c|c|c||}
\hline
\text{\textbf{e\_ref}} & \text{\textbf{e\_min}} & \text{\textbf{e\_max}} & \text{\textbf{sqrt\_ts}} & \text{\textbf{counts}} & \text{\textbf{dnde}} & \text{\textbf{dnde\_ul}} & \text{\textbf{dnde\_errp}} & \text{\textbf{dnde\_errn}} \\
\text{\textbf{[TeV]}} & \text{\textbf{[TeV]}} & \text{\textbf{[TeV]}} &  &  & $\mathbf{[\text{\textbf{TeV}}^{-1\,}\text{\textbf{cm}}^{-2\,}\text{\textbf{s}}^{-1}]}$ & $\mathbf{[\text{\textbf{TeV}}^{-1\,}\text{\textbf{cm}}^{-2\,}\text{\textbf{s}}^{-1}]}$ & $\mathbf{[\text{\textbf{TeV}}^{-1\,}\text{\textbf{cm}}^{-2\,}\text{\textbf{s}}^{-1}]}$ & $\mathbf{[\text{\textbf{TeV}}^{-1\,}\text{\textbf{cm}}^{-2\,}\text{\textbf{s}}^{-1}]}$ \\
\hline
3.19 & 2.08 & 4.90 & 3.36 & 40973 & 4.08e-14 & 8.35e-14 & 1.33e-14 & 1.27e-14 \\
8.66 & 4.90 & 15.32 & 3.15 & 15812 & 4.42e-15 & 7.01e-15 & 1.59e-15 & 1.50e-15 \\
23.50 & 15.32 & 36.04 & 5.29 & 3363 & 1.42e-15 & 2.00e-15 & 3.64e-16 & 3.33e-16 \\
47.94 & 36.04 & 63.76 & 4.83 & 835 & 4.73e-16 & 1.02e-15 & 1.52e-16 & 1.33e-16 \\
84.80 & 63.76 & 112.78 & 4.00 & 454 & 1.89e-16 & 4.91e-16 & 7.76e-17 & 6.54e-17 \\
130.07 & 112.78 & 150.00 & 0.00 & 145 & 1.20e-23 & 2.21e-16 & 2.48e-17 & 1.20e-23 \\
\hline
\end{tabular}
\caption{Spectral points of \mbox{HESS J1702-420A}.}
\end{subtable}
\bigskip

\begin{subtable}{1\textwidth}
\begin{tabular}{||c|c|c|c|c|c|c|c|c||}
\hline
\text{\textbf{e\_ref}} & \text{\textbf{e\_min}} & \text{\textbf{e\_max}} & \text{\textbf{sqrt\_ts}} & \text{\textbf{counts}} & \text{\textbf{dnde}} & \text{\textbf{dnde\_ul}} & \text{\textbf{dnde\_errp}} & \text{\textbf{dnde\_errn}} \\
\text{\textbf{[TeV]}} & \text{\textbf{[TeV]}} & \text{\textbf{[TeV]}} &  &  & $\mathbf{[\text{\textbf{TeV}}^{-1\,}\text{\textbf{cm}}^{-2\,}\text{\textbf{s}}^{-1}]}$ & $\mathbf{[\text{\textbf{TeV}}^{-1\,}\text{\textbf{cm}}^{-2\,}\text{\textbf{s}}^{-1}]}$ & $\mathbf{[\text{\textbf{TeV}}^{-1\,}\text{\textbf{cm}}^{-2\,}\text{\textbf{s}}^{-1}]}$ & $\mathbf{[\text{\textbf{TeV}}^{-1\,}\text{\textbf{cm}}^{-2\,}\text{\textbf{s}}^{-1}]}$ \\
\hline
2.77 & 2.08 & 3.68 & 18.62 & 31603 & 5.86e-13 & 7.16e-13 & 3.62e-14 & 3.58e-14 \\
4.25 & 3.68 & 4.90 & 12.13 & 9370 & 1.98e-13 & 2.36e-13 & 1.92e-14 & 1.87e-14 \\
6.51 & 4.90 & 8.66 & 11.01 & 10687 & 4.98e-14 & 5.78e-14 & 5.18e-15 & 5.06e-15 \\
11.52 & 8.66 & 15.32 & 6.60 & 5125 & 9.82e-15 & 1.14e-14 & 1.68e-15 & 1.64e-15 \\
23.50 & 15.32 & 36.04 & 6.89 & 3363 & 2.40e-15 & 3.44e-15 & 4.02e-16 & 3.89e-16 \\
63.76 & 36.04 & 112.78 & 3.20 & 1289 & 1.78e-16 & 3.74e-16 & 6.31e-17 & 5.94e-17 \\
130.07 & 112.78 & 150.00 & 0.49 & 145 & 2.48e-17 & 3.34e-16 & 5.62e-17 & 2.48e-17 \\
\hline
\end{tabular}
\caption{Spectral points of \mbox{HESS J1702-420B}.}
\label{points}
\end{subtable}

\end{center}
\caption{Spectral points obtained from the 3D analysis with \texttt{gammapy} (Section~\ref{results}). The points were obtained by rescaling the amplitude of the reference spectral model within each energy bin, re-optimizing at the same time all free nuisance parameters of the model. The column names follow the convensions defined by the open-source ``data formats for gamma-ray astronomy'' community (\url{https://gamma-astro-data-formats.readthedocs.io/en/latest/}).
}
\label{tab:fp}
\end{table*}

\begin{figure}

\centering
\includegraphics[width=\linewidth]{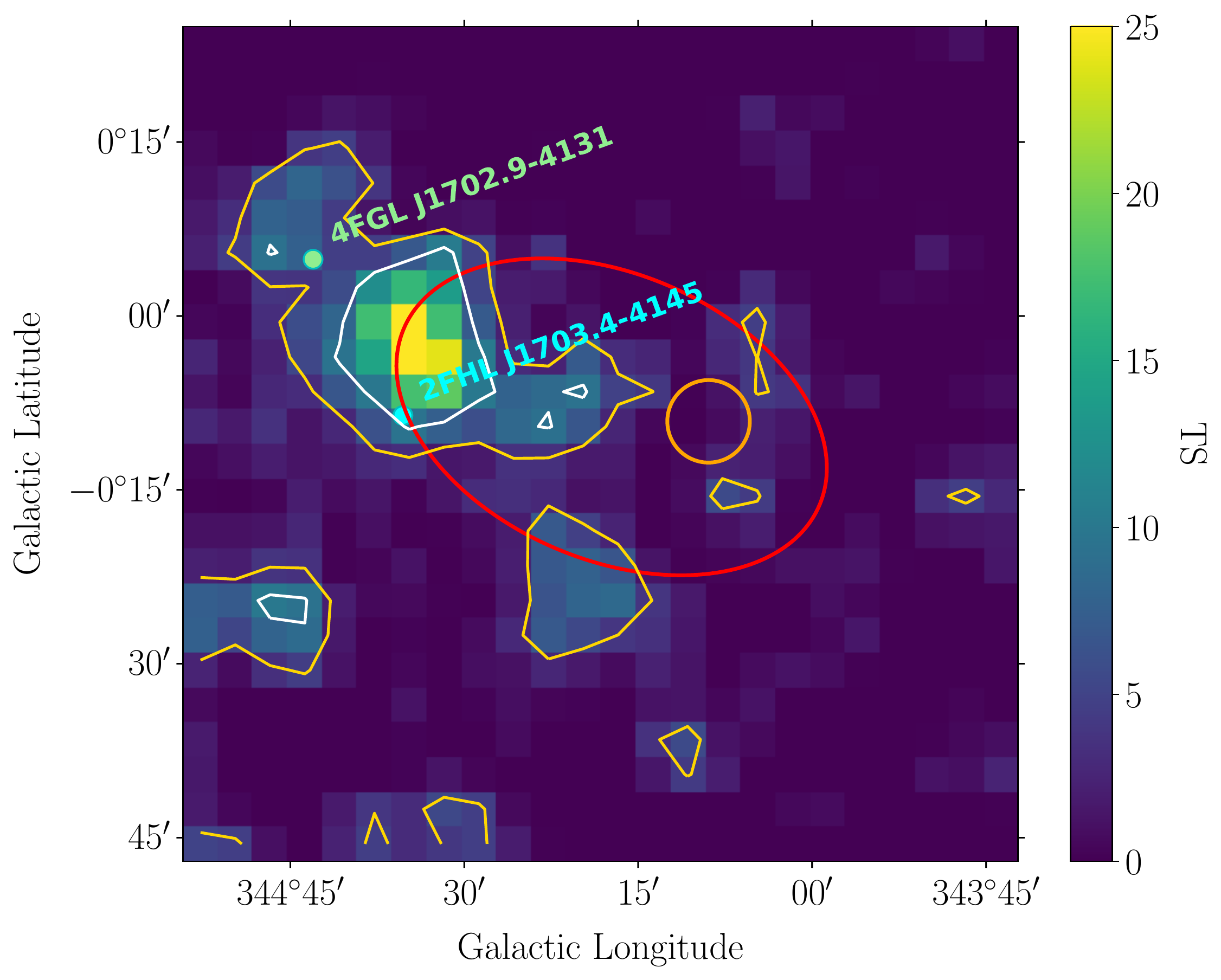}
\caption{ Map of test statistic (TS) for the presence of an additional source in the RoI, with respect to a source model containing only the galactic and isotropic \mbox{$\gamma$-ray} diffuse sources. The color bar, contours and markers are identical to Figure~\ref{fig:fermits}.}
\label{fig:fermicounts}
\end{figure}

\begin{figure*}
\centering
\includegraphics[width=0.8\linewidth]{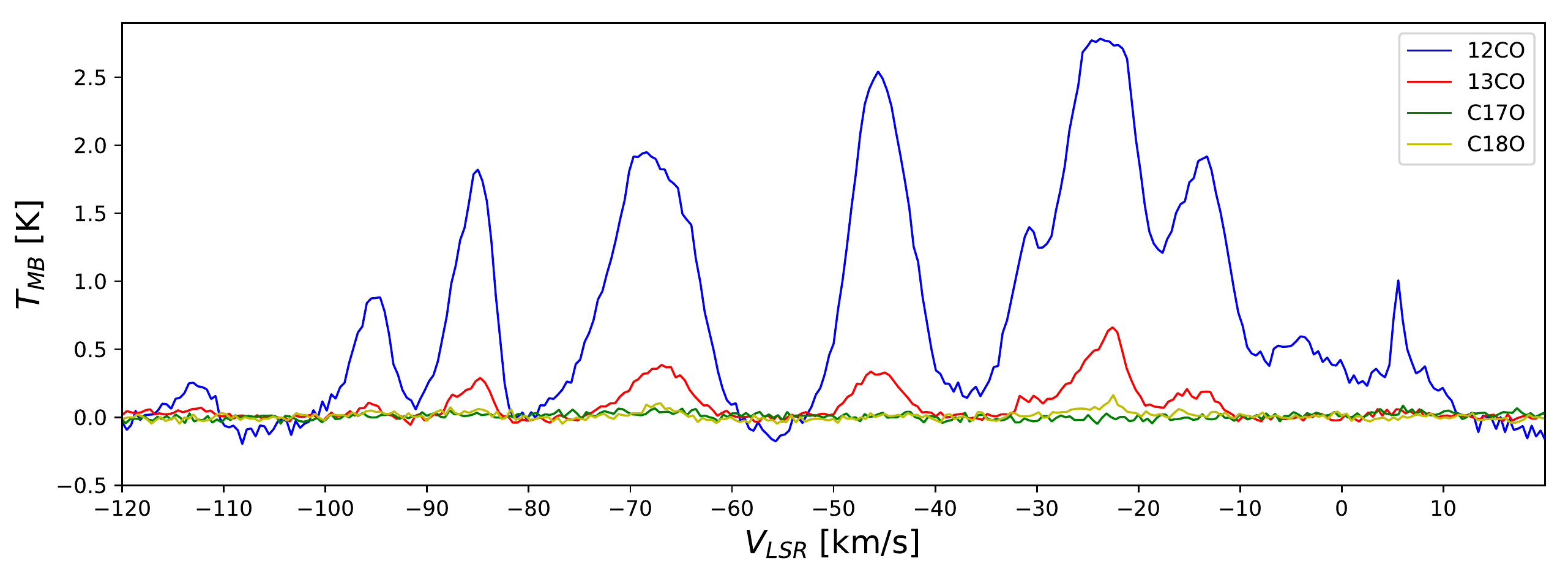}
\caption{Brightness temperature peaks obtained by integrating the data from the Mopra radio survey~\citep{mopra} within a $0.3^{\,\text{o}}$-side square window centered at the best-fit position of \mbox{HESS J1702-420A} (shown in red in Figure~\ref{clouds-integrated}).}
\label{fig:clouds}
\end{figure*}

\begin{figure*}

\begin{subfigure}[t]{0.33\textwidth}
\centering
\includegraphics[width=\linewidth]{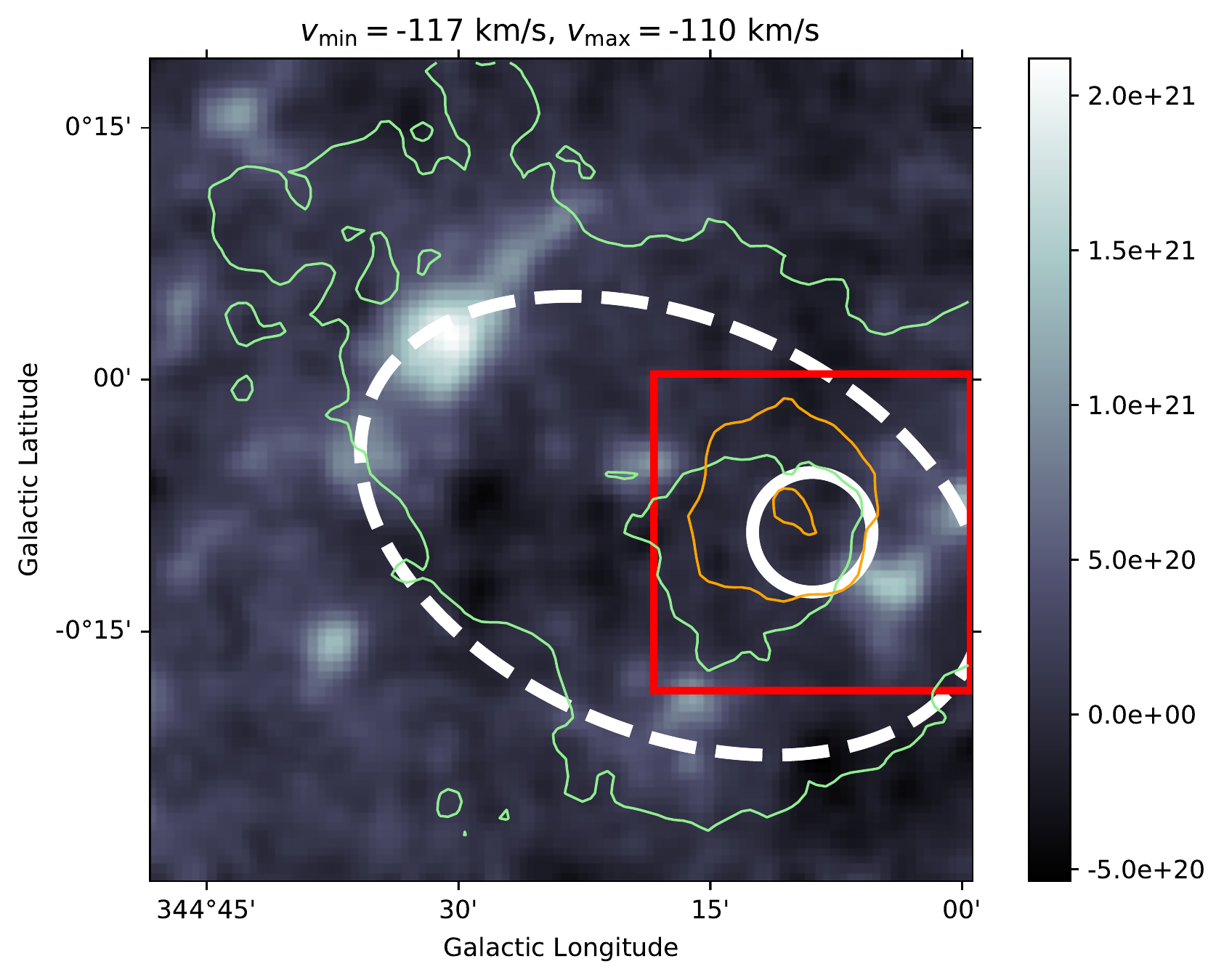}
\end{subfigure}
\begin{subfigure}[t]{0.33\textwidth}
\centering
\includegraphics[width=\linewidth]{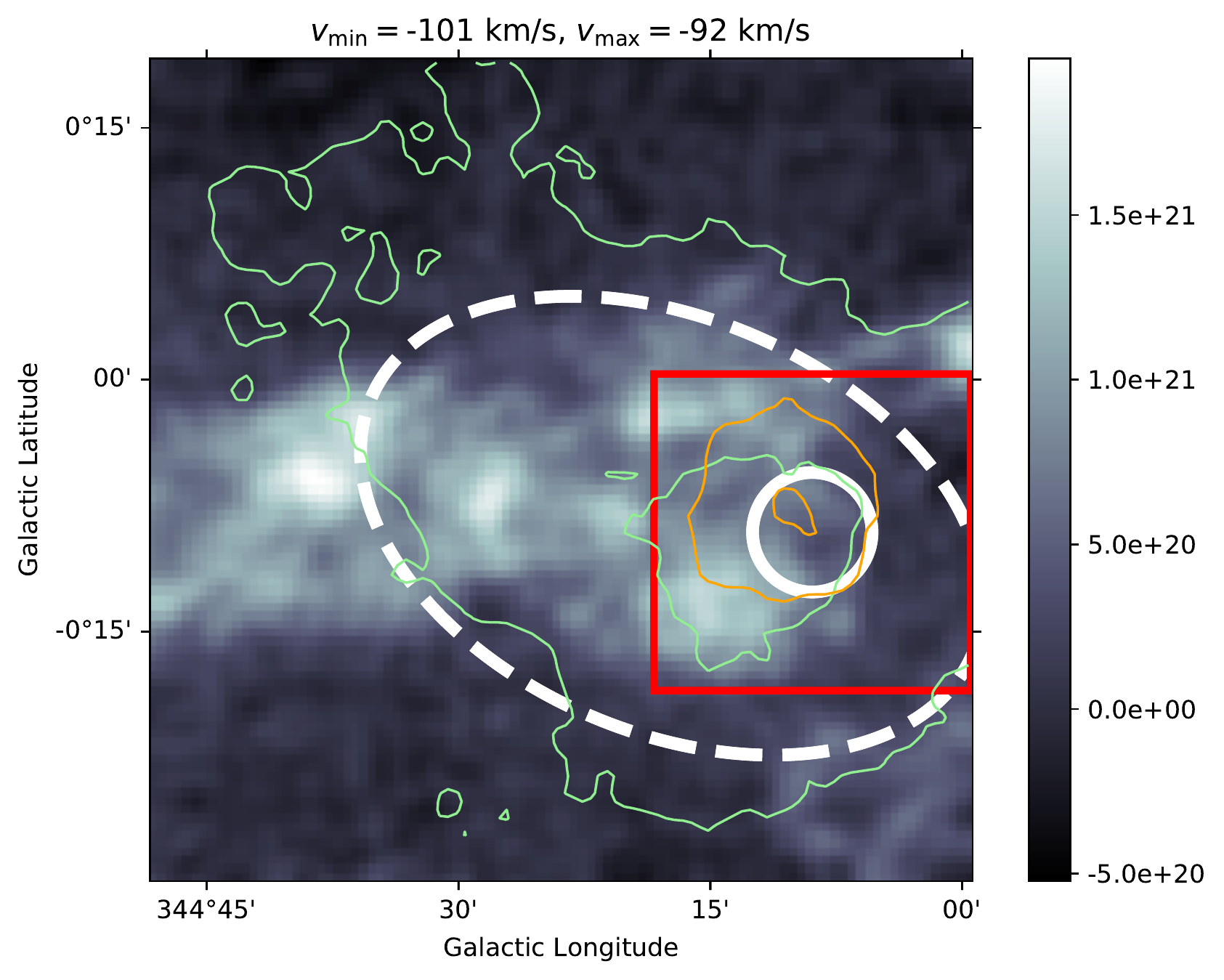}
\end{subfigure}
\begin{subfigure}[t]{0.33\textwidth}
\centering
\includegraphics[width=\linewidth]{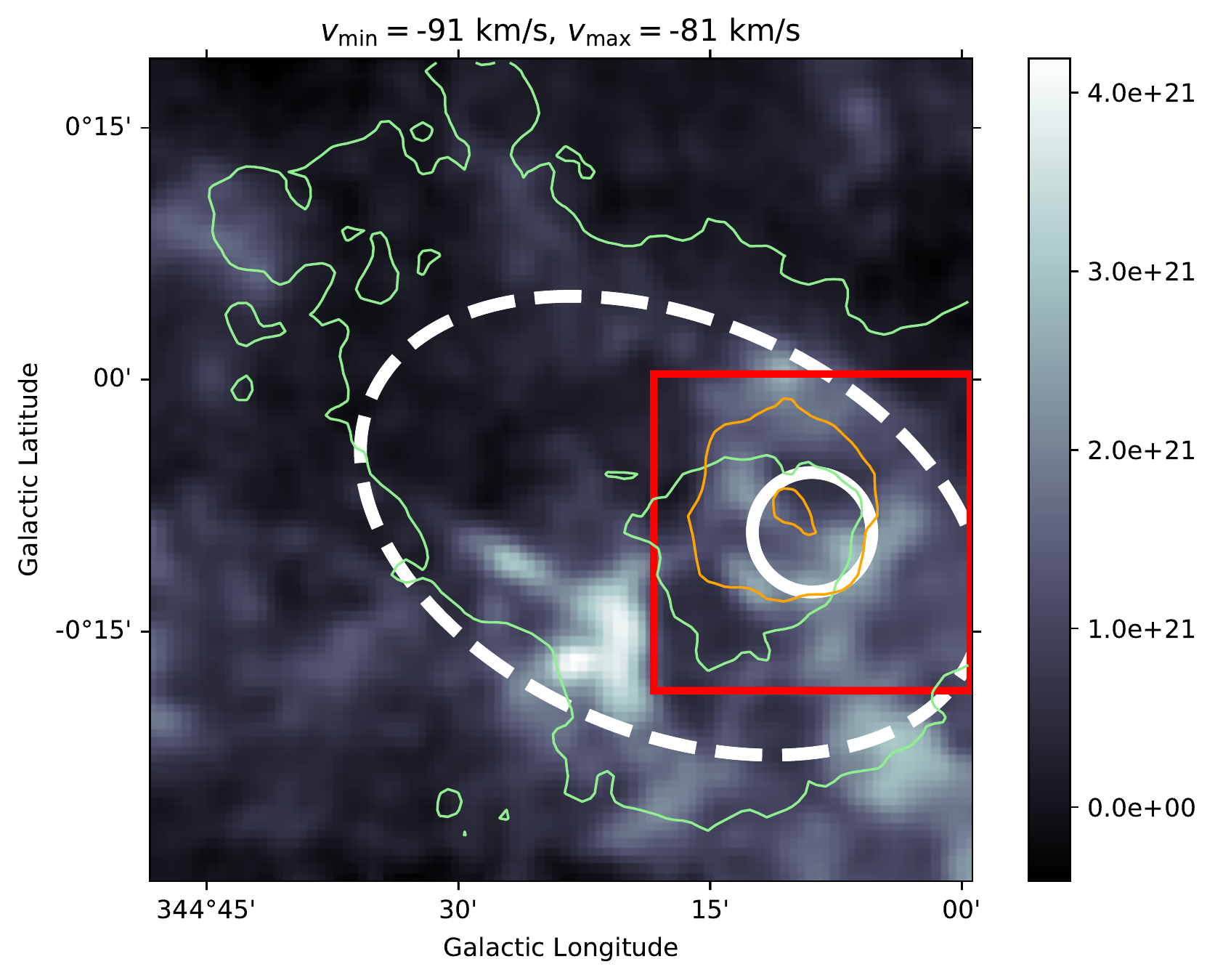}
\end{subfigure}
\begin{subfigure}[t]{0.33\textwidth}
\centering
\includegraphics[width=\linewidth]{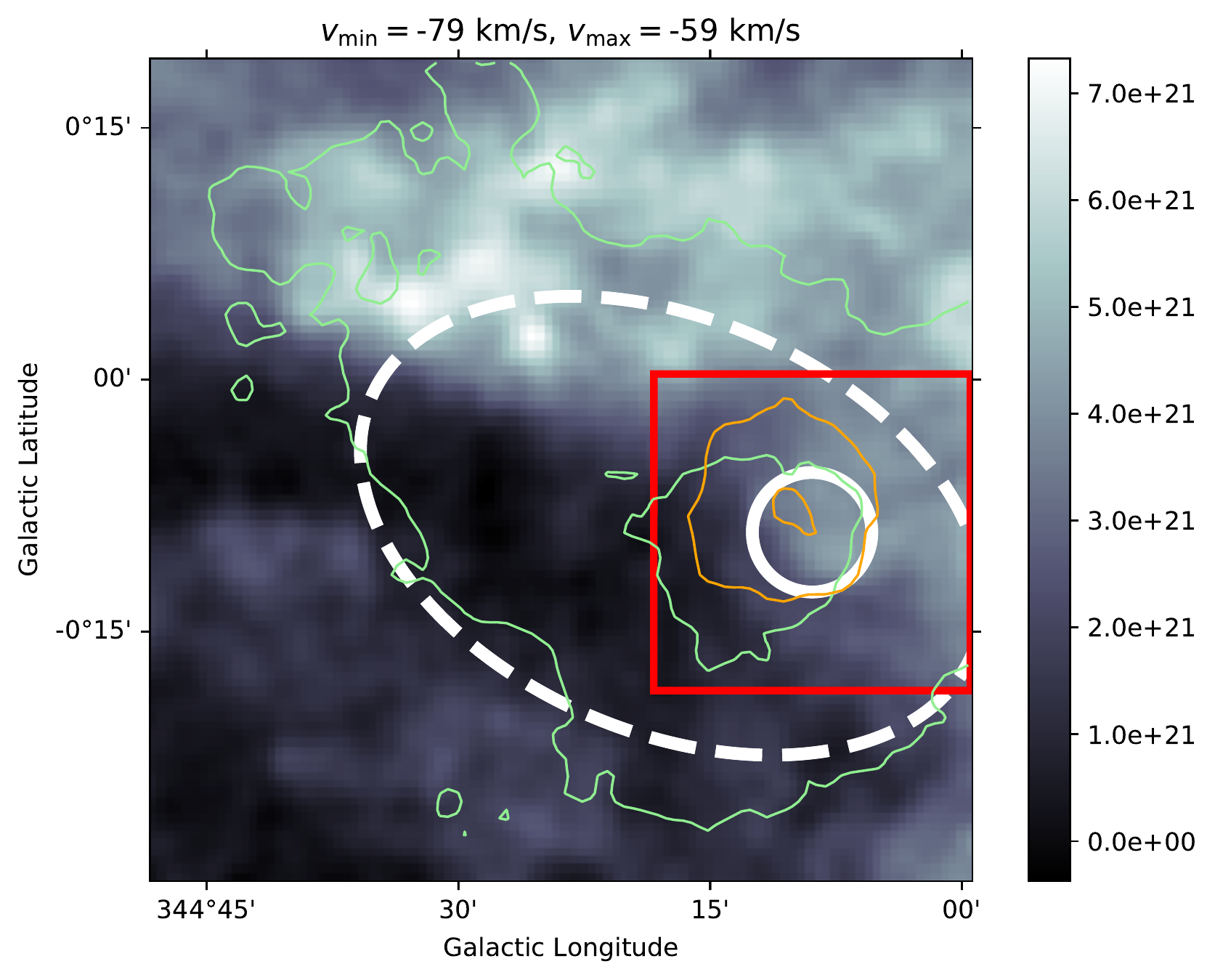}
\end{subfigure}
\begin{subfigure}[t]{0.33\textwidth}
\centering
\includegraphics[width=\linewidth]{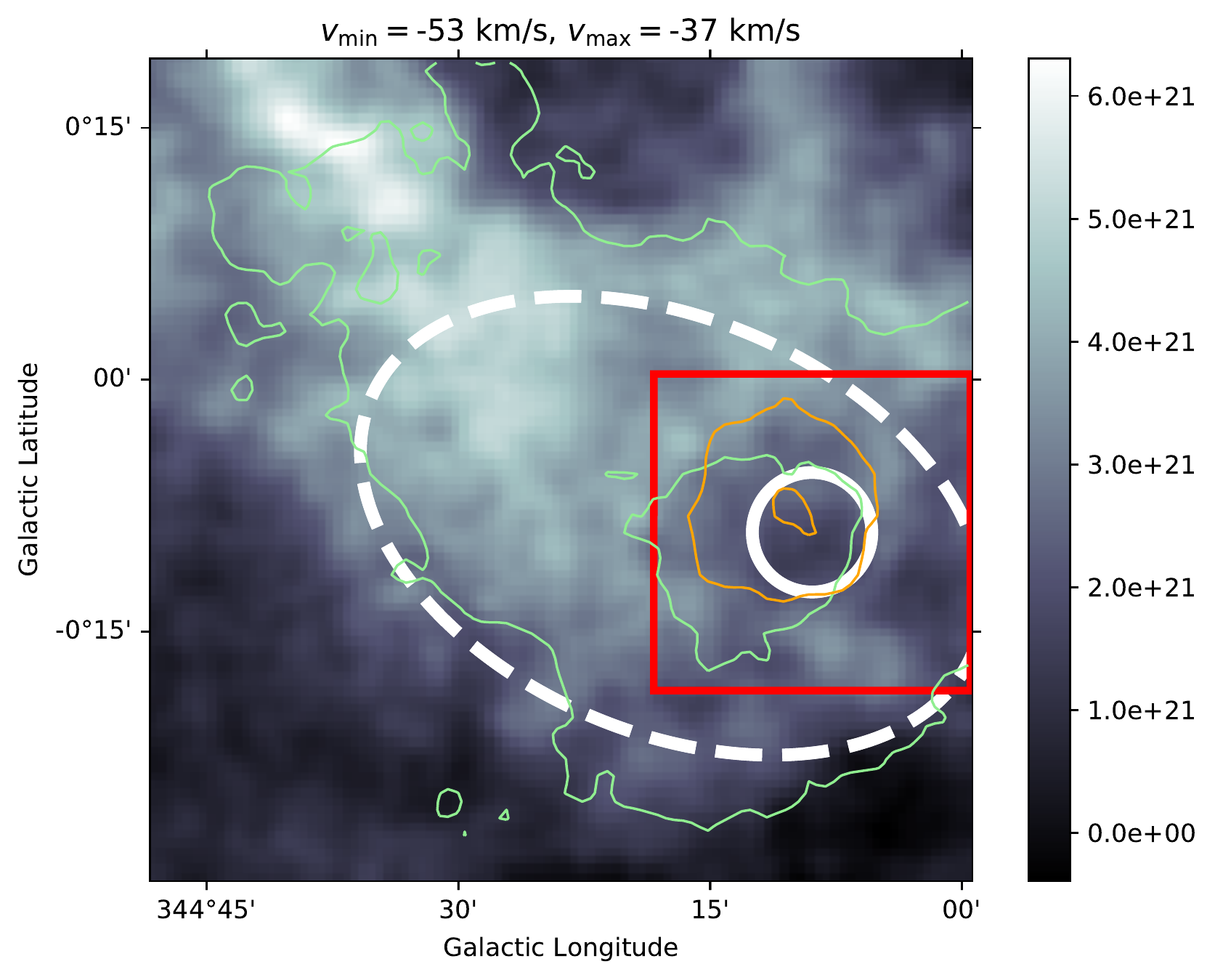}
\end{subfigure}
\begin{subfigure}[t]{0.33\textwidth}
\centering
\includegraphics[width=\linewidth]{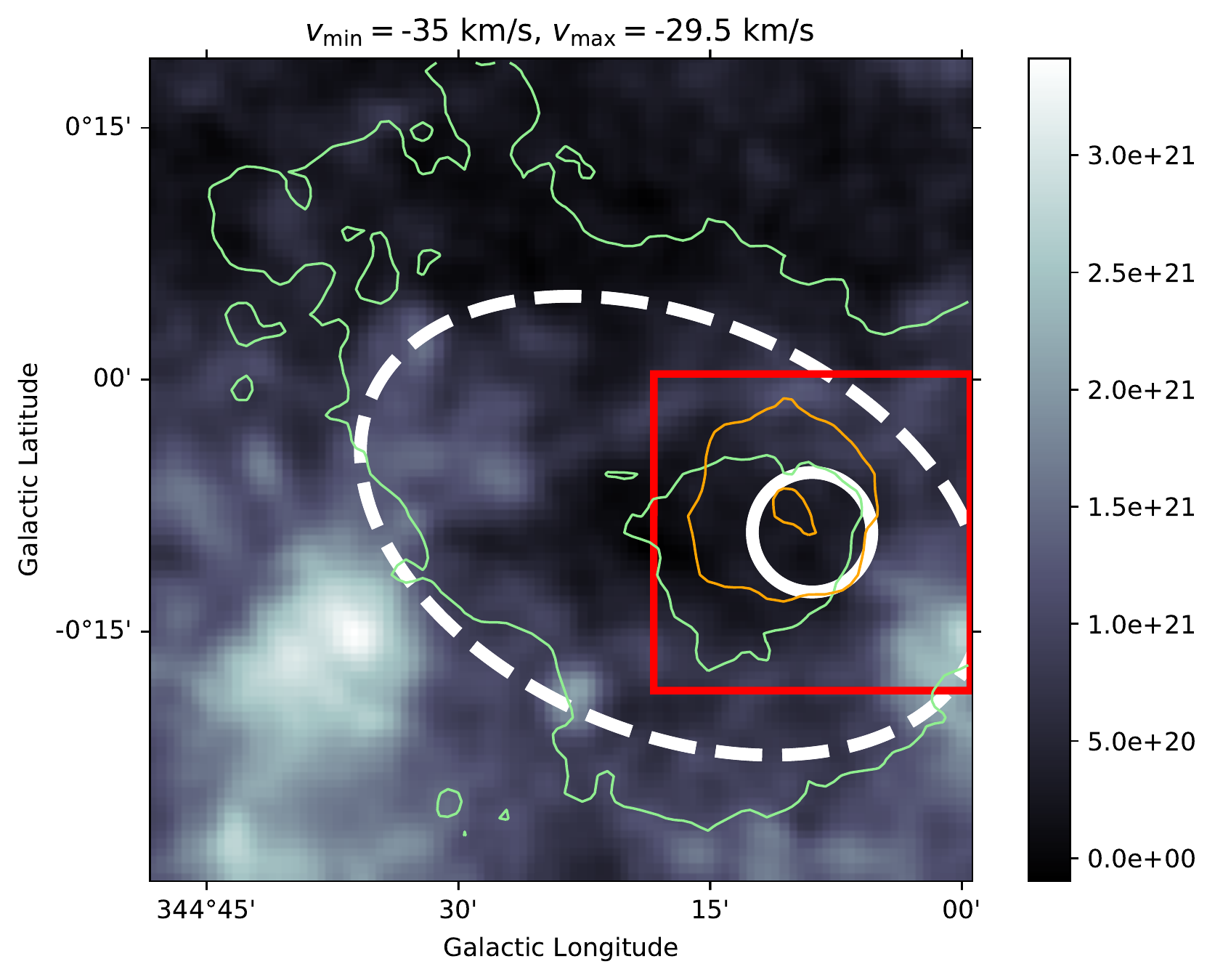}
\end{subfigure}
\begin{subfigure}[t]{0.33\textwidth}
\centering
\includegraphics[width=\linewidth]{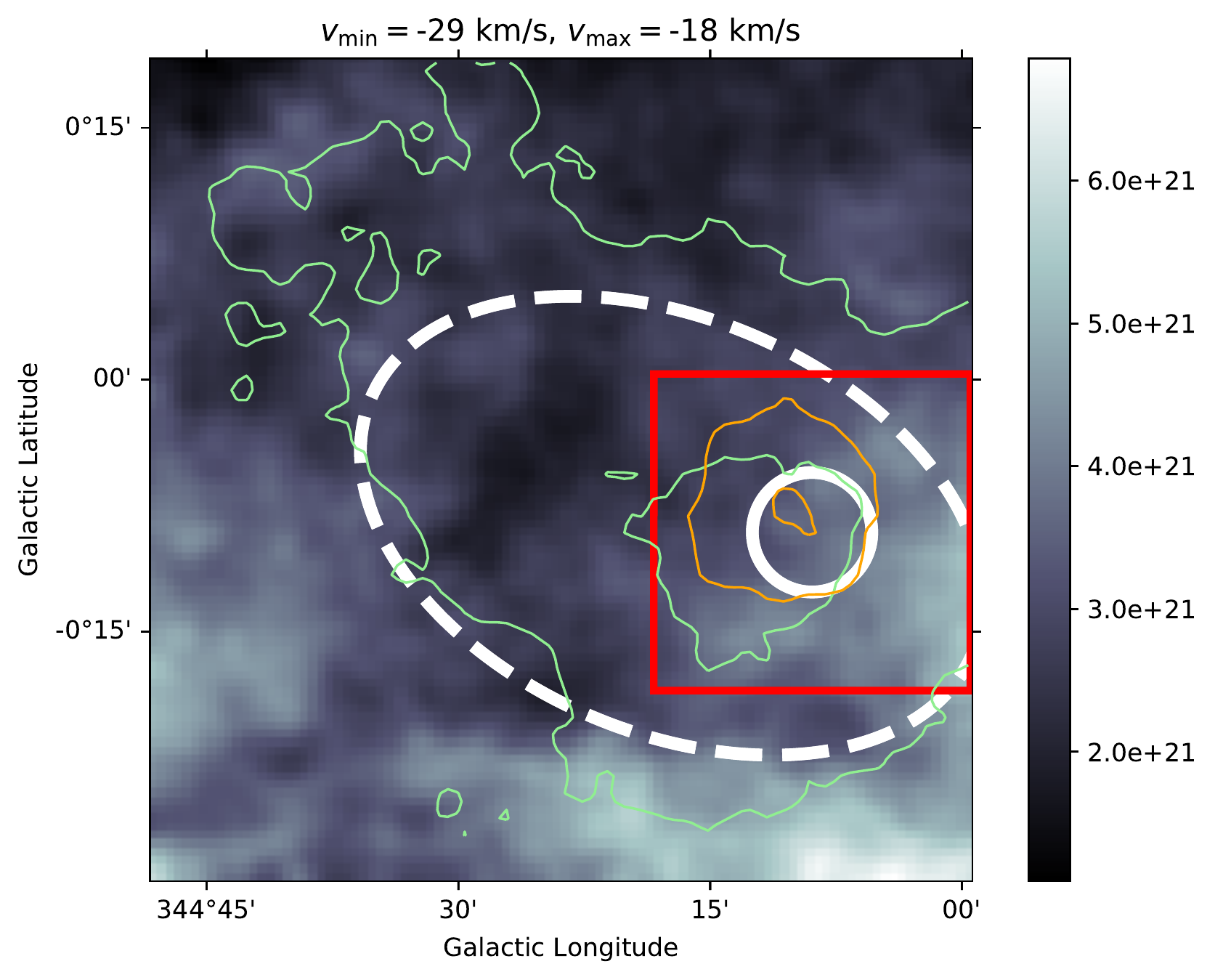}
\end{subfigure}
\begin{subfigure}[t]{0.33\textwidth}
\centering
\includegraphics[width=\linewidth]{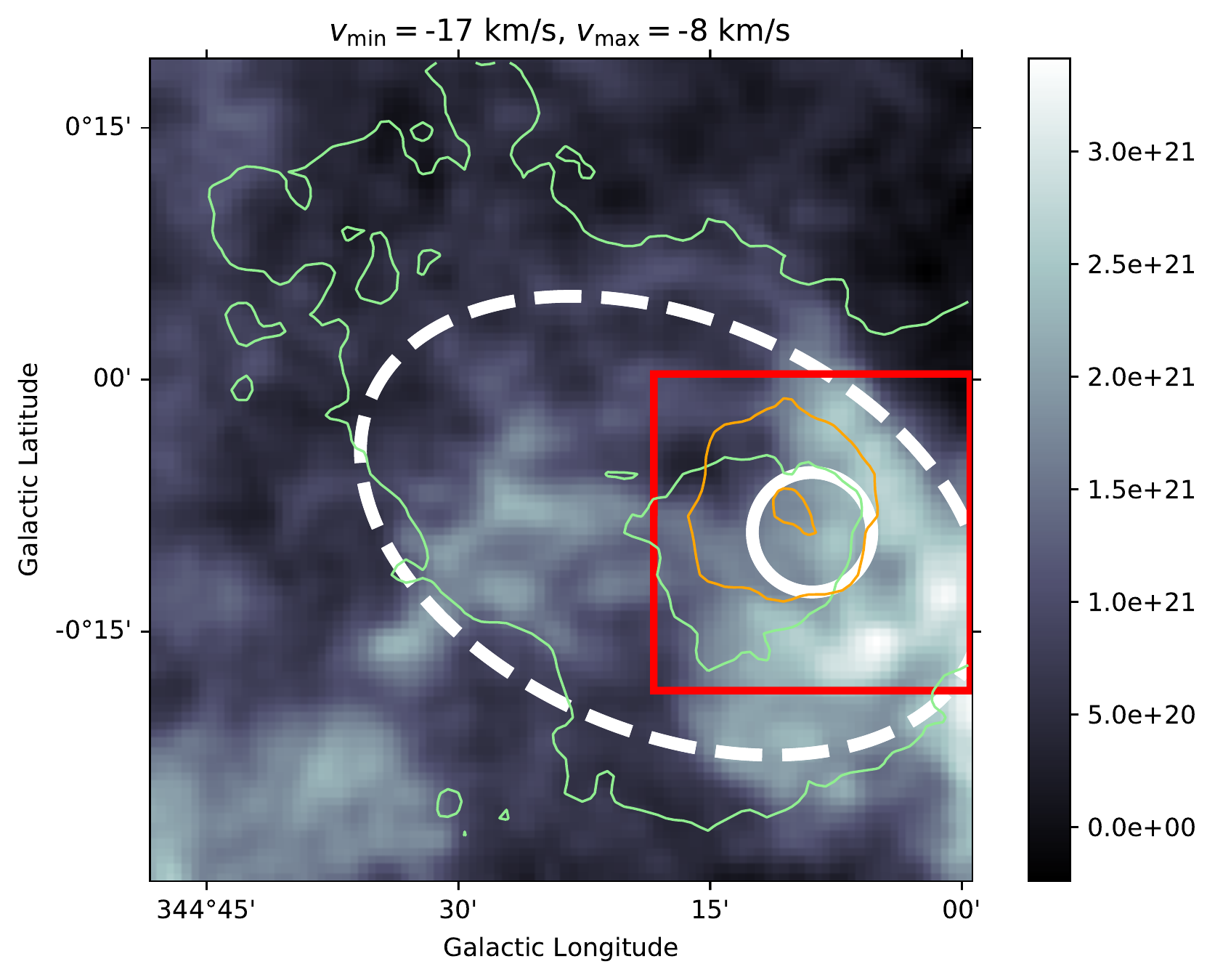}
\end{subfigure}
\begin{subfigure}[t]{0.33\textwidth}
\centering
\includegraphics[width=\linewidth]{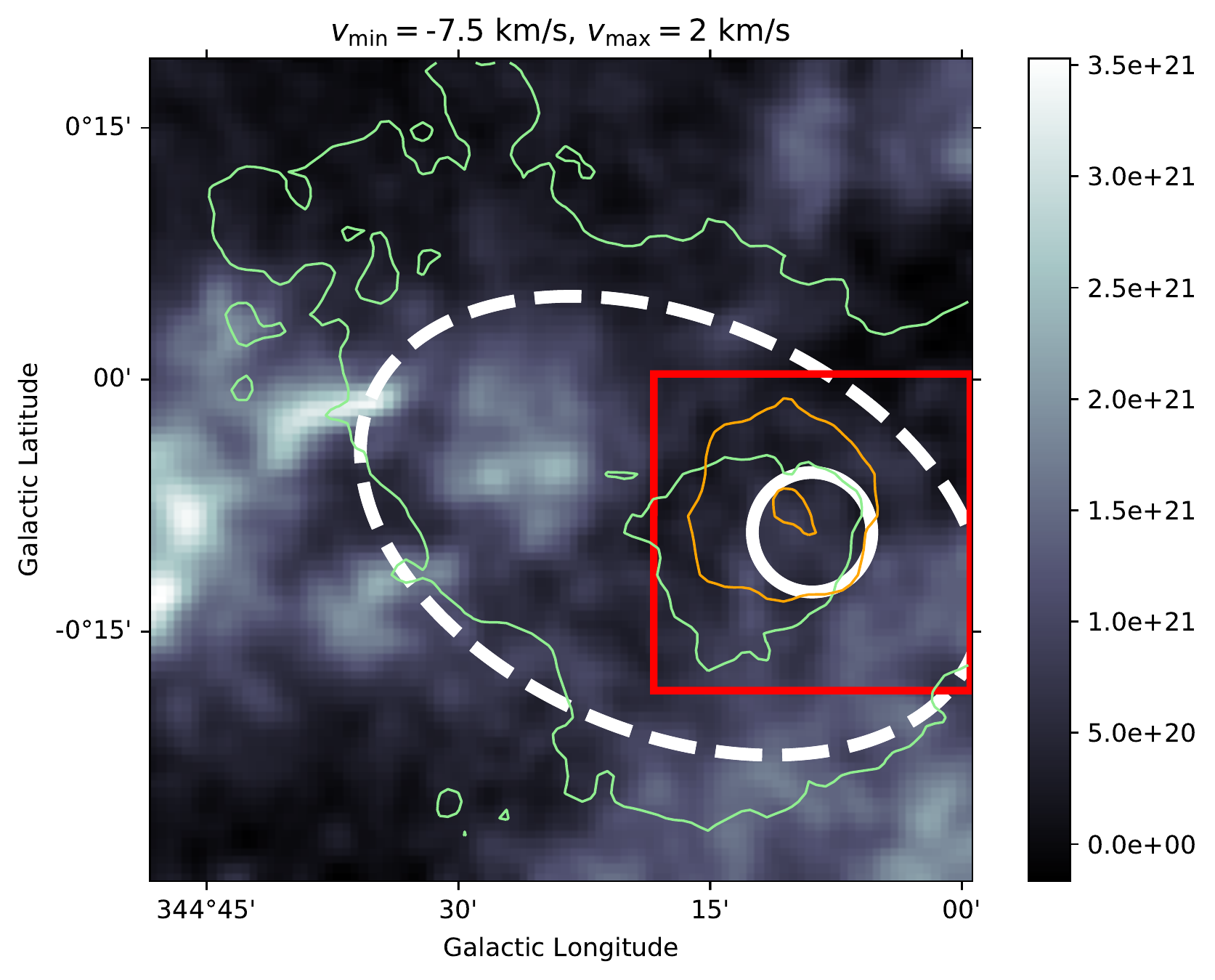}
\end{subfigure}
\begin{subfigure}[t]{0.33\textwidth}
\centering
\includegraphics[width=\linewidth]{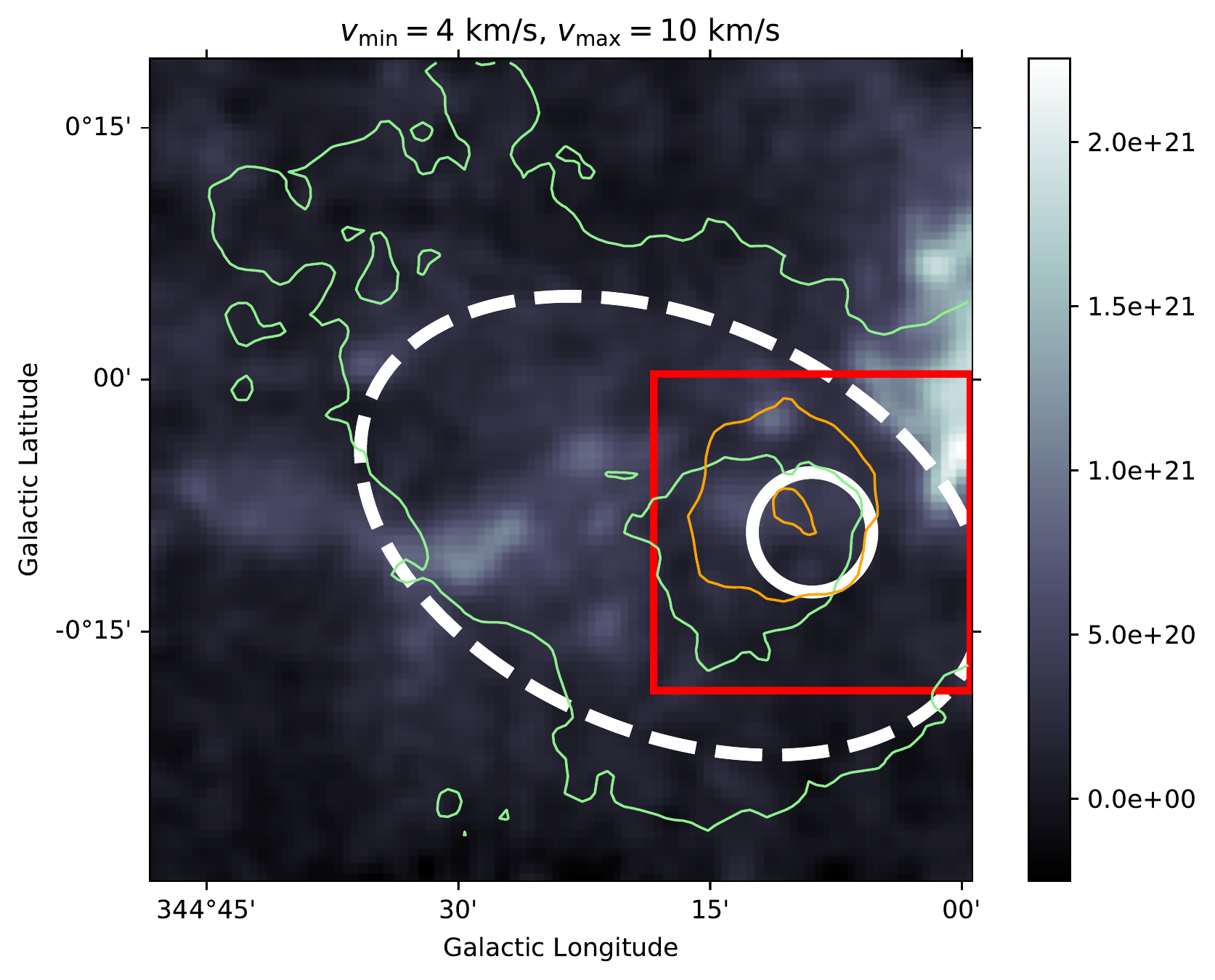}
\end{subfigure}

\caption{Column density maps of molecular hydrogen in the direction of \mbox{HESS J1702–420}, obtained by integrating the brightness temperature profile of $^{12}$CO$(J=1\rightarrow0)$ data from the Mopra radio survey  within the velocity intervals indicated above each panel (corresponding to the peaks in Figure~\ref{fig:clouds}). The brightness temperature values were converted to $H_2$ column density assuming the conversion factor $X_{CO} = 1.5\times 10^{20}\,\mathrm{cm}^{-2}\,(\mathrm{K}\,\mathrm{km}\,\mathrm{s}^{-1})^{-1}$ \citep{strong}. The green (orange) contours indicate the 5 and $12\sigma$ (3 and $5\sigma$) significance levels of the TeV $\gamma$-ray flux above $2\,\text{TeV}$ ($40\,\text{TeV}$). The dashed ellipse and solid circle represent the $1\sigma$ morphologies of \mbox{HESS J1702-420B} and \mbox{HESS J1702-420A}, respectively. Finally, the red square --- centered at the best-fit position of \mbox{HESS J1702-420A} --- indicates the extraction region used to produce the profile reported in Figure~\ref{fig:clouds}}
\label{clouds-integrated}
\end{figure*}

\begin{figure}
\centering
\includegraphics[width=\linewidth]{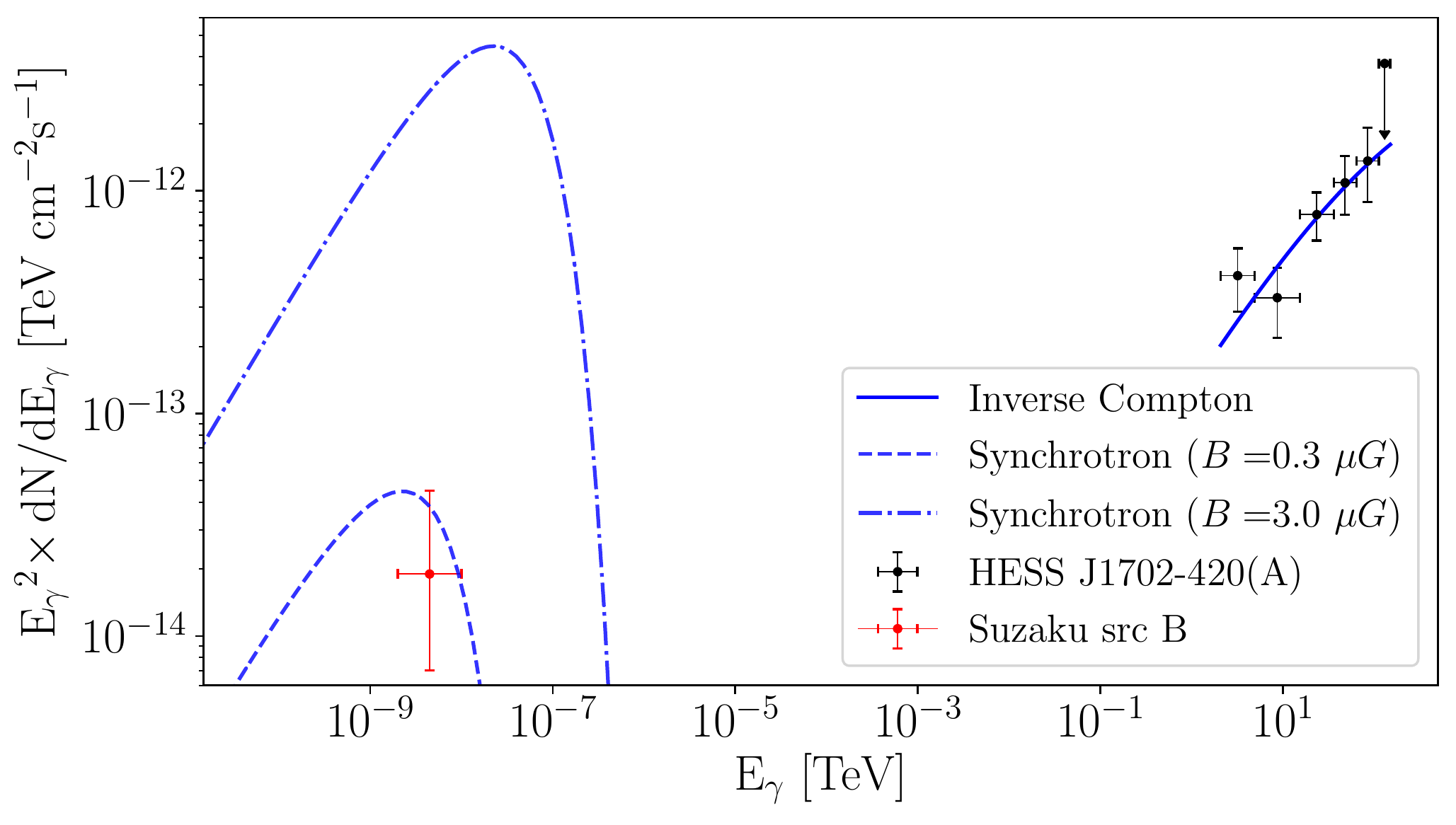}
\caption{multi wavelength modeling of \mbox{HESS J1702-420A}, under the assumption of a one-zone leptonic scenario powered by a power law distribution of electrons. The synchrotron emission was computed assuming a magnetic field value of 0.3 and $3\,\mu G$. The black error bars represent the VHE flux points of \mbox{HESS J1702-420A}, while the red one indicates the flux of the unidentified \emph{Suzaku} src B~\citep{fujinaga}. }
\label{fig:suzaku}
\end{figure}

\begin{table*}[ht]
\begin{center}
\begin{tabular}{||c|c|c||}
    \hline
    \text{\textbf{Near distance}}&$\mathbf{n_\text{\textbf{H}}}$&$\mathbf{W_p(E>1\,\text{\textbf{GeV}})}$\\
    \text{\textbf{[kpc]}}&$\boldsymbol{[\times100\,\text{\textbf{cm}}^{-3\,}]}$&\text{\textbf{[erg]}}\\
    \hline
0.25 & 1.8 & $7.9\,10^{45}$ \\
0.5 & 5 & $1.1\,10^{46}$ \\
1.6 & 1 & $5.8\,10^{47}$ \\
2.6 & 3.3 & $4.7\,10^{47}$ \\
4 & 1.4 & $2.6\,10^{48}$ \\
5.1 & 0.5 & $1.2\,10^{49}$ \\
5.7 & 0.4 & $1.9\,10^{49}$ \\
6 & 0.2 & $4.1\,10^{49}$ \\
    \hline
\end{tabular}
\end{center}
\caption{For each one of the molecular clouds on the line of sight of \mbox{HESS J1702-420}, we report the distance and density from~\citep{lau}, and the proton energetics that would be necessary to power the observed VHE emission of \mbox{HESS J1702-420B} in each case.}
\label{tab:clouds}
\end{table*}

\end{document}